\documentclass[reprint,aps,prb,amsmath,amssymb,dvipdfmx,twocolumn]{revtex4-2}
\usepackage{graphicx,bm,braket,color,hyperref,comment,physics,multirow}
\usepackage{fancybox}
\usepackage{empheq}
\usepackage{ulem}
\hypersetup{colorlinks=true,citecolor=blue,linkcolor=blue,urlcolor=blue}
\newcommand{\q}{{\bm q}}
\newcommand{\p}{{\bm p}}
\newcommand{\Q}{{\bm Q}}
\newcommand{\m}{{\bm m}}

\newcommand{\vecr}{{\bm r}}
\newcommand{\veck}{{\bm k}}

\newcommand{\dr}[1]{\multirow{2}{*}{{#1}}}

\begin{document}


\title{Triple-$\mathcal{Q}$ partial magnetic orders induced by quadrupolar interactions: \\Triforce order scenario for UNi$_4$B}

\author{Takayuki Ishitobi}
\author{Kazumasa Hattori}%
\affiliation{%
 Department of Physics, Tokyo Metropolitan University,\\ 1-1, Minami-osawa, Hachioji, Tokyo 192-0397, Japan
}%

\date{\today}

\begin{abstract}
We theoretically investigate possible symmetry-broken states in ${\rm UNi_4B}$, constructing a localized pseudo triplet crystalline-electric field model.  For a long time, its low-temperature symmetry-broken phase in ${\rm UNi_4B}$ has been considered to be a magnetic toroidal order forming atomic-scale vortices lattice with disordered sites at each center of the vortices.  However, recent observation of current-induced magnetizations offers a reinvestigation about the validity of this order parameter because of the contradiction in their anisotropy.  Our model takes into account the quadrupole degrees of freedom, whose importance is recently evidenced by the sound-velocity softening.  We find that the quadrupole moments play an important role in determining the magnetic structure in the ordered states.  For a wide range of parameter space, we obtain two triple-$\mathcal{Q}$ magnetic orders in our 36-site mean-field analysis: toroidal order and another one with the same number of disordered sites as in the toroidal order.  We name the latter  ``triforce'' order after its magnetic structure.  Importantly, the triforce order possesses exactly the same spin structure factor as the toroidal order does, while the phase factors in the superposition of the triple-$\mathcal{Q}$ structure are different.  We show that the triforce order is consistent with the observed current-induced magnetization when the realistic crystal structure of ${\rm UNi_4B}$ is taken into account.  We compare the predictions of the triforce order with the experimental data available at present in detail and also discuss possible applications of the present mechanism of triple-$\mathcal{Q}$ orders to anisotropic correlated systems.

\end{abstract}

\pacs{Valid PACS appear here}
\maketitle


\section{Introduction}
\label{sec:Introduction}
Antiferromagnetic (AFM) orders with noncollinear and noncoplanar spin configurations have attracted much attention, as they can lead to rich phenomena such as magnetoelectric (ME) effects \cite{LandauLifshitzContinuousMedia, Dzyaloshinskii1959, Astrov1960}, anomalous Hall effects \cite{Nagaosa2010}, and non-reciprocal transport \cite{Tokura2018}.  
Periodic topological spin textures, for example, skyrmion lattices \cite{Fert2017, Nagaosa2013, Rossler2006, Muhlbauer2009, Nayak2017, Kezsmarki2015, Tokunaga2015, Seki2012, Yu2010, Hayami2021_Itinerant} and hedgehog lattices \cite{Kanazawa2016, Fujishiro2019, Okumura2020, Aoyama2021}, can generate emergent electromagnetic fields \cite{Nagaosa2012_Emergent, Nagaosa2012_Gauge, Xiao2010, Berry1984, Volovik1987}, which result in anomalous transport phenomena such as topological Hall effects \cite{Ye1999, Bruno2004, Onoda2004, Binz2008, Nakazawa2019, Smejkal2022}.  Various non-collinear and non-coplanar AFM spin structures, including the topological ones, appear as multiple-$\mathcal{Q}$ orders, which are characterized by multiple modulation vectors $\mathcal{Q}$ \cite{Khanh2022}.  Multiple-$\mathcal{Q}$ orders possess phase degrees of freedom in their complex coefficients of the superposed waves.  The importance of the phase has been emphasized recently in a line of discussion about their stability and topology \cite{Shimizu2021_Moire, Shimizu2021_Chiral, Hayami2021_Phase, Hayami2021_Lock}.  
 
In addition to the multiple-$\mathcal{Q}$ spin texture, multipole degrees of freedom also exhibit multiple-$\mathcal{Q}$ orders \cite{Mannix1999, Caciuffo2003, Tokunaga2005, Tokunaga2006, Magnani2008, Kubo2005, Walker1994, McEwen1995, McEwen1998, Hanzawa2019}.  Multipole moments represent anisotropic charge and/or magnetic densities.  This makes their single-site properties and interactions anisotropic.  It has been pointed out recently that such anisotropy of multipole moments stabilizes the multiple-$\mathcal{Q}$ orders \cite{Kubo2005, Hanzawa2019, Tsunetsugu2021, Ishitobi2021, Hattori2022}.  When two or more types of multipole moments are active, couplings between them are in a nontrivial form in comparison to isotropic spin-spin couplings.  For example, a non-collinear multiple-$\mathcal{Q}$ magnetic order is realized inside an antiferro-quadrupole ordered phase due to the couplings between the magnetic dipole and electric quadrupole moments \cite{Yamauchi1999, Tanaka1999, Yamauchi2003, Chen2010, Hirai2020}.  The ordered quadrupole moments act as a site-dependent single-site anisotropy on the magnetic dipole moments and cause the non-collinear magnetic orders.  In the case of conventional AFM orders for large effective spin systems, quadrupole moments are induced and can affect the order of the transition \cite{Libero1993,De_Lima2000,Allen1968}.  This suggests that even induced quadrupole moments can play an important role in the phase transition. 

The concept of multipole has been utilized in inter-site multipoles beyond atomic degrees of freedom.  Such augmented multipoles include, e.g., cluster multipoles \cite{Ederer2007, Suzuki2017, Suzuki2019, Yanagi2023} and bond multipoles \cite{Shannon2006, Hikihara2008, Hayami2019, Hayami2020}.  The augmented multipoles are useful for understanding macroscopic symmetry in symmetry-broken phases and are directly related to possible responses under external fields \cite{Hayami2018, Watanabe2018, Yatsushiro2021, Oiwa2022}.  Among many responses, linear ME effects are one of the main subjects in both perspectives of fundamental and applied physics and have been studied particularly in multiferroic materials \cite{Khomskii2006, Khomskii2009}.  It has been considered that both time-reversal and spatial inversion symmetry must be broken to realize ME effects.  However, recently ME effects in metals \cite{Edelstein1990} have been reinvestigated \cite{Fujimoto2007, Yanase2014, Hayami2014_Parity, Thole2018} and summarized in systematic classifications \cite{Hayami2018, Watanabe2018}.  
In metals, ME effects can emerge even though the time-reversal symmetry is preserved.  Such ME effects in metals are called magneto-current (MC) effects, where electric currents break the time-reversal symmetry.

\begin{figure}[b]
\begin{center}
\includegraphics[width=0.48\textwidth]{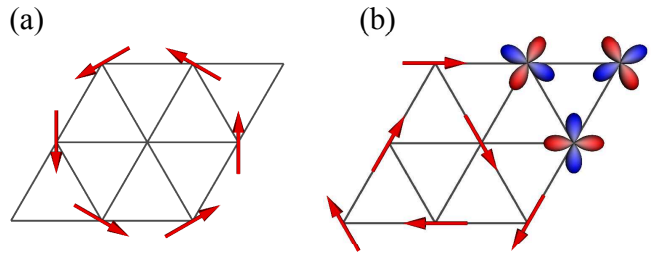}
\end{center}
\caption{Schematic configurations of (a) toroidal order and (b) triforce order proposed in this paper.  Arrows represent magnetic dipole moments.  Sites without an arrow are disordered without both magnetic and quadrupole moments in (a), while quadrupole moments emerge at magnetically disordered sites in (b). }
\label{fig:OrderedStates}
\end{figure}

 ${\rm UNi_4B}$ is one of the first candidates for bulk metals showing MC effects \cite{Saito2018, Hayami2014_Toroidal}, as is trigonal tellurium \cite{Furukawa2017}.  The U ions form an almost regular triangular lattice in ${\rm UNi_4B}$.  The magnetic moments of the U ions  antiferromagnetically order below $T_{\rm N} \sim 20$ K \cite{Mentink1994}.  From the neutron scattering experiment, a triple-$\mathcal{Q}$ magnetic order with one-third of U ions remaining paramagnetic has been proposed \cite{Mentink1994}.  In the triple-$\mathcal{Q}$ order, the ordered moments form vortices on the triangular $ab$ plane.  See Fig. \ref{fig:OrderedStates}(a), where the unit of the vortex is shown.  This vortex structure is equivalent to a ferroic ordering of toroidal moments parallel to the $c$ axis \cite{Hayami2014_Toroidal, Suzuki2019}.  In the toroidal ordered states, the in-plane transverse ME effect is theoretically predicted \cite{Hayami2014_Toroidal} and is indeed experimentally confirmed \cite{Saito2018}. However, the theoretical predictions and the observed ME effects are not completely consistent.  The experiment shows that not only in-plane but also out-of-plane currents induce the in-plane magnetization \cite{Saito2018}.  
  Remarkably, the in-plane and out-of-plane current-induced magnetization need different irreducible representations of $D_{6h}$ in the order parameter \cite{Hayami2018,Watanabe2018}. 
 This inconsistency suggests that the magnetic and/or crystal structures should be reconsidered.  Recent neutron and resonant x-ray scattering, and $^{11}$B-NMR experiments clarify that the crystal structure of ${\rm UNi_4B}$ is not hexagonal $P6/mmn$ (No. 191, $D_{6h}^{1}$) but orthogonal $Cmcm$ (No. 63, $D_{2h}^{17}$) \cite{Haga2008,Tabata2021,Willwater2021}.  Nevertheless, the toroidal order with the orthogonal distortion cannot explain the ME effects.  Thus, the magnetic structure should be reconsidered, and this is the main subject of this study. 
 Note that the in-plane and out-of-plane current-induced magnetization need different irreducible representations even in $D_{2h}$. This excludes orders with the $D_{2h}$ center, such as the toroidal order, except for accidental cases. This is a remarkable constraint on the magnetic structure.
  
  Recently, Yanagisawa {\it et al.} have reported a softening in the $C_{66}$ mode, and this does not stop even below $T_{\rm N}$ in their ultrasound experiment and proposed a crystalline electric field (CEF) model \cite{Yanagisawa2021}.  They have demonstrated that the softening stops at $T^{\ast}\simeq 0.3$ K, where the specific heat also shows a broad anomaly \cite{Movshovich1999}.  These observations indicate that the $E_{2g}$ quadrupole moments are active even in the ordered phase, and they gradually freeze via crossover or order at $\sim T^{\ast}$. 
The presence of the softening is inconsistent with the Kondo screening mechanism for the partial disorder in the early stage of the study about ${\rm UNi_4B}$ \cite{Lacroix1996}.  Thus, the effects of the quadrupole moments on the magnetic order, including partial one, are worthwhile to be considered.  We will clarify the interplay between the magnetic and the quadrupole moments in this paper. 
  
  In the present study, motivated by the CEF model proposed in Ref.~\onlinecite{Yanagisawa2021}, we investigate the effects of quadrupole degrees of freedom on the ordered magnetic structure in ${\rm UNi_4B}$.  A localized model with both magnetic dipole and electric quadrupole degrees of freedom is introduced and analyzed by means of a 36-site mean-field approximation.  The numerical results and symmetry-based arguments show that the quadrupolar interactions play a crucial role in determining the ordered magnetic structure.  Interestingly, we find that a triple-$\mathcal{Q}$ order shown in Fig. \ref{fig:OrderedStates}(b), which we call ``triforce order'' named after its unit cell structure \cite{[The ``triforce'' is a fictional symbol and icon of Nintendo's video games: {\it The legend of Zelda} series. The word ``triforce'' is used in the graph theory:~] Fox2020}, is more favorable than the toroidal order in many aspects observed in the experiments.  We compare the physical quantities in the triforce order with the experimental ones and propose several experiments which can semi-directly check the triforce order scenario. 
 
 This paper is organized as follows.  In Sec. \ref{sec:model}, we introduce the local CEF Hamiltonian with the multipole degrees of freedom at the U ions and the interactions between the magnetic and the quadrupole moments.  The Landau free energy characteristic of this system is also discussed.  In Sec. \ref{sec:results}, we analyze the model within the mean-field approximation and discuss its phase diagrams.  In Sec. \ref{sec:discussions}, we examine the triforce order as the order parameter for ${\rm UNi_4B}$ and discuss the existing experimental data.  Possible extensions of the present mechanism for triple-$\mathcal{Q}$ orders are also discussed.  Finally, Sec. \ref{sec:summary} summarizes this paper. Throughout this paper, we use the unit with the Boltzmann constant $k_{\rm B}=1$ and the Planck constant $\hbar=1$.

\section{Model}
\label{sec:model}
In this section, we will introduce a localized moment model on a triangular lattice, with the site point group symmetry $D_{6h}$ and the lattice constant set to unity. Here, we neglect the effect of the orthogonal distortion in our model calculations since it is small \cite{Saito2018, Tabata2021, Willwater2021}. The perturbative effects of the realistic crystal structure, such as the orthogonal distortion, will be discussed in Sec. \ref{sec:realcrys}.
The CEF scheme is based on the model derived in the recent ultrasonic experiments \cite{Yanagisawa2021}.  The interaction parameters are chosen in such a way that they reproduce the observed thermodynamic quantities.  We will discuss the Landau theoretical analysis and show the importance of dipole-quadrupole couplings for determining stable magnetic structures. 

\subsection{CEF scheme and multipole operators}
\label{subsec:CEF}
\begin{figure}[t]
\begin{center}
\includegraphics[width=0.48\textwidth]{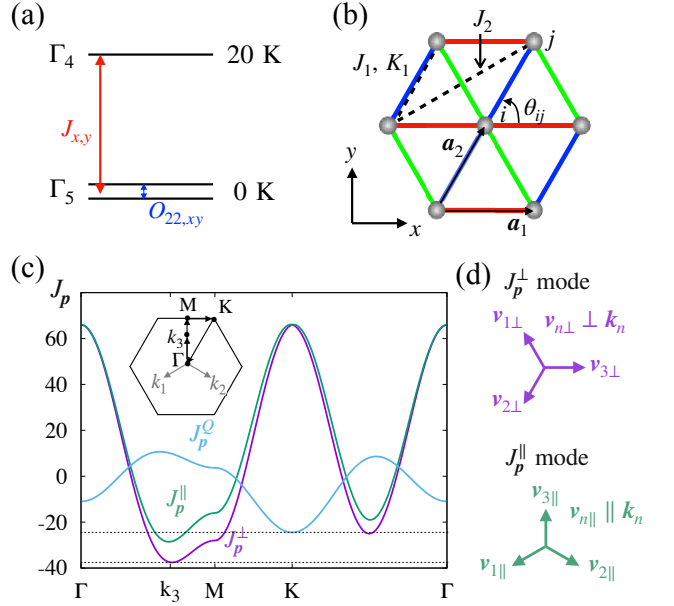}
\end{center}
\caption{(a) CEF level scheme used in this study.  Red and blue arrows represent the finite matrix elements for the magnetic dipoles $J_{x,y}$ and the electric quadrupoles $O_{22,xy}$, respectively.  (b) Exchange interactions $J_{1,2}^{M,Q}$ and $K_{1}^{M}$ are indicated along the corresponding bond, where the superscripts are omitted in the figure. The primitive translation vectors ${\bm a}_1$ and ${\bm a}_2$ are indicated by arrows, and the bond angle $\theta_{ij}$ relative to the $x$ axis is also defined.  (c) Eigenvalues of the exchange interactions for $(J_1^M,J_2^M,J_1^Q,J_2^Q,K_1^M)=(0,11,1.5,-3.33,-1.5)$ K along the high-symmetry lines shown in the inset.  $J_\p^{\parallel}$ ($J_\p^{\perp}$) is the  eigenvalue of the exchange $\hat{J}_\p^M$.  $J^{Q}_\p$ is that for the quadrupoles and degenerate owing to the isotropic nature of the exchange coupling constants.  (d) Eigenvectors at $\p={\bm k}_n$ ($n=1,2,3$) are parallel (perpendicular) to the wave vector $\p$ for $J_\p^{\parallel}$ ($J_\p^{\perp}$).  
}
\label{fig:CEF_Interactions}

\end{figure}

We first discuss the local states at the U ions.  Recently, Yanagisawa {\it et al.,} have carried out the ultrasonic experiment and proposed a CEF scheme \cite{Yanagisawa2021}.  
They claim that the valence of U ions is U$^{4+}$ ($5f^2$), and the atomic ground states are those for the total angular momentum $J=4$ with the nine-fold degeneracy.  They split into several CEF states.  The CEF ground state is a $\Gamma_5$ non-Kramers doublet, and the first excited state is a $\Gamma_4$ singlet with its excitation energy $E_4 \sim 20$ K.  The other states are separated more than 600 K above in the energy and safely ignored at low temperatures.  We take into account the $\Gamma_5$ and $\Gamma_4$ states forming a pseudo triplet as a minimal model.  In this pseudo triplet, three magnetic dipolar and five electric quadrupolar moments are active.  In the basis of $(\ket{\Gamma_{5+}}, \ket{\Gamma_{5-}}, \ket{\Gamma_4})^{\rm T}$, where $\pm$ in $\Gamma_{5\pm}$ represents the eigenvalues of the $z$ component of the total angular momentum $J_z$ and the superscript ${\rm T}$ represents the transpose, we define the operators for the dipole $\{J_x, J_y, J_z\}$, the $E_{2g}$ quadrupole $\{O_{22},O_{xy}\}$, and the $A_{1g}$ quadrupole $O_{20}$ as
\begin{align}
	O_{22} &=\bar{Q}\mqty(
	0 & 1 & 0 \\
	1 & 0 & 0 \\
	0 & 0 & 0 ), \ 
	O_{xy} =\bar{Q} \mqty(
		0 & -i & 0 \\
		i & 0 & 0 \\
		0 & 0 & 0 ), \label{eq:Ope_Q}\\
	J_x &=\frac{\bar{J}_{ab}}{\sqrt{2}} 
	\mqty(
		0 & 0 & 1 \\
		0 & 0 & 1 \\
		1 & 1 & 0 ), \ 
		J_y =\frac{\bar{J}_{ab}}{\sqrt{2}} 
	\mqty(
		0 & 0 & -i \\
		0 & 0 & i \\
		i & -i & 0 ), \label{eq:Ope_Jxy}\\
	J_z &= \bar{J}_c 
	\mqty(
		1 & 0 & 0 \\
		0 & -1 & 0 \\
		0 & 0 & 0 ) \label{eq:Ope_Jz}, \
	O_{20} = \bar{Q^\prime} 
	\mqty(
		-\frac{1}{2} & 0 & 0 \\
		0 & -\frac{1}{2} & 0 \\
		0 & 0 & 1 )
,\end{align}
where $\bar{Q}=5.27$, $\bar{J}_{ab}=2.34$, $\bar{J}_c=0.71$, and $\bar{Q^\prime}=0.831$ are determined by the detail of the CEF scheme proposed in Ref. \onlinecite{Yanagisawa2021}.  We use the Cartesian coordinates $x$, $y$, and $z$, which are parallel to $a$, $b$, and $c$ axes in the usual hexagonal notation, respectively. 
We note that the $xy$ plane is the magnetic easy plane, while the $z$ axis is the hard axis.  This is due to the difference between the magnitudes of the matrix elements $\bar{J}_{ab}>\bar{J}_c$.  Note that $J_x$ and $J_y$, which correspond to the primary order parameter in ${\rm UNi_4B}$, have their matrix elements between the ground doublet $\Gamma_{5\pm}$ and the excited singlet $\Gamma_4$, while $J_z$, $O_{22}$, and $O_{xy}$ are finite only in $\Gamma_{5\pm}$ as shown in Fig. \ref{fig:CEF_Interactions}(a).  
Hereafter, we focus on the in-plane components $J_{x,y}$ and $O_{22,xy}$, ignoring $J_z$ and other quadruple moments. 
This is a quite natural starting point to construct a minimal model for this system; the primary-order parameters $J_{x,y}$ in the magnetic sector and the ground-state components $O_{22,xy}$ in the quadrupole sector are retained. 
The $A_{1g}$ quadrupole $O_{20}$ simply represents the excitation energy from the $\Gamma_5$ to $\Gamma_4$.
We use normalized operators such as $\tilde{O}_{22}=O_{22}/\bar{Q}$, where the tilde will be omitted in the following.  This means $\bar{Q}$ and $\bar{J}_{ab}$ are included in the definition of interactions introduced in Sec.~\ref{subsec:Exchange}.

For later purposes, we introduce important wave vectors in this study. 
The important wave vectors in the first Brillouin zone are $\Gamma$: $\p=(0,0)\equiv\veck_0$, K: $k_{\rm K}(1/2,\sqrt{3}/2)\equiv\veck_{\rm K}$, $(k_{\rm K}/\sqrt{3})(-\sqrt{3}/2,-1/2)\equiv\bm{k}_1$, $(k_{\rm K}/\sqrt{3})(\sqrt{3}/2,-1/2)\equiv\bm{k}_2$, and $(k_{\rm K}/\sqrt{3})(0,1)\equiv \bm{k}_3$, where $k_{\rm K}\equiv 4\pi/3$. This comes from the fact that the ordering vectors of the magnetic moments in ${\rm UNi_4B}$ are $\bm{k}_{n}$ ($n=1,2,3$), which are parallel to $(-\sin \omega_n, \cos \omega_n)$ with $\omega_n=2n\pi/3$. In terms of the reciprocal lattice vectors $\bm{g}_1\equiv 2\pi(1,-1/\sqrt{3})$ and ${\bm g}_2\equiv 2\pi(0,2/\sqrt{3})$, $\veck_{\rm K}={\bm g}_1/3+2{\bm g}_2/3=(\frac{1}{3}~\frac{2}{3})$, ${\veck}_1=-{\bm g}_1/3-{\bm g}_2/3=(\bar{\frac{1}{3}}~\bar{\frac{1}{3}})$, $\veck_2={\bm g}_1/3=(\frac{1}{3}~0)$, and $\veck_3={\bm g}_2/3=(0~\frac{1}{3})$. 
Note that cubic mode-mode couplings are possible among these wave vectors. For example, $\veck_3-\veck_1=\veck_{\rm K}$ and also trivially $\veck_3-\veck_3=\veck_{0}$ hold. From these relations, one can expect that the quadrupole moments at the K and the $\Gamma$ points play a role in determining the magnetic structure through the cubic couplings, as will be discussed in Sec.~\ref{subsec:FreeEnergy}.  

\subsection{Exchange interactions}
\label{subsec:Exchange}
Here we consider minimal inter-site exchange interactions between $\bm{M}=(J_x,J_y)^{\rm T}$ and $\bm{Q}=(O_{22},-O_{xy})^{\rm T}$ in the triangular plane, implicitly assuming ferroic configuration along the $z$ axis. The interplane couplings do not play a major role in the phase transition in UNi$_4$B, and we neglect them for simplicity. We note that the most important term to realize the planar magnetic orders at the ordering vectors ${\bm k}_{n}~(n=1,2,3)$ in UNi$_4$B is not the nearest-neighbor magnetic coupling $J_1^M$ but the next-nearest one $J_2^M$. Magnetic interactions between the further neighbor sites do not play an important role in the discussion of the magnetic order in UNi$_4$B. They can be regarded as renormalization of $J_{1,2}^M$ in the expression of the magnetic susceptibility at the ordering vector $\veck_n$. 
The importance of $J_2^M$ is evident because $J_1^M$ causes the 120$^\circ$ order for $J_1^M>0$ or a trivial ferromagnetic one for $J_1^M<0$, but they are not realized in UNi$_4$B. This is also consistent with the N\'{e}el and the Curie-Weiss temperatures as discussed in Sec.~\ref{subsec:Parameters}. 
We also take into account simple isotropic quadrupole interactions $J_1^Q$ and $J_2^Q$ between the nearest-neighbor and the second-neighbor sites, respectively.

These four couplings consist of the main part in our minimal model for UNi$_4$B in this paper, and the exchange Hamiltonian reads as
\begin{align}
	H_{\rm int}&=\sum_{X=M,Q}\sum_{n=1,2}J_n^X\sum_{(i,j)_n} \bm{X}_i\cdot \bm{X}_j
		+H_{\rm ani}. \label{eq:Hint_minimal}
\end{align}
Here, $(i,j)_{1(2)}$ represents the summations for the nearest-neighbor (next-nearest-neighbor) pairs. 
See also Fig.~\ref{fig:CEF_Interactions}(b). Here the last term in  $H_{\rm ani}$ is introduced in order to make the magnetic moment at ${\bm k}_{n}$ parallel or perpendicular to ${\bm k}_{n}$ and given by the nearest-neighbor anisotropic coupling: 
\begin{align}
	H_{\rm ani}&=K_1^M\sum_{(i,j)_1 } ( M_{ij}^nM_{ji}^n - M_{ij}^tM_{ji}^t ), \label{eq:Hint_minimalK}
\end{align} 
where $M_{ij}^{n}=\bm{M}_i \cdot \bm{n}_{ij}$ and  $M_{ij}^{t}={\bm M}_i \cdot {\bm t}_{ij}$ are the projections to the bond-parallel and the bond-perpendicular directions with $\bm{n}_{ij}=(\cos \theta_{ij}, \sin \theta_{ij})^{\rm T}$, and 
$\bm{t}_{ij}=(-\sin \theta_{ij}, \cos \theta_{ij})^{\rm T}$, respectively, where $\theta_{ij}$ is the angle for the $i$-$j$ bonds as shown in Fig.~\ref{fig:CEF_Interactions}(b).  
This term does not play a major role in determining the phase transition but mainly controls the magnetic configurations. Although there might be many other coupling constants in the real UNi$_4$B, the model above is the simplest in the following senses. First, this consists of the shortest magnetic interactions which lead to the magnetic orders at ${\bm k}_n$ with the moment direction perpendicular to ${\bm k}_n$. Second, this consists of the simplest quadrupole interactions within the same range as those in the magnetic sector. Terms not present in Eq.~(\ref{eq:Hint_minimal}) can affect the results in this paper quantitatively, but deriving the exact interaction Hamiltonian is beyond the scope of this paper.

We now consider the eigenmodes of the interactions matrices $\hat{J}^M_\p$ and $\hat{J}^Q_\p$ in the Fourier space. By straightforward calculations, we obtain  
\begin{align}
H_{\rm int}=\sum_{\p} \Big[\bm{M}_{-\p} \cdot (\hat{J}_\p^M \bm{M}_{\p})+\bm{Q}_{-\p} \cdot(\hat{J}_\p^Q \bm{Q}_{\p}) \Big],\label{eq:Hint}
\end{align}
\begin{align}
 \hat{J}_\p^X &= 
 \begin{pmatrix}
 J_{\p}^{X,A_{1g}}+J_{\p}^{X,E_{2g,22}} & 	 J_{\p}^{X,E_{2g,xy}}\\
 J_{\p}^{X,E_{2g,xy}} & 	 J_{\p}^{X,A_{1g}}-J_{\p}^{X,E_{2g,22}}
 \end{pmatrix}. \label{eq:Jmat}
\end{align}
where, $X=M$ or $Q$.
See the detailed profile of $\hat{J}^X_\p$ shown in Appendix A. The form (\ref{eq:Jmat}) is common 
to any two-dimensional irreducible representations in $D_{6h}$ point group.

Figure \ref{fig:CEF_Interactions}(c) shows the eigenvalues of $\hat{J}^{X}_\p$ for a typical parameter set along the high-symmetry lines shown in the inset.  Let us concentrate on the magnetic part.  The two eigenvalues of $\hat{J}^M_\p$ are degenerate at the $\Gamma$ and the K points due to the presence of the $C_3$ rotational symmetry at these points.  Thus, the interactions there are isotropic, and the eigenvectors are arbitrary.  In contrast, $\hat{J}_\p^M$ at $\p=\bm{k}_{1,2,3}$ has two distinct eigenvalues $J_\p^\parallel$ and $J_\p^\perp$. The eigenvectors at $\p=\bm{k}_{1,2,3}$ are locked by the direction of $\p$.  One is parallel to $\bm{k}_n$, while the other is perpendicular: 
\begin{align}
	{\bm v}_{n\parallel}=\begin{pmatrix}
		-\sin\omega_n\\
		\cos\omega_n
	\end{pmatrix}, \ 
	{\bm v}_{n\perp}=\begin{pmatrix}
		\cos\omega_n\\
		\sin\omega_n
	\end{pmatrix}, \ {\rm with}\ \omega_n=\frac{2n\pi}{3}
\label{eq:v_n}
.\end{align} 
As we mentioned before, the anisotropic coupling $K_1^M$ controls the direction of the magnetic moment at $\p={\bm k}_n$; 
The eigenmode for the smaller eigenvalue is ${\bm v}_{n\parallel}$ (${\bm v}_{n\perp}$) for $K^M_1>0$ ($<0$).  Table \ref{table:Jq} summarizes the eigenvalues and eigenvectors at $\bm{p}= \veck_\Gamma$, $\veck_{\rm K}$, and ${\bm k}_n$.  The eigenvectors along the M-K line are locked by the directions of the nearest ${\bm k}_{1,2,3}$, not $\p$ itself, due to the mirror symmetry.

\begin{table}[t]
\begin{center}
\caption{Eigenvalues and eigenvectors of $\hat{J}^M_\p$.  The exchange parameters $J^M_{i}$ ($K_i^M$) are the $i$th neighbor isotropic (anisotropic) interactions for the magnetic dipole moments. The eigenvectors at $\bm{k}_n$ depend on the sign of $K^M_1$, while those at $\veck_0$ and $\veck_{\rm K}$ can be any linear combinations of the two degenerate eigenmodes. We thus denote ``deg.'' (degenerate) for $\veck_0$ and $\veck_{\rm K}$. Note that this list is applicable also to the quadrupole exchange interactions by $J_n^M\to J_n^Q$ etc.}
\label{table:Jq}
\begin{tabular*}{0.45\textwidth}{lccccc}
\hline \hline$\bm{p}$ & & eigenvalues & & \multicolumn{2}{c}{eigenvectors} \\
& & & & \hspace{3mm} $K^M_{1}<0$ \hspace{3mm}& \hspace{3mm} $K^M_{1}>0$ \hspace{3mm}\\
\hline  $\veck_\Gamma$ & & $6(J^M_1+J^M_2)$ & &  deg.  & deg. \\
 $\veck_{\rm K}$ & & $-3(J^M_1-2J^M_{2})$ & & deg. & deg.\\
 $\bm{k}_n$ & &  $-3J^M_2+ 3|K^M_1|$ & & $\bm{v}_{n\parallel}$  & $\bm{v}_{n\perp}$\\
   & &  $-3J^M_2- 3|K^M_1|$ & & $\bm{v}_{n\perp}$ & $\bm{v}_{n\parallel}$\\
\hline \hline
\end{tabular*}
\end{center}
\end{table}

\subsection{Parameters}
\label{subsec:Parameters}
Before discussing the properties of the model [Eq.~(\ref{eq:Hint_minimal})], we introduce constraints on the model parameters, $E_4$, $J^{M,Q}_{1,2}$, $K^M_{1}$, appropriate to ${\rm UNi_4B}$.  

First, the CEF excitation energy $E_4$ is set to $E_4=20$ K as proposed in Ref. \onlinecite{Yanagisawa2021}. We will use this value of $E_4$ throughout the present study. We note that the ordering wave vectors in ${\rm UNi_4B}$ are $\bm{k}_{n}(n=1,2,3)$, which are {\it not} at the high-symmetry points.  
We do not discuss the reason why the ordering vector is at ${\bm k}_{1,2,3}$ in detail here. We use this fact as a starting point of our analysis. A possible origin for this will be discussed in Sec. \ref{sec:discussions}, where we analyze the realistic crystal structure of ${\rm UNi_4B}$.  Within our model, the eigenvalues $J_{\bm k_{1,2,3}}^{\perp, \parallel}$ are not exactly at the extremum. Thus, the magnetic orders at $\p=\veck_n$ are considered to be realized by some commensurate locking. 
The physical origin for this is the realistic crystal structure of UNi$_4$B, as we mentioned above. Nevertheless,  
 the $J_{\bm k_{1,2,3}}^{\perp, \parallel}$ must be minimum among the values listed in Table~\ref{table:Jq}. These conditions lead to the conclusion that $J_1^M$ is smaller than $J_2^M$ in its magnitude, contrary to the naive expectation concerning their distance. 
 We consider this is not unphysical since this is indeed supported also by the following constraints (\ref{eq:TN}) and (\ref{eq:TCW_M}), which arise from the N\'{e}el and the Curie-Weiss temperatures observed in the experiments \cite{Mentink1994}.

Next, we discuss the constraints arising from the observed N\'{e}el temperature $T_{\rm N}=20$ K and Curie-Weiss temperatures $\theta^{M}_{\rm CW}=-65$ K estimated in the magnetic susceptibility measurement \cite{Mentink1994}, and 
$\theta^Q_{\rm CW}=11$ K in the ultrasonic experiment \cite{Yanagisawa2021}. In the mean-field approximations, the above three scales are related to the exchange interactions in the corresponding sectors:
\begin{align}
	-J^M_{\bm k_{1,2,3}}=3(J^M_2+|K^M_1|) \sim 1.87 \times T_{\rm N} &= 37.5~{\rm K}, \label{eq:TN} \\
	-J^M_{\Gamma} = -6(J^M_1+J^M_2) \sim \theta^M_{\rm CW} &= -65~{\rm K}, \label{eq:TCW_M} \\
	-J^Q_{\Gamma}=-6(J^Q_1+J^Q_2) \sim \theta^Q_{\rm CW} &= 11~{\rm K} \label{eq:TCW_Q}
.\end{align}
The numerical factor $1.87$ in Eq.~(\ref{eq:TN}) is introduced so that the magnetic transition temperature in the mean-field approximation is $\sim 20$ K. 

Finally, we discuss the anisotropic interactions.  The neutron scattering experiments \cite{Mentink1994} suggest that the ordered magnetic moment is perpendicular to the ordering wave vector $\p$, which means $K^M_1<0$ in our model.  

Under these constraints, two parameters remain undetermined, and we take $J_1^M$ and $J_1^Q$ as the control parameters. In the actual microscopic calculations in Sec.~\ref{sec:results}, it suffices that one only considers the small $J_1^M$ limit, and the results exhibiting the magnetic orders at $\veck_n$ can be understood by analyzing the $J_1^M\to 0$ limit. Thus, although the model itself contains several parameters, the practical parameter is indeed only $J_1^Q$, and this controls the effect of the quadrupole moments on the magnetic order as discussed in Sec.~\ref{sec:II-LandauF-Struc}.

\subsection{Coupling between dipole and quadrupole moments and Landau theory}
\label{subsec:FreeEnergy}
In this subsection, we will discuss Landau free energy for this system. The analysis here is important to understand the microscopic mean-field results in Sec.~\ref{sec:results}. We will demonstrate that third-order couplings between the magnetic dipole $\bm{M}$ and quadrupole moments $\Q$ are the key to the stability of magnetic orderings.  We will show that each magnetic order favors a specific third-order coupling consisting of fields at $\bm{p}= \veck_\Gamma$, $\veck_{\rm K}$, and ${\bm k}_n$. 
\subsubsection{Single-site Landau free energy}
Let us start by discussing the coupling between the dipoles and the quadrupoles.  We define mean fields acting on the magnetic dipoles ${\bm M}$ and the quadrupoles $\Q$, as $\bm{h}=(h_x,h_y)^{\rm T}$ and $\tilde{\bm{h}}=(\tilde{h}_{22},-\tilde{h}_{xy})^{\rm T}$, respectively.  In polar coordinates $\bm{h}=h(\cos \theta, \sin \theta)^{\rm T}$ and $\tilde{\bm{h}}=\tilde{h}(\cos \phi, \sin \phi)^{\rm T}$, the single site mean-field Hamiltonian is given 
in the basis of $(\ket{\Gamma_{5+}}, e^{i\phi}\ket{\Gamma_{5-}}, e^{i\theta}\ket{\Gamma_4})^{\rm T}$ as, 
\begin{align}
H_{{\rm MF}} &= 
	\mqty(
		0 & \tilde{h} & h \\
		\tilde{h} & 0 & he^{i(2\theta+\phi)} \\
		h & he^{-i(2\theta+\phi)} & E_4 )
\label{eq:MF}
.\end{align}
Here, the field-direction anisotropy arises in the form of $2\theta+\phi$. 
In the absence of the quadrupole interaction ($\tilde{h}=0$), $\phi$ is an arbitrary phase factor in the definition of $|\Gamma_{5-}\rangle$, and one can set $\phi=-2\theta$.  Thus, the eigenvalues of $H_{\rm MF}$ are independent on $\theta$, and the magnetic anisotropy vanishes.  In the presence of quadrupole interactions ($\tilde{h}\neq 0$), the eigenvalues of $H_{\rm MF}$ depend on $2\theta+\phi$. 
This indicates that the configuration of $\bm{Q}$ strongly affects that of $\bm{M}$. 
Note that this effect is important even when the primary order parameters are not ${\bm Q}$ but magnetic dipole moments ${\bm M}$. In the following, we will show that multiple-$\mathcal{Q}$ magnetic structures can be stabilized by this coupling.

To investigate the dipole-quadrupole coupling in more detail, we perform Landau expansion and obtain the effective free energy.  To avoid confusion between classical variables and quantum operators, we will use ``${\bm m}$'' and ``${\bm q}$'' instead of ``${\bm M}$'' and ``${\bm Q}$'' as the classical dipole and the quadrupole fields, respectively.  
First, there are trivial ``$\phi^4$'' terms in the free energy per site $F^{\rm loc}_{24}=F^{\rm loc}_{2m}+F^{\rm loc}_{4m}$ in the magnetic dipole sector as, 
\begin{align}
	F^{\rm loc}_{2\rm{m}}&=\frac{a}{2N}\sum_\vecr \sum_{\mu} m_\mu(\vecr)m_\mu(\vecr), \label{eq:F_loc2}\\
	F^{\rm loc}_{4\rm{m}}&=\frac{b}{4N}\sum_\vecr \left[\sum_{\mu} m_\mu(\vecr)m_\mu(\vecr)\right]^2, \label{eq:F_loc4}
\end{align}
where $N$ is the number of the sites in the triangular lattice.
Here, we have introduced the dipole field at the real space position $\bm{r}$: $m_{\mu}(\bm{r})$ ($\mu=x,y$), which corresponds to ${\bm M}$ in Eqs.~(\ref{eq:Hint_minimal}) and (\ref{eq:Hint_minimalK}).  We have ignored the intersite effects in the fourth-order terms since they are in general  irrelevant in the sense of renormalization group.  The third-order term per site in the free energy arising from the single-site CEF potential is  
\begin{align}
	F^{\rm loc}_3 = -\frac{c}{3N}\sum_\vecr
	\Big\{ &\left[m_x^2(\vecr)-m_y^2(\vecr)\right]q_{22}(\vecr)\nonumber\\
	&+2m_x(\vecr)m_y(\vecr)q_{xy}(\vecr) \Big\}
	\label{eq:F_loc3}
,\end{align}
where $q_{\nu}(\bm{r})$ ($\nu=22,xy$) is the quadrupole field, and $q_{22}$ ($q_{xy}$) corresponds to $O_{22}$ ($O_{xy}$). 
See Appendix \ref{sec:AppLocLandau} for the expression of the coefficient $c>0$ and the detail of the derivation.  
In the polar coordinate, 
\begin{align}
\bm{m}(\vecr)&=
\begin{bmatrix}
m_x(\vecr)\\
m_y(\vecr)
\end{bmatrix} \ \
=m(\vecr)
\begin{bmatrix}
\cos\theta(\vecr)\\
\sin\theta(\vecr)
\end{bmatrix}, \\ 
\bm{q}(\vecr)&=
\begin{bmatrix}
q_{22}(\vecr)\\
-q_{xy}(\vecr)
\end{bmatrix}
=q(\vecr)
\begin{bmatrix}
\cos\phi(\vecr)\\
\sin\phi(\vecr)
\end{bmatrix},\label{eq:polar}
\end{align}
$F^{\rm loc}_3$ reads 
\begin{align}
	F^{\rm loc}_3 &= -\frac{c}{3N}\sum_\vecr m^2(\vecr)q(\vecr)\cos[2\theta(\vecr)+\phi(\vecr)]\label{eq:def_m_q}
.\end{align}
The anisotropy arises in the form of $2\theta(\vecr)+\phi(\vecr)$ as expected from the mean-field Hamiltonian~(\ref{eq:MF}).  

\subsubsection{Landau free energy in momentum space} \label{subsec:LandauF}
Let us introduce Fourier transforms $m^{\mu}_\p$ defined as
\begin{align}
	m_\mu(\bm{r})&= \sum_\p m^\mu_\p e^{i\p\cdot \bm{r}},\  
	m^\mu_{\bm{p}}= \frac{1}{N}\sum_\vecr m_\mu(\vecr) e^{-i\p\cdot \bm{r}},
\label{eq:fourie}
\end{align}
and similar ones for $q^\nu_{\p}$.
Since they are real in the real space, $(m^\mu_\p)^*=m^\mu_{\bar{\p}}$ and  $(q^\nu_\p)^*=q^\nu_{\bar{\p}}$ with $\bar{\p}\equiv-\p$.  In the momentum-space representation, $F_3^{\rm loc}$ reads as
\begin{align}
	F^{\rm loc}_3 &= -\frac{c}{3}\sum_{\bm{G}}\sum_{\p,\p',\p''}f_3(\p,\p',\p'')
	\delta_{\p+\p'+\p'',\bm{G}},\label{eq:F3locFT}
	\end{align}
	where $\bm{G}$ is the reciprocal lattice vectors.
$f_3$ decomposes into  several terms reflecting different physical processes. Here, we are interested in those processes including the magnetic dipole fields at $\p=\veck_{1,2,3}$ since they correspond to the primary order parameters in this study. For later purpose, it is useful to introduce a simplified notation and the polar coordinate for $\p=\veck_{n}~(n=1,2,3)$ such that  
\begin{align}
{\bm m}_n&\equiv
\begin{pmatrix}
	m_n^x\\
	m_n^y
\end{pmatrix}=m_ne^{i\delta_n}
\begin{pmatrix}
	\cos \omega_n\\
	\sin \omega_n
\end{pmatrix}\ \ {\rm with}\ \omega_n=\frac{2n\pi}{3}.\label{eq:mn_def_FT}
\end{align}
We have introduced a common phase factor $e^{i\delta_n}$ for both $x$ and $y$ components with $0< \delta_n\le 2\pi$ and $m_n\ge 0$. $\omega_n$ is the angle variable corresponding to the eigenvector ${\bm v}_{n\perp}$ [Eq.~(\ref{eq:v_n})].  This choice of the mode is sufficient for our discussion below since $m_\p$ for $\p=\veck_{1,2,3}$ is the primary order parameter and the anisotropic interactions determine the unique eigenvector with the lower energy: $\bm{v}_{\perp}$ for $K_{1,3}^M<0$ (see Table~\ref{table:Jq}). 
 For $\p=-\veck_{1,2,3}$, ${\bm m}_{\bar{n}}=m_ne^{-i\delta_n}(\cos \omega_n, \sin \omega_n)^{\rm T}$.

In the following, we will write the third-order couplings consisting of $\bm{m}_n$ and those coupled with them. For notational simplicity, we use the abbreviations in such a way that the wave vector $\p$ is represented by a subscript $\gamma$ and the fields are expressed as $m_\gamma^\mu$ and $q_\gamma^\nu$. Here, $\gamma=1,2,3$, $\bar{1},\bar{2},\bar{3}$, K, K$^\prime$, and $0$ indicate ${\bm k}_{1,2,3}$, $-{\bm k}_{1,2,3}$, $\veck_{\rm K}$, $-\veck_{\rm K}$, and $\veck_0$, respectively. For the quadrupole fields, $q^\nu_{\veck_{1,2,3}}=q^\nu_{1,2,3}$, $q^\nu_{\veck_0}=q^\nu_{0}$, and  $q^\nu_{\veck_{\rm K}}=q^\nu_{\rm{K}}$.

There are four relevant processes in $f_3(\p,\p',\p'')$ including  the primary order parameters $m^\mu_{1,2,3}$ as 
	\begin{align}
	f_3(\p,\p',\p'') &=f_{3,123} + f_{3,{\rm K}} + f_{3,\Gamma} + f_{3,111}+\cdots. \label{eq:f3}
\end{align}
 By introducing ``quadrupole'' consisting of $m_\gamma^{\mu}$, 
$\mathcal{M}_{\gamma\gamma'}^{22}\equiv m_\gamma^x m_{\gamma'}^x-m_\gamma^y m_{\gamma'}^y$ and 
$\mathcal{M}_{\gamma\gamma'}^{xy}\equiv m_\gamma^x m_{\gamma'}^y+m_\gamma^y m_{\gamma'}^x$,
the four terms in Eq.~(\ref{eq:f3}) are given as 
	\begin{align}
	f_{3,123} &= 2( \mathcal{M}_{12}^{22}q^{22}_3 + \mathcal{M}_{12}^{xy}q^{xy}_3 + {\rm c.c.} ) + {\rm c.p.}, \label{eq:f3-123}\\
	f_{3,{\rm K}} &=  2( \mathcal{M}_{1\bar{3}}^{22}q^{22}_{\rm K} + \mathcal{M}_{1\bar{3}}^{xy}q^{xy}_{\rm K} + {\rm c.c.} ) + {\rm c.p.}, \\
	f_{3,\Gamma} &=  2(\mathcal{M}_{1\bar{1}}^{22}q^{22}_0 + \mathcal{M}_{1\bar{1}}^{xy}q^{xy}_0  ) + {\rm c.p.}, \label{eq:f3-11b0}\\
	f_{3,111} &=   (\mathcal{M}_{11}^{22}q^{22}_1 + \mathcal{M}_{11}^{xy}q^{xy}_1 + {\rm c.c.} ) + {\rm c.p.} \label{eq:f3-111}
\end{align}
  The abrreviation ``c.p.'' means cyclic permutations $123\to 231$ and $312$.
Equations (\ref{eq:f3-123})--(\ref{eq:f3-111}) represent mode-mode coupling processes among the primary dipole moments ${\bm m}_{1,2,3}$  and the quadrupole moments at $\veck_{1,2,3}$, $\veck_{\rm K}$, and $\veck_\Gamma$ with the quasi-momentum conservation.

Now, we derive the fourth-order renormalization by integrating out all the quadrupole fields. This can be done by taking into account the quadratic terms for the quadrupole fields ${\bm q}_\gamma\equiv (q_\gamma^{22},-q_{\gamma}^{xy})^{\rm T}$, 
\begin{align}
	F_2^{\rm Q} &= \frac{1}{2}\sum_{\p} \sum_{\nu\nu'} q^\nu_{-\p} a^{\rm Q}_{\p \nu\nu'} q^{\nu'}_{\p} 
	\label{eq:F2Q}
.\end{align}
The important terms in Eq.~(\ref{eq:F2Q}) are those for $\p=\veck_{1,2,3}, \veck_{\rm K}$, and $\veck_{\Gamma}$, since they are coupled with $\bm{m}_n$ in Eq.~(\ref{eq:F3locFT}). 
They are not primary order parameter and thus gapped. This allows us to regard $a_{\p \nu\nu'}^{\rm Q}$ as a diagonal matrix depending on $\p$ in the zeroth-order approximation. This means one can approximate $F_2^{\rm Q}$ as 
\begin{align}
	F_2^{\rm Q} &\simeq \frac{1}{2}a_0^{\rm Q}|{\bm q}_0|^2+a^{\rm Q}\sum_{n=1,2,3}|{\bm q}_n|^2+a_{\rm K}^{\rm Q}|{\bm q}_{\rm K}|^2+\cdots, \label{eq:F2Q2}
	\end{align}
with $a_0^{\rm Q}, a^{\rm Q}, a_{\rm K}^{\rm Q}>0$.

By minimizing $F_2^{\rm Q}+F_3^{\rm loc}$ in terms of ${\bm q}_0$, ${\bm q}_{1,2,3}$, and ${\bm q}_{\rm K}$, with keeping Eqs.~(\ref{eq:f3-123})--(\ref{eq:f3-111}) and Eq.~(\ref{eq:F2Q2}), and then substituting the stationary values $\bm{q}_\gamma=\bar{\bm{q}}_\gamma$ into  $F_2^{\rm Q}+F_3^{\rm loc}$,  the following fourth-order terms appear:
\begin{align}
	\delta F_{4{\rm m}}=
	-\Bigl\{ &\frac{4 c^2
	}{9 a^{\rm Q}}
 \Big[\frac{m_3^4}{4}+m_1^2 m_2^2+m_1 m_2 m_3^2\cos \left(\delta_{23}-\delta_{31}\right)
 \Big]\nonumber\\
+&\frac{4 c^2
	}{9 a_{\rm K}^{\rm Q}}
 \Big[m_1^2 m_2^2-m_1m_2m_3^2\cos(\delta_{23}-\delta_{31}) \Big]\nonumber\\
 +&\frac{2 c^2
	}{9 a_{0}^{\rm Q}}
 \Big[m_3^4-m_1^2 m_2^2 \Big] \Bigr\} +{\rm c.p.} 
 \label{eq:dF4}
 \end{align}
Here, we have introduced $\delta_{ij}\equiv \delta_i-\delta_j$, and the stationary values $\bar{\bm{q}}_\gamma$ are 
\begin{align}
	\bar{{\bm q}}_0&=\frac{c}{3a_0^{\rm Q}}\begin{pmatrix}
		2m_3^2-m_1^2-m_2^2\\
		\sqrt{3} (m_1^2-m_2^2)
	\end{pmatrix},\label{eq:Avq0}\\
  \bar{{\bm q}}_{\rm K}&=\frac{c}{3a_{\rm K}^{\rm Q}}\begin{pmatrix}
		2e^{i\delta_{12}}m_1m_2-e^{i\delta_{23}}m_2m_3-e^{i\delta_{31}}m_3m_1\\
		\sqrt{3}(e^{i\delta_{23}}m_2m_3-e^{i\delta_{31}}m_3m_1)
	\end{pmatrix},\label{eq:AvqK}\\
\bar{{\bm q}}_1&=\frac{c}{3a^{\rm Q}}\Big[2e^{-i(\delta_2+\delta_3)}m_2m_3+e^{-i2\delta_1}m_1^2\Big]
\begin{pmatrix}
		-\frac{1}{2}\\[1mm]
		\frac{\sqrt{3}}{2}
	\end{pmatrix},\label{eq:Avq1}\\
\bar{{\bm q}}_2&=\frac{c}{3a^{\rm Q}}\Big[2e^{-i(\delta_3+\delta_1)}m_3m_1+e^{-i2\delta_2}m_2^2\Big]
\begin{pmatrix}
		-\frac{1}{2}\\[1mm]
		-\frac{\sqrt{3}}{2}
	\end{pmatrix},\label{eq:Avq2}\\
\bar{{\bm q}}_3&=\frac{c}{3a^{\rm Q}}\Big[2e^{-i(\delta_{1}+\delta_2)}m_1m_2+e^{-i2\delta_{3}}m_3^2\Big]
\begin{pmatrix}
		1\\
		0
	\end{pmatrix}.	\label{eq:Avq3}
\end{align}
Note that the stationary directions of $\bm{q}_n$ are $\bar{\q}_n \parallel {\bm m}_n \parallel {\bm v}_{n\perp}$ for $n=1,2,3$.  See Eq.~(\ref{eq:v_n}) for the definition of $\bm{v}_{1,2,3\perp}$.

\subsubsection{Stability of triple-$\mathcal{Q}$ states}
\label{sec:II-LandauF-Struc}
We now discuss the effective free energy for the primary order parameters ${\bm m}_n~(n=1,2,3)$.  The Fourier transform of $F_{2,4{\rm m}}^{\rm loc}$ [Eqs. (\ref{eq:F_loc2}) and (\ref{eq:F_loc4})] consisting of ${\bm m}_n$ are given by
\begin{align}
	F_{2\rm m}^{\rm loc} &= \sum_{n=1,2,3}a^{\rm m} m_n^2	+\cdots,
	\label{eq:F2m}\\
		F^{\rm loc}_{4\rm m}
		=&\frac{b}{4}\sum_{\p_1,\p_2,\p_3,\bm{G}} \sum_{\mu,\mu'}m_{\p_1}^{\mu}m_{\p_2}^{\mu}m_{\p_3}^{\mu'}m_{\bm{G}-\p_1-\p_2-\p_3}^{\mu'} \label{eq:F4},\\
		=&\frac{3b}{2}(m_1^2+m_2^2+m_3^2)^2+ \cdots\label{eq:F4locFin}
,\end{align}
where $a^{\rm m}\equiv a+J_{\veck_n}^{\perp}$ with $J_{\veck_n}^{\perp}$ being $n$ independent and the ellipsis indicates terms including no $\bm{m}_n$.  For $T\sim T_{\rm N}$, the modes with smaller $F_{4{\rm m}}^{\rm loc}+\delta F_{4{\rm m}}$ realize. 

First, we calculate the free energy for a single-$\mathcal{Q}$ state. Let us set the ordering wave vector to $\p=\veck_3$ and define $m\equiv \sqrt{2}m_3$. The free energy reads as
\begin{align}
	F^{\rm single} &= \frac{1}{2}a^{\rm m}m^2 + \frac{1}{4} \qty[ \frac{3b}{2} - \frac{c^2}{9}\qty(\frac{2}{a^{\rm Q}_0}+\frac{1}{a^{\rm Q}} ) ]m^4
	\label{eq:F_single2}
.\end{align}
From Eqs.~(\ref{eq:Avq0})--(\ref{eq:Avq3}), the induced quadrupoles are 
\begin{align}
	\bar{\bm q}_0=\frac{cm^2}{3a_0^{\rm Q}}\begin{pmatrix}
		1\\
		0
	\end{pmatrix},\ 	
	\bar{\bm q}_{{\rm K},1,2}=\begin{pmatrix}
		0\\
		0
	\end{pmatrix},\ 
	\bar{\bm q}_3=\frac{cm^2}{6a^{\rm Q}}\begin{pmatrix}
		e^{-i2\delta_3}\\
		0
	\end{pmatrix}
	\label{eq:induced_q_single}
,\end{align}
where the phase factor $\delta_3$ is arbitrary.

Next, we examine triple-$\mathcal{Q}$ states. To capture essential points in the microscopic mean-field results shown in Sec.~\ref{sec:results},
 we concentrate on symmetric triple-$\mathcal{Q}$ states with $m_1=m_2=m_3$. 
These triple-$\mathcal{Q}$ states possess the $C_3$ rotational symmetry along the $c$ axis. We find two such solutions. See Appendix~\ref{sec:App_Minimization} for the detail of the derivations.  For $a^{\rm Q}<a^{\rm Q}_{\rm K}$, $\delta_{1,2,3}=\delta$ where $\delta$ is arbitrary and the free energy is given as
 \begin{align}
		F^{\text{triple-}(1)} &= \frac{1}{2}a^{\rm m}m^2 + \frac{1}{4} \qty[ \frac{3b}{2} - \frac{c^2}{3a^{\rm Q}} ]m^4
	\label{eq:F_toroidal2}
,\end{align}
where $m_{1,2,3}=m/\sqrt{6}$ has been used. These triple-$\mathcal{Q}$ configurations include the toroidal order shown in Fig.~\ref{fig:OrderedStates}(a), which realizes for $\delta=\pi/2$. As for the induced quadrupole moments, we obtain  
\begin{align}
	\bar{\bm q}_{0,{\rm K}}=\begin{pmatrix}
		0\\
		0
	\end{pmatrix},\ 
	\bar{\bm q}_n=\frac{cm^2}{6a^{\rm Q}}e^{-i2\delta}\begin{pmatrix}
		\cos\omega_n\\
		\sin\omega_n
	\end{pmatrix},
	\label{eq:induced_q_toroidal}
\end{align}
with $n=1,2,3$. 

\begin{figure}[t]
\begin{center}
\includegraphics[width=0.35\textwidth]{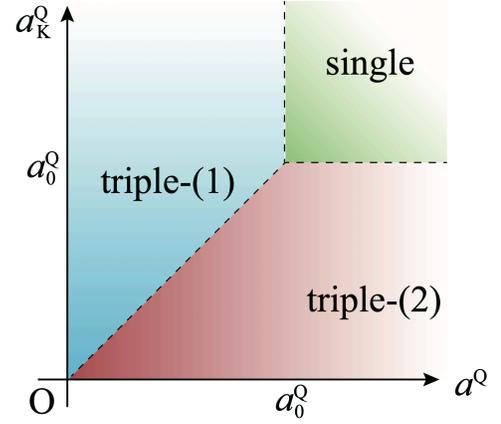}
\end{center}
\caption{Stable phases for $T\lesssim T_{\rm N}$ as functions of $a^{\rm Q}$ and $a^{\rm Q}_{\rm K}$. When microscopic parameters vary, $a_0^{\rm Q}$ changes in addition to the changes in $a^{\rm Q}$ and $a^{\rm Q}_{\rm K}$.}
\label{fig:PhaseLandau}

\end{figure}
For $a^{\rm Q}>a^{\rm Q}_{\rm K}$, triple-$\mathcal{Q}$ states with $(\delta_1,\delta_2,\delta_3)=(\delta,\delta,\delta\pm 2\pi/3)$ and the equivalent permutations for $\{123\}$ are realized, where $\delta$ is arbitrary.  See the discussion in Appendix \ref{sec:App_sixth}.  The free energy is given as
 \begin{align}
	F^{\text{triple-}(2)} &= \frac{1}{2}a^{\rm m}m^2 + \frac{1}{4} \qty[ \frac{3b}{2} - \frac{c^2}{9}
	\qty(\frac{2}{a^{\rm Q}_{\rm K}}+\frac{1}{a^{\rm Q}} ) ]m^4 
	\label{eq:F_triforce2}
.\end{align}
Again, $m_{1,2,3}=m/\sqrt{6}$ has been introduced. The induced quadrupole moments are 
\begin{align}
	\bar{\bm q}_{0}&=\begin{pmatrix}
		0\\
		0
	\end{pmatrix},\ 
	\bar{\bm q}_{\rm K}=\frac{cm^2}{6a^{\rm Q}_{\rm K}}\begin{pmatrix}
		1\\
		\mp i
	\end{pmatrix},\label{eq:q_phase_triforce1}\\
	\bar{\bm q}_{1(2)}&=\mp i\frac{cm^2e^{-i2\delta}}{\sqrt{3}a^{\rm Q}}
	\bm{v}_{1(2)\perp},\ 
	\bar{\bm q}_{3}=e^{\pm i\pi/6}\frac{cm^2e^{-i2\delta}}{\sqrt{3}a^{\rm Q}}\bm{v}_{3\perp}. \label{eq:q_phase_triforce2}
\end{align}
For the other domains, one can derive the expressions from Eqs.~(\ref{eq:Avq0})--(\ref{eq:Avq3}).  These triple-$\mathcal{Q}$ orders include the triforce order shown in Fig.~\ref{fig:OrderedStates}(b), which is realized for $\delta=0$.

Now, let us compare the three free-energies Eqs.~(\ref{eq:F_single2}), (\ref{eq:F_toroidal2}), and (\ref{eq:F_triforce2}), which are all conventional $\phi^4$ type.  Interestingly, the value of the local fourth-order term is  the same and given by $3bm^4/8$.  Thus, the lowest free energy solution is determined solely by the the magnitude of 
the fourth-order term in $m$ that arises from the third-order $m$-$q$ coupling in Eqs.~(\ref{eq:F_single2}), (\ref{eq:F_toroidal2}), and (\ref{eq:F_triforce2}), as long as we consider the solution near the second-order transition temperature at $a^{\rm m}=0$.  

We show which state among the three realizes at the transition temperature $T_{\rm N}$ as functions of $a^{\rm Q}$ and $a^{\rm Q}_{\rm K}$ in Fig.~\ref{fig:PhaseLandau}. It is easy to derive the phase boundaries from Eqs.~(\ref{eq:F_single2}), (\ref{eq:F_toroidal2}), and (\ref{eq:F_triforce2}): the single-$\mathcal{Q}$--triple-$\mathcal{Q}$(1) phase boundary along $a^{\rm Q}_0=a^{\rm Q}$, the single-$\mathcal{Q}$--triple-$\mathcal{Q}$(2) phase boundary along $a^{\rm Q}_0=a^{\rm Q}_{\rm K}$, and that between the two triple-$\mathcal{Q}$ along $a^{\rm Q}=a^{\rm Q}_{\rm K}$.  
These results show that the quadrupole interactions determine the magnetic structure at least near the second-order transition.  We will numerically examine these aspects in Sec.~\ref{sec:results}. We emphasize that the discussion in this section relies only on the phenomenological Landau free energy for ${\bm m}_n$, without assuming the microscopic exchange parameters $\hat{J}^M_{\bm p}$ and $\hat{J}^Q_{\bm p}$.

\section{Results}
\label{sec:results}
In this section, we will show the results of microscopic mean-field calculations.  We minimize the free energy numerically, assuming $6\times 6$ sites parallelogram magnetic unit cell in a triangular lattice.  First, we will show the phase diagram in temperature $T$ and the interaction $J^Q_1$ plane in Sec.~\ref{subsec:PhaseDiagram}.  Then, in Sec.~\ref{subsec:OrderedPhases}, the nature of each ordered state is explained.

\subsection{$T$--$J_1^Q$ phase diagram}
\label{subsec:PhaseDiagram}
\begin{figure}[t]
\begin{center}
\includegraphics[width=0.49\textwidth]{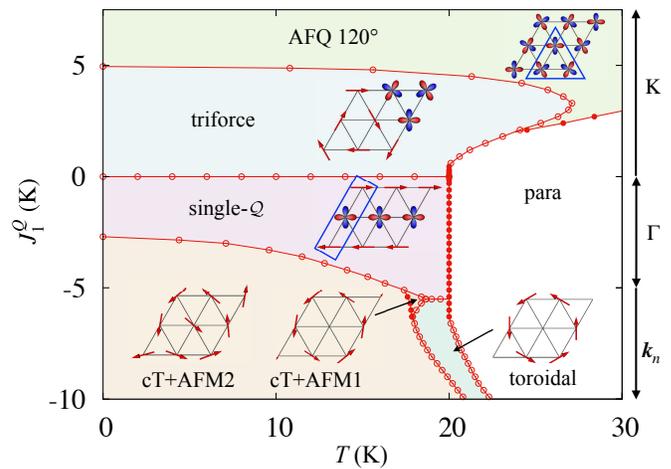}
\end{center}
\caption{$T$--$J_1^Q$ phase diagram for $J_1^M=0$ and $J_2^M=11$ K with constraints (\ref{eq:TN})-(\ref{eq:TCW_Q}).  The phase boundaries drawn by the filled circles represent second-order transitions, while the open circles mean first-order ones.  The wave vectors at which $J^Q_\p$ has minima in each region are indicated on the right.  For each phase, schematic configurations of magnetic dipole and electric quadrupole moments are illustrated.  The quadrupole moments at the sites with finite dipole moments are not shown for simplicity.  The blue triangular (rectangular) frame indicates the ordered unit cell for the single-$\mathcal{Q}$ (AFQ $120^\circ$) state.  For the others, the unit cell is $3\times 3$.}
\label{fig:PhaseDiagram}

\end{figure}
We have discussed in Sec.~\ref{sec:II-LandauF-Struc} that the third-order couplings between the magnetic dipole and the electric quadrupole moments play important roles in determining the stability of magnetic orderings.  The magnetic moments at $\bm{k}_{1,2,3}$ couple to the quadrupole moments at $\bm{k}_{1,2,3}$, $\veck_{\rm K}$, and $\veck_\Gamma$, via the third-order coupling Eq.~(\ref{eq:f3}).  In our setup described in Sec. \ref{subsec:Parameters}, there are two free parameters. Let us examine the cases for fixed $J_1^M$ and vary $J^Q_1$ with keeping the constraints (\ref{eq:TN})--(\ref{eq:TCW_Q}). The variations in $J^Q_1$ can control the effects of the quadrupole moments on the magnetic orders.  We will examine such effects arising from $J_1^Q$ on the phase diagrams in the following.

To make our presentation simple, let us concentrate on the case with a simple parameter set. 
Namely, we set $J_1^M=0$ since $J_1^M$ is not relevant to the appearance of the magnetic orders at $\p=\veck_n$. This simplification does not alter the qualitative aspects that will be shown in this section. 
The cases for finite $J_1^M$ and for other parameter sets without the experimental constraints are discussed in Appendix \ref{app:ResultsOther}. 
Figure \ref{fig:PhaseDiagram} shows the $T$--$J_1^Q$ phase diagram for $J_1^M=0$ under the constraints Eq.~(\ref{eq:TCW_Q}).  The ordered patterns of each phase and the unit cell smaller than nine sites (blue frame) are illustrated.  
Note that the minimum eigenvalue of $\hat{J}^Q_\p$, $J^Q_\p$, is at the K point for $J^Q_1>0$, at the $\Gamma$ point for $\-11/2<J_1^Q<0$, and at $\bm{k}_{1,2,3}$ for $J^Q_1<-11/2$ in the unit of Kelvin.  The horizontal phase boundaries between triforce $\leftrightarrow$ single-$\mathcal{Q}$ and single-$\mathcal{Q}$ $\leftrightarrow$ toroidal phases at high temperatures $\sim 20$ K correspond to the critical $J^Q_1$ at which the positions of the minimum in $J^Q_\p$ changes.  The detail of each phase will be explained in Sec. \ref{subsec:OrderedPhases}. 

The ``triforce'' phase [Fig.~\ref{fig:OrderedStates}(b)] is named after its magnetic structure \cite{Fox2020} and is stable in a wide region of $J^Q_1>0$. As shown in Fig. \ref{fig:PhaseDiagram}, the magnetic unit cell of the triforce order consists of six non-collinearly ordered magnetic sites and three quadrupole ordered ones.  We will discuss the detail of this phase in Sec. \ref{subsubsec:triforce}.  When $J^Q_1$ is larger, a quadrupole order is realized, which is labeled by AFQ $120^\circ$, the three-sublattice (A,B,C) $120^\circ$ structure of quadrupole moments: the angles of the sublattice quadrupole moments [Eq.~(\ref{eq:polar})] are $\phi_{\rm A}=0$, $\phi_{\rm B}=2\pi/3$, and $\phi_{\rm C}=4\pi/3$.  Such $120^\circ$ structure in triangular lattice systems is known to be stable for large antiferroic nearest-neighbor interactions \cite{Lee1984, Ramirez1994}.  The detail of AFQ $120^\circ$ phase will be discussed in Sec. \ref{subsubsec:AFQ}.  

 When $-11/2<J_1^Q<0$, a single-$\mathcal{Q}$ phase is favored.  Similar to the triforce phase, two-thirds of the sites are magnetically ordered, while there are finite quadrupole moments at the other one-third.   
 However, three differences from the triforce phase exist.  First, the unit cell for the single-$\mathcal{Q}$ order contains three sites, while that for the triforce phase does nine sites.  Second, the magnetic moments order collinearly, while those for the triforce phase are non-collinear.  Third, the quadrupole moments have large ferroic components.  The third point is the reason why this phase is favored when $J^Q_\p$ has a minimum at the $\Gamma$ point. 
 
  For $J^Q_1<-11/2$, a toroidal order is realized.  Similar to the triforce phase, the magnetic unit cell consists of six non-collinearly ordered magnetic sites and three disordered sites.  
Interestingly, the pure toroidal phase is unstable and replaced by another magnetically-ordered phase without magnetically-disordered sites at low temperatures.  This is in stark contrast to the cases for the larger $J_1^Q$, where the triforce and single-$\mathcal{Q}$ phases are stable even at zero temperature.  

In the triforce, the single-$\mathcal{Q}$, and the toroidal phases, one-third of the whole lattice sites are magnetically disordered.  When considering the stability against lowering $T$, the former two are stable, while the toroidal phase is unstable.  In the triforce and the single-$\mathcal{Q}$ phases, the quadrupole moments order at the magnetically disordered sites.  Thus, the two phases can be stable down to zero temperature, at least from the point of view of the entropy.  In the toroidal phase, however, the disordered sites are ``truly'' disordered without any ordered moments.  The local entropy at the disordered sites must be released by, e.g., another phase transition.  Although the second transition can be any orderings lifting the degeneracy at the disordered sites, magnetic orders are quite natural since the magnetic interaction between the disordered sites ($J^M_2=11$ K) is larger than that of quadrupolar one ($J^Q_2\sim 4$ K).  Indeed, several AFM orders at the disordered sites take place for  $J_1^Q<0$, as shown in Fig.~\ref{fig:PhaseDiagram}.  
Note that taking the large $J^M_2$ is the most direct and natural way to realize the ordering vector at ${\bm k}_{1,2,3}$.  In this sense, the toroidal order tends to be unstable since the bonds connected by $J_2^M$ contain the disordered sites.  In contrast, the triforce and single-$\mathcal{Q}$ phases can be stable since the magnetic interactions between magnetically disordered sites are $J^M_{1}$ and $K^M_{1}$, which are not necessarily large for the ordering vector at ${\bm k}_{1,2,3}$ realized. For sufficiently large $J_1^M>0$, 
a magnetic 120$^\circ$ structure is realized as expected. However, we note that as far as the ordering wave vectors are at the $\veck_{n}$, the three phases appearing in the phase diagram for $J_1^M=0$ are stable. The condition for realizing the magnetic 120$^\circ$ structure is $J_1^M>3J_2^M+|K^M_1|$ when one assumes the transition is continuous. In addition, a stripe order with ${\bm p}={\bm g}_1/2$ or the equivalent M points appears for $(J_2^M-|K^M_1|)/2<J_1^M<3J_2^M+|K^M_1|$. 
The detail of the $J_1^M$ dependence is discussed in Appendix \ref{sec:App_finite_J1M}.

\subsection{Properties of ordered phases}
\label{subsec:OrderedPhases}
In this subsection, we will discuss the detail of the ordered phases appearing in the phase diagram shown in Fig.~\ref{fig:PhaseDiagram}.  We will start by analyzing the triforce phase since this phase has many properties consistent with the experimental data, as will be discussed in the following and also in Sec. \ref{sec:discussions}.  Throughout this section, we will use $\bm{M}$ as the expectation value for the magnetic dipole moments and $\bm{Q}$ for the electric quadrupole moment to distinguish the quantities calculated in the microscopic mean-field calculations and the Landau theory in Sec.~\ref{subsec:FreeEnergy}, where we have used ${\bm m}$ and $\bm{q}$.

\subsubsection{Triforce order}
\label{subsubsec:triforce}
\begin{figure}[t]
\begin{center}
\includegraphics[width=0.48\textwidth]{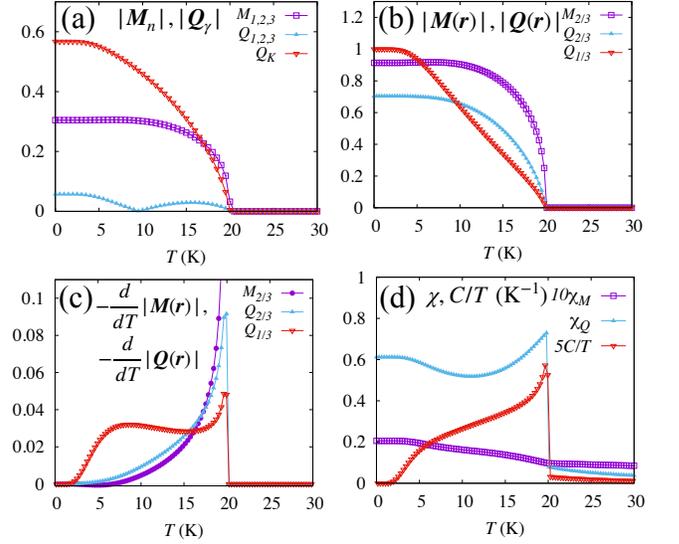}
\end{center}
\caption{$T$ dependence of the order parameters and the thermodynamic quantities. 
The interaction parameters are $(J_1^M,J_2^M,J_1^Q,J_2^Q,K_1^M)=(0,11,0.15,-1.98,-1.5)$ K. 
(a)  Order parameters $M_{1,2,3}$ and $Q_{1,2,3,{\rm K}}$ in the $\p$ space, and (b) those in the real space. The magnetic moments are finite only at $2N/3$ sites with the same magnitudes and are denoted by $M_{2/3}$.  At these $2N/3$ sites, the magnitudes of the quadrupole moments are also uniform and denoted by $Q_{2/3}$.  For the remaining $N/3$ sites, which are magnetically disordered, the magnitude of the quadrupole moments is uniform and labeled as $Q_{1/3}$.
(c)  $T$ derivative of the order parameters in the real space $dM_{2/3}/dT$ and $dQ_{2/3,1/3}/dT$.  (d) Magnetic (quadrupole) susceptibilities $\chi_M~(\chi_Q)$ and specific heat $C$ divided by $T$, $C/T$. }
\label{fig:triforce_OP_Chi}
\end{figure}

\begin{figure}[t]
\begin{center}
\includegraphics[width=0.45\textwidth]{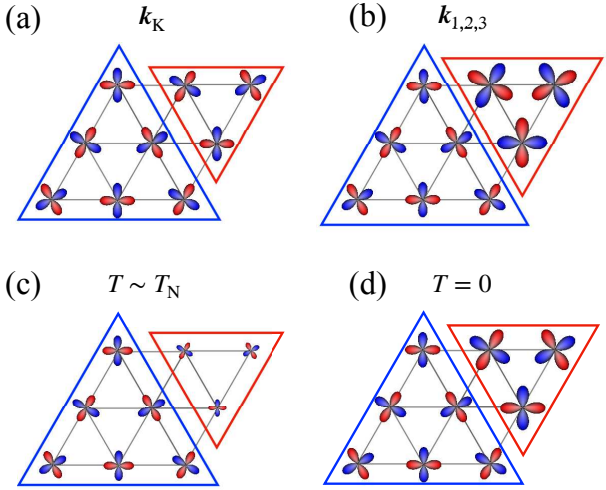}
\end{center}

\caption{Schematic profile of the quadrupole moments in the triforce phase.  (a) K point component ${\bm Q}_{\rm K}$, which corresponds to the $120^\circ$ structure of quadrupole moments and (b) $\bm{k}_{1,2,3}$ component ${\bm Q}_{1,2,3}$.  (c) Quadrupole configurations near the transition temperature $T_{\rm N}$ and (d) those at $T=0$ K.}
\label{fig:triforce_Quadrupole}

\end{figure}

First, we explain the magnetic and the quadrupole structure of the triforce order. The magnetic moment ${\bm M}(\bm{r})$ and the quadrupole one $\Q(\bm{r})$ at the position $\bm{r}$ in the triforce order are given by
\begin{align}
	{\bm M}(\bm{r}) &= \sum_{n=1,2,3} {\bm M}_{n} \cos({\bm k_n}\cdot \bm{r}+\delta_n), \label{eq:triforce_M}\\
	\Q(\bm{r}) &= \sum_{n=1,2,3} \Q_{n} \cos({\bm k_n}\cdot \bm{r}+\delta_n^{\prime}) \nonumber \\
	&+\Q_{\rm K}'\cos (\bm{k}_{\rm K}\cdot \bm{r}) + \Q_{\rm K}''\sin (\bm{k}_{\rm K}\cdot \bm{r})
	\label{eq:triforce_Q}
,\end{align}
where ${\bm \delta}\equiv\{\delta_1,\delta_2,\delta_3\}=\{0,0,2\pi/3\}$ and 
 { $\bm{\delta}'=\{-\pi/2,-\pi/2,\pi/6 \}$. }  These phase factors are consistent with the result in Sec.~\ref{sec:II-LandauF-Struc}.  The arbitrary phase factor $\delta$ in ${\bm \delta}=\{\delta,\delta,\delta+2\pi/3\}$ defined  above Eq.~(\ref{eq:F_triforce2}) is now fixed to $\delta=0$.  See Appendixes \ref{sec:App_triple2} and \ref{sec:App_sixth}.   
Here, ${\bm M}_{n}=M\bm{v}_{n\perp}$ and $\Q_{n}=Q\bm{v}_{n\perp}$ are perpendicular to $\bm{k}_{n}$ ($n=1,2,3$).   See Eq.~(\ref{eq:v_n}) for the definition of ${\bm v}_{1,2,3\perp}$.  Note that we take a convention that $M$ and $Q$ can take negative values in order to allow $\pi$ rotation of ${\bm M}_n$ and ${\bm Q}_n$.  Indeed, the sign of $Q$ changes as varying temperature, as will be discussed later and shown in Figs.~\ref{fig:triforce_OP_Chi}(a) and \ref{fig:triforce_Quadrupole}.  $\Q(\bm{r})$ includes the components at $\veck=\veck_{\rm K}$, $\Q_{\rm K}'=(Q_{\rm K},0)^{\rm T}$, and $\Q_{\rm K}''=(0,-Q_{\rm K})^{\rm T}$.  The Fourier modes ${\bm M}_{{1,2,3}}$ are exactly the same as those in the toroidal order [Eq.~(\ref{eq:Toroidal})].  The difference lies only on the phase factors; for the toroidal order $\delta_{1,2,3}=\pi/2$ [see Eq.~(\ref{eq:Toroidal})].  

\begin{figure}[t]
\begin{center}
\includegraphics[width=0.48\textwidth]{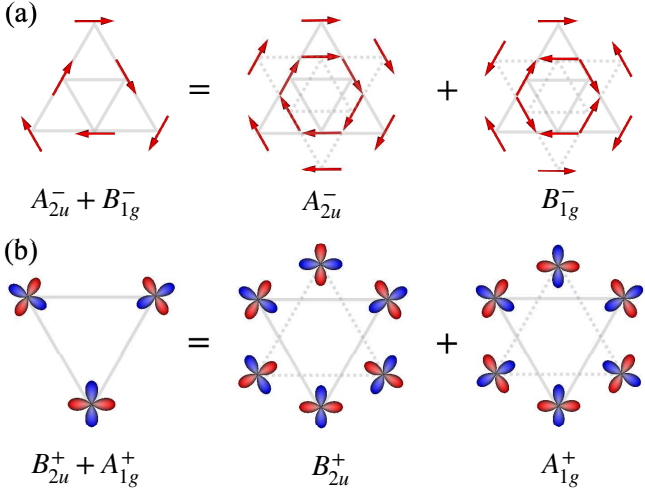}
\end{center}

\caption{Cluster multipole decomposition of (a) magnetic dipole moments and (b) electric quadrupole moments for the triforce order.  The moments at the vertices of the dotted triangles are virtual ones introduced in the cluster construction \cite{Suzuki2019}.}
\label{fig:ClusterMultipole}

\end{figure}

As illustrated in Fig. \ref{fig:PhaseDiagram}, the unit cell consists of an inverted triangle formed by the three nearest-neighbor sites, a larger triangle formed by the three third-nearest-neighbor sites, and a nearest-neighbor inverted triangle by the quadrupolar order.  Within each triangle, the magnetic or quadrupole moments form the $120^\circ$ structure.   
 We call it ``triforce'' order, named after the arrangement of the magnetic moments in the unit cell \cite{Fox2020}.

Next, we consider the symmetry of the triforce phase.  To this end, we use the cluster multipole decomposition, which is useful for the description of the global symmetry in a given ordered state \cite{Suzuki2019}. 
 We can choose the cluster center at a $C_3$ rotational symmetric point, which is the highest symmetry point.  There are two types of such $C_3$ symmetric points: the center of the nearest-neighbor magnetic triangle or that of the quadrupole triangle, and the choice does not affect the result for macroscopic symmetry.
 Figure \ref{fig:ClusterMultipole} shows the cluster multipole decomposition of (a) the magnetic moments and (b) the quadrupole moments in the triforce phase.  In Fig.~\ref{fig:ClusterMultipole}(b), the only quadrupole moments on the magnetically disordered sites are shown for simplicity.  Note that there are finite quadrupole moments also at the magnetically ordered sites.  The configuration of the magnetic moments is decomposed into $A_{2u}^-$ magnetic toroidal dipole and $B_{1g}^-$ magnetic octupole moments in the $D_{6h}$ symmetry.  Here, the superscripts ``$\pm$'' in the irreducible representations (irreps) describes the time-reversal parity, and the subscript ``$g$'' and  ``$u$'' for the spatial inversion parity as in the standard notation.  The configuration of the quadrupole moments consists of $A_{1g}^+$ electric monopole and $B_{2u}^+$ electric octupole.  They can be interpreted as induced moments: $(A_{2u}^-)^2,~(B_{1g}^-)^2=A_{1g}^+$, and $A_{2u}^-\otimes B_{1g}^-=B_{2u}^+$.  These moments are important when we discuss the experimental data in Sec. \ref{sec:discussions}.  Note that  
 the cluster multipole decomposition contains both even and odd parity components.  This is because Eqs. (\ref{eq:triforce_M}) and (\ref{eq:triforce_Q}) have both $\cos(\veck_n\cdot \vecr)$ and $\sin(\veck_n\cdot \vecr)$ parts irrespective of any choices of the origin taken.

We now discuss the temperature dependence of the order parameters and several thermodynamic quantities in the triforce phase. 
Figures \ref{fig:triforce_OP_Chi}(a)--\ref{fig:triforce_OP_Chi}(c) show temperature dependence of the order parameters for $J^Q_1=0.15$ K and the other parameters are the same as in Fig.~\ref{fig:PhaseDiagram}.  The amplitudes of the order parameters $M_n\equiv |\bm{M}_n|$ in the $\p$-space are shown in Fig.~\ref{fig:triforce_OP_Chi}(a).  There is a single second-order transition at $T_{\rm N}\sim 20$ K.  The magnetic dipole moments $M_{1,2,3}=M$ at $\p=\bm{k}_{1,2,3}$ are the primary order parameters, which are proportional to $(T_{\rm N}-T)^{1/2}$ below $T_{\rm N}$, while the quadrupoles $Q_{\rm K}$ at the K point and $Q_{1,2,3}=|Q|$ at $\p=\bm{k}_{1,2,3}$ are induced as the secondary order parameters, which are proportional to $T_{\rm N}-T$ below $T_{\rm N}$. These $T$ dependencies are the conventional mean-field type and consistent with the Landau analysis in Sec.~\ref{subsec:FreeEnergy}.  The primary dipole and the induced quadrupole moments at the K point increase monotonically as lowering $T$.  In contrast, the quadrupole $Q_{1,2,3}$ changes its sign at approximately $\sim10$ K as shown in Fig.~\ref{fig:triforce_OP_Chi}(a).

The reason for the sign change in $Q_{1,2,3}$ can be understood by illustrating the quadrupole moments for $\bm{p}=\veck_{\rm K}$ and $\bm{k}_{1,2,3}$ separately in the real space.  Figures \ref{fig:triforce_Quadrupole}(a) and \ref{fig:triforce_Quadrupole}(b) show the schematic configuration of each contribution.  In the triforce phase, $\Q_{\rm K}$ and $\Q_{1,2,3}$ contribute cooperatively at the six of nine sites (the larger triangle), while interference destructively at the three of nine sites (the smaller triangle).  Thus, the magnitudes of the quadrupole moments differ in the two groups.  At high temperature, two-thirds are larger, as shown in Fig. \ref{fig:triforce_Quadrupole}(c), since the quadrupole moments are directly induced by the on-site magnetic moments.  In contrast, at low temperature, the quadrupoles at one-third of the sites become larger, as shown in Fig.~\ref{fig:triforce_Quadrupole}(d), since their amplitudes should be their eigenvalues in the ground state at the non-magnetic sites.  In other words, this comes from a constraint of vanishing entropy at $T=0$.

Figure \ref{fig:triforce_OP_Chi}(b) shows the magnitudes of the order parameters in the real space.  We denote $|\bm{M}(\vecr)|$ and $|\Q(\vecr)|$ at the magnetically ordered sites by $M_{2/3}$ and $Q_{2/3}$ in Fig. \ref{fig:triforce_OP_Chi}(b), respectively. They increase as $T$ decreases in accord with the usual mean-field behavior.  In contrast, $|\Q(\vecr)|$  at the remaining one-third of the sites $(\equiv Q_{1/3})$ shows unusual behavior with slightly convex downward $T$ dependence in the intermediate temperature region.  There, $-dQ_{1/3}/dT$ has a peak at $\sim 7$ K [Fig. \ref{fig:triforce_OP_Chi}(c)].  This characteristic temperature dependence of the order parameter affects various physical quantities [Fig. \ref{fig:triforce_OP_Chi}(d)].  The uniform quadrupole susceptibility $\chi_Q$ and the specific heat coefficient $C/T$ have a shoulder at $\sim 5$ K, which reflect that the quadrupole moments at the non-magnetic sites begin to freeze at around 5 K.

To close this subsubsection, we discuss the susceptibilities shown in Fig. \ref{fig:triforce_OP_Chi}(d).  We note that the magnetic susceptibility increases even below $T_{\rm N}$ since the magnetically-disordered sites remain.  The isotropy in the susceptibility reflects the presence of the $C_3$ rotational symmetry in the triforce phase.  
The quadrupole susceptibility increases at low temperatures, which reflects the fact that the quadrupole moments are not frozen at one-third of the sites.  
Interestingly, the quadrupole susceptibility is discontinuous at $T_{\rm N}$.  
This is a general mean-field nature of susceptibility of the secondary order parameters $q$ \cite{Kubo2004, Hattori2014}.  Let us consider a minimal Ising-type Landau free energy with $m$ and $q$, 
\begin{align}
F = \frac{\alpha}{2}m^2 + \frac{\beta}{4}m^4 -(\tilde{h}+\gamma m^2)q + \frac{\delta}{2}q^2 
,\end{align}
where $\tilde{h}$ is the field that couples with $q$.  Here, $\alpha, \beta>0, \gamma>0$, and $\delta>0$ are coefficients.  By minimizing $F$ in terms of $q$ and $m$, we have
\begin{align}
q = \frac{\tilde{h}}{\delta} - \frac{\gamma \tilde{\alpha}}{\delta \tilde{\beta}} \theta(-\tilde{\alpha})
,\end{align}
where $\theta (-\tilde{\alpha})$ is the step function, $\tilde{\alpha}=\alpha-2\gamma \tilde{h}/\delta$, and $\tilde{\beta}=\beta-2\gamma^2/\delta$.  
One can easily find that $q$ is continuous but $\chi_Q \equiv \partial q/\partial \tilde{h}$ is discontinuous at the transition point $\tilde{\alpha}=0$ even for $\tilde{h}\rightarrow 0$. The explicit form is given by
\begin{align}
\chi_Q = \left\{ 
\begin{array}{ll}
\displaystyle \frac{1}{\delta} & (\alpha > 0)\\
\displaystyle \frac{\beta}{\delta \tilde{\beta}} & (\alpha < 0)
\end{array} 
\right.
.\end{align}
Replace $q$ by $Q_0$ and $\alpha$ by $T-T_{\rm N}$ for the triforce order.  The discontinuity in $\chi_Q$ at $T_{\rm N}$ is common to the other phases, although we will not show them in this study. 

\subsubsection{Toroidal order}
\label{subsubsec:Toroidal}
Historically, the toroidal order has been considered to be realized in UNi$_4$B \cite{Mentink1994, Hayami2014_Toroidal}. 
The toroidal order breaks the inversion symmetry, and thus, the order parameter is classified in the odd-parity cluster multipoles \cite{Hayami2018}.
In our model based on the CEF scheme proposed in Ref.~\onlinecite{Yanagisawa2021}, it appears as the high-temperature phase for $J_1^Q<0$ in the phase diagram  (Fig.~\ref{fig:PhaseDiagram}). Let us first discuss the structure of the toroidal order.  The pure toroidal structure at high temperatures is shown in Fig. \ref{fig:OrderedStates}(a) and represented as
\begin{align}
	{\bm M}(\bm{r}) &= \sum_{n=1,2,3} {\bm M}_{n} \sin({\bm k_n}\cdot \bm{r}), \label{eq:Mr_toroidal}\\
	\Q(\bm{r}) &= -\sum_{n=1,2,3} \Q_{n} \cos({\bm k_n}\cdot \bm{r})
	\label{eq:Toroidal}
,\end{align}
where ${\bm M}_{n}=M\bm{v}_{n\perp}$ and $\Q_{n}=Q\bm{v}_{n\perp}$, as predicted in Eq.~(\ref{eq:induced_q_toroidal}).  As mentioned in Sec. \ref{subsubsec:triforce}, ${\bm M}_{n}$ are exactly the same as those in the triforce phase.  We note that the phase factors 
$\delta_n$ in ${\bm M}(\bm{r}) = \sum_{n=1,2,3} {\bm M}_{n} \cos({\bm k_n}\cdot \bm{r}+\delta_n)$ cannot be determined in the mean-field approximation. We here fix $\delta_n=-\pi/2$ in Eq.~(\ref{eq:Mr_toroidal}), which corresponds to $A_{2u}^-$ toroidal dipole configuration shown in Fig.~\ref{fig:PhaseDiagram}. In the mean-field approximation, the phases $\delta_n$'s are arbitrary as long as $\delta_1=\delta_2=\delta_3$. This means that there exist other phases with the same free energy.  For example, an even parity $B_{1g}^-$ magnetic octupole state possesses the same free energy, which is written as ${\bm M}(\bm{r}) = \sum_{n=1,2,3} {\bm M}_{n} \cos({\bm k_n}\cdot \bm{r})$.  This accidental degeneracy is lifted by, e.g., six-fold local anisotropy proportional to $\cos (6\theta)$, which exists in general but not in the pseudo triplet model [Eq.~(\ref{eq:MF})]. See Appendix \ref{sec:App_sixth} for the related analysis. 

Since there are three disordered sites in the magnetic unit cell, further symmetry breakings take place at lower temperatures.   
In the vicinity of the single-$\mathcal{Q}$ phase, there is a small parameter region where magnetic moments emerge at the two of the three disordered sites, while the other site remains disordered.  This phase is labeled by cT+AFM1, where cT means ``canted toroidal''.  The ordered moments that emerge in this phase are anti-parallel with each other, and the magnetic moments are slightly modulated at the six sites forming a toroidal hexagon.  In this phase, there is a mirror symmetry, which interchanges the two sites where the magnetic moments emerge in cT+AFM1 indicated by the shorter arrows in Fig.~\ref{fig:PhaseDiagram}.  
 
 As $T$ decreases further,  magnetic moments appear at the remaining disordered sites as shown in Fig. \ref{fig:PhaseDiagram}.  This phase has finite ferromagnetic moments and no symmetry except for the simultaneous horizontal mirror and time-reversal operations.  We label this phase by cT+AFM2.  For the smaller $J_1^Q$, this phase transition takes place directly from the high-temperature pure toroidal phase.  Note that the transition in this case is of first order. 
 	Another route to this phase is from the single-$\mathcal{Q}$ phase through a first-order transition (see Fig.~\ref{fig:PhaseDiagram}).  
 
\begin{figure}[t]
\begin{center}
\includegraphics[width=0.48\textwidth]{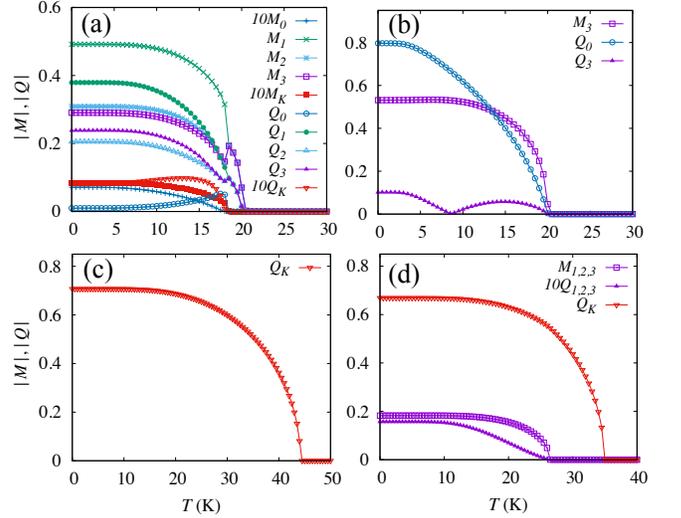}
\end{center}
\begin{flushleft} 
\caption{Temperature dependence of the order parameters in the $\p$-space for (a) $J_1^Q=-5.7$ K (toroidal), (b) $J_1^Q=-0.15$ K (single-$\mathcal{Q}$), (c) $J_1^Q=5.25$ K (AFQ $120^\circ$), and (d) $J_1^Q=3.75$ K (AFQ $120^\circ$ and triforce). The other parameters are the same as in Fig.~\ref{fig:PhaseDiagram}.}
\label{fig:OP_Others}
\end{flushleft} 
\end{figure}
 Figure \ref{fig:OP_Others}(a) shows the temperature dependence of the magnitudes of the order parameters in the $\p$ space for $J^Q_1=-5.7$ K and the other parameters are the same as in Fig.~\ref{fig:PhaseDiagram}. In the toroidal phase above $\sim 18.5$ K, the magnitudes of the magnetic moments $M_{1,2,3}$ take the same value, which reflects the $C_3$ rotational symmetry.  The first-order transition into cT+AFM1 breaks the $C_3$ symmetry and leads to $M_1>M_{2,3}$.  In the cT+AFM2 phase below $\sim 18$ K, the magnitudes of $M_{1,2,3}$ are all different, and finite $M_0$ and $M_{\rm K}$ emerge.  This reflects the low symmetry of this phase.

\subsubsection{Single-$\mathcal{Q}$ order}
\label{subsubsec:SQ}
Let us now focus on the single-$\mathcal{Q}$ phase appearing in the phase diagram shown in Fig. \ref{fig:PhaseDiagram}.  The ordered moments ${\bm M}(\bm{r})$ and $\Q(\bm{r})$ at the position $\bm{r}$ in the single-$\mathcal{Q}$ phase for the ordering vector e.g., $\p=\bm{k}_3$ are given by
\begin{align}
	{\bm M}(\bm{r}) &= {\bm M}_3 \sin(\bm{k}_3 \cdot \bm{r}), \label{eq:M_single}\\
	\Q(\bm{r}) &= -\Q_3 \cos(\bm{k}_3 \cdot \bm{r}) + \Q_0 \label{eq:Q_single}
,\end{align}
where ${\bm M}_3=(M_3, 0)^{\rm T}$, $\Q_3=(Q_3, 0)^{\rm T}$, and $\Q_0=(Q_0, 0)^{\rm T}$. 
The magnetic unit cell contains three sites. Collinear antiferromagnetic moments emerge at the two of the three sites, while the remaining site is non-magnetic. At low temperatures, the quadrupole moments at the non-magnetic sites grow.  

Let us comment on the symmetry. 
The symmetry of the single-$\mathcal{Q}$ state for $\p={\bm k}_3$ is $\sim yM_x$ when expressed in the real-space coordinate $(x,y,z)$ and the magnetic dipole $(M_x,M_y,M_z)$.  This is decomposed into two irreps, $A_{2u}^-:~xM_y-yM_x$ magnetic toroidal dipole and $E_{2u,xy}^-:~xM_y+yM_x$ magnetic quadrupole.  They induce the $E_{2g,22}$ electric quadrupole (${\bm Q}_0$) through the relation $A_{2u}^- \otimes E_{2u,xy}^-=E_{2g,22}^+$.

Figure \ref{fig:OP_Others} (b) shows the temperature dependence of $M_3$, $Q_3$, and $Q_0$ in the single-$\mathcal{Q}$ phase for $J_1^Q=-0.15$ K.  The primary order parameter is $M_3$, while $Q_0$ and $Q_{3}$ are induced as the secondary ones. As for the other domains for $\p=\veck_{1(2)}$, the primary order parameter is $M_{1(2)}$, and $Q_0$ and $Q_{1(2)}$ are induced.  
As in the triforce phase, the quadrupole moments at the magnetically disordered sites develop at low temperature, and this leads to increases in $Q_{0,3}$ down to $\sim3$ K.  One can also see the sign change in $Q_3$ as $T$ varies, which arises in a similar manner to the triforce phase.

\subsubsection{AFQ $120^\circ$ order}
\label{subsubsec:AFQ}
Finally, we briefly discuss AFQ $120^\circ$ phase realized for large $J^Q_1$ in Fig. \ref{fig:PhaseDiagram}.  
When $J_1^Q$ is large, the quadrupole interactions become dominant in the interaction energy, the pure quadrupole order is realized.  The ordered moment $\Q(\bm{r})$ at the position $\bm{r}$ in AFQ $120^\circ$ phase is given by 
\begin{align}
	\Q(\bm{r}) &= Q\begin{bmatrix}
		 \cos(\bm{k}_{\rm K} \cdot \bm{r}+\delta)\\
		 \sin(\bm{k}_{\rm K} \cdot \bm{r}+\delta)
	\end{bmatrix} 
,\end{align}
 where $Q$ is the magnitude of the quadrupole moment, and $\delta$ is an arbitrary phase factor.  This is a $120^\circ$ structure of quadrupole moments consisting of $Q_{\rm K}$ at $\p=\veck_{\rm K}$ and $\veck_{\rm K'}=-\veck_{\rm K}$.  Figure \ref{fig:OP_Others}(c) shows the temperature dependence of $Q_{\rm K}$ for $J_1^Q=5.25$ K.  The angle of the quadrupole moments can freely rotate as long as their relative angles are fixed at $120^\circ$, as in AFM Heisenberg magnets in the triangular lattice \cite{Lee1984,Ramirez1994}. 

In the intermediate region of $J_1^Q$, the phase transition from AFQ $120^\circ$ to the triforce order occurs as lowering $T$ as shown in Fig.~\ref{fig:OP_Others}(d) for $J_1^Q=3.75$ K.  
We note that, in this regime, $Q_{1,2,3}$ increases monotonically as lowering $T$ owing to the large $J_1^Q$. See the difference between the data in Figs.~\ref{fig:OP_Others}(d) and \ref{fig:triforce_OP_Chi}(a).

\section{Discussions}
\label{sec:discussions}
We have discussed that our model consisting of $\Gamma_4-\Gamma_5$ CEF states exhibits various triple-$\mathcal{Q}$ phases in addition to the single-$\mathcal{Q}$ ordered phases.  In this section, we will compare the theoretical results with the experimental data in UNi$_4$B in detail. Our main conclusion is that the triforce order is better in explaining the overall results in the experiments than the toroidal order. We review the experimental data of UNi$_4$B, focusing first on the neutron scattering in Sec.~\ref{sec:neutron}. 
Next, we will examine the impact of the realistic crystal structure in Sec.~\ref{sec:realcrys}.  This turns out to be quite important to explain the data of the current-induced magnetization in UNi$_4$B, which is discussed in Sec.~\ref{subsec:ME}. The triforce order in combination with the realistic crystal structure can explain the 
anisotropy in the current-induced magnetization in UNi$_4$B, while the others fail.
Thermodynamic properties are also discussed in Sec.~\ref{subsec:Comparison}. 
In Sec.~\ref{subsec:Prediction}, we will propose several experiments that can examine the triforce order scenario in UNi$_4$B.  Finally, in Sec.~\ref{subsec:Interest}, we will discuss possible theoretical extensions of the mechanism for the triple-$\mathcal{Q}$ magnetic order, which is triggered by the coupling with the quadrupole moments.

\subsection{Neutron scattering experiments}\label{sec:neutron}
First, we discuss the ordering wave vectors and magnetic moments in our results, comparing with those observed in the neutron scattering experiments \cite{Mentink1994, Willwater2021}.  There are clear magnetic Bragg peaks in the experimental data at $\veck=\veck_{1,2,3}$.  Thus, the AFQ $120^\circ$ phase is inconsistent with the experimental data.  In our calculations, there are mainly three magnetic ordered phases: triforce, toroidal, and single $\mathcal{Q}$.  The ordering wave vectors ${\bm k}_{n}$ and the magnetic moment ${\bm M}_n \parallel {\bm v}_{n\perp}$ are the same in the triforce and the toroidal orders, both of which agree with the neutron scattering experiments.  The single-$\mathcal{Q}$  order is also consistent when multiple domains of single-$\mathcal{Q}$ states are considered.  Note that analyses of spin structure factors in the neutron scattering experiments are not a powerful way to distinguish a multiple-$\mathcal{Q}$ state from multiple-domain states of single-$\mathcal{Q}$ orders for ${\bm p}={\bm k}_{1,2,3}$.  Although various moments at high-harmonic wave vectors can be induced in general, the magnetic part includes those at ${\bm k}_{1,2,3}$ for the present case because $2{\bm k}_{1,2,3}$ is equivalent to $-{\bm k}_{1,2,3}$.  This fact makes the analysis of the order parameter in ${\rm UNi_4B}$ nontrivial.  Thus, all the three states cannot be ruled out by the neutron scattering data.  To identify the magnetic order in UNi$_4$B, we need to examine other aspects of these phases.  

We should also comment about the observed weak reflections at $\veck=(\frac{h}{6}\frac{k}{6}0)=h\bm{g}_1/6+k\bm{g}_2/6$ in the paramagnetic phase \cite{Mentink1994, Tabata2021, Willwater2021}; The unit cell in the crystal with $3\times 4$ U sites and the magnetic unit cell with $3\times 3$ U sites as shown in Fig.~\ref{fig:OrderedStates} mismatch.  The interpretation of these results is discussed in Sec.~\ref{sec:realcrys}.

\subsection{Realistic crystal structure}
\label{sec:realcrys}
We here discuss how the realistic crystal structure of ${\rm UNi_4B}$ influences the ordered phases obtained in this study based on the regular triangular lattice model.  Recent experiments \cite{Haga2008, Tabata2021, Willwater2021} show that the space group symmetry of UNi$_4$B is $Cmcm$ (No.~63,\ $D^{17}_{2h}$) in the paramagnetic phase and there are two crystallographically distinct U sites.  Sites labeled by $8f$  form honeycomb structure, and those labeled by $4c$ lie in the center of the honeycomb hexagon \cite{Tabata2021, Willwater2021}.  In total, there are four types of U ions: those surrounded by 0, 2, 4, and 6 B atoms, which are $4c(1)$, $8f(1)$, $8f(2)$, and $4c(2)$ sites, respectively. 
Although the neutron data are also explained by the space group $Pmm2$ (No.~25 $C_{2v}^1$), we assume $Cmcm$ since there is no significant difference for discussing the magnetic structure \cite{Willwater2021}.  The inequivalence of the two sites leads to different CEF potential at $4c$ and $8f$ sites, which has been neglected in this study.  We will discuss two aspects expected when the CEF schemes are modulated differently at the $4c$ and $8f$ sites.

\subsubsection{Odd-parity moments}
First, we note that the $8f$ sites have no inversion symmetry.  This means that odd-parity multipole moments can be active at the $8f$ sites.  Our model is based on the assumption that the effects of this local inversion symmetry breaking are negligible, which is valid when the electrons at U ions are well localized.  If strong hybridizations between $f$ and $d$ or $s$ electrons are present, the effects owing to such odd-parity multipole moments become important \cite{Hayami2014_Toroidal}. 

The assumption of the weak anisotropy at the $8f$ sites is justified by analyzing the experimental results.  It is reported that the paramagnetic unit cell contains $3\times 4$ U ions \cite{Tabata2021, Willwater2021}, while the magnetic orders proposed so far consist of $3\times 3$ as in the triforce or toroidal orders.  Thus, when the unit cell in the ordered state is $3\times 3$, a mismatch between the magnetic and the crystal structure occurs.  For example, an identical magnetic moment is assumed even at the different $8f$ ($4c$) sites or at the same class of $8f$ sites with the different principal axis.  
This mismatch leads to a magnetic configuration with a longer modulation period.  However, the magnetic reflection of such a longer modulation is not reported \cite{Mentink1994,Willwater2021}, and the proposed magnetic structure has a $3\times 3$ periodicity.  In the latest experiment \cite{Willwater2021}, the magnetic unit cell has $3\times 6$ sites, but the proposed configuration is $3\times 3$ structure.  This indicates that the anisotropy at the $8f$ sites plays a minor role in determining the magnetic structure.

\subsubsection{Site-dependent CEF potential}
Second, we discuss real-space modulation in the CEF level schemes.  The CEF levels are different at the crystallographically different sites in general.  The CEF excitation gap $E_4$ at the two different $4c$ ($8f$) sites seem to be similar due to the above discussion about the small magnitudes of the longer period modulation. As for the difference in the CEF levels at the $4c$ and the $8f$ sites, it can be in general noticeable, although the difference cannot be estimated from the neutron data. The presence of site-dependent CEF levels can be a possible reason why the ordering vectors are at ${\bm k}_{1,2,3}$, which are not at the high-symmetry points for the triangular lattice model.  In the presence of site-dependent CEF, the unit cell contains three U sites if the difference between the two kinds of $4c$ ($8f$) sites is ignored: a $4c$ site and two $8f$ sites.  See also Fig.~\ref{fig:LocalField}(a).  The $3\times 3$ orders contain three such unit cells.  This corresponds to the ordering vector at the K point $\tilde{\bm k}_{\rm K}$ in the folded Brillouin zone reflecting the larger paramagnetic unit cell.  In the folded Brillouin zone, $\tilde{\bm k}_{\rm K}$ is at one of the high-symmetry points.  Thus, the model parameters do not necessarily need to be fine-tuned when assuming that the CEF and/or the interaction parameters are different at the $4c$ and the $8f$ sites.  

The ordered structure is also affected by the site-dependent CEF level. The most remarkable effect occurs when the CEF ground state is different at the $4c$ and the $8f$ sites.  For example, if the CEF ground state is $\Gamma_4$ singlet at the $4c$ sites, the toroidal order can be stable at low temperatures, at least from the viewpoint of the entropy.  However, this is inconsistent with the observed Curie-Weiss softening in the ultrasonic experiments, which suggests that $\Gamma_5$ doublet is the CEF ground state \cite{Yanagisawa2021}.  

\begin{figure}[t]
\begin{center}
\includegraphics[width=0.48\textwidth]{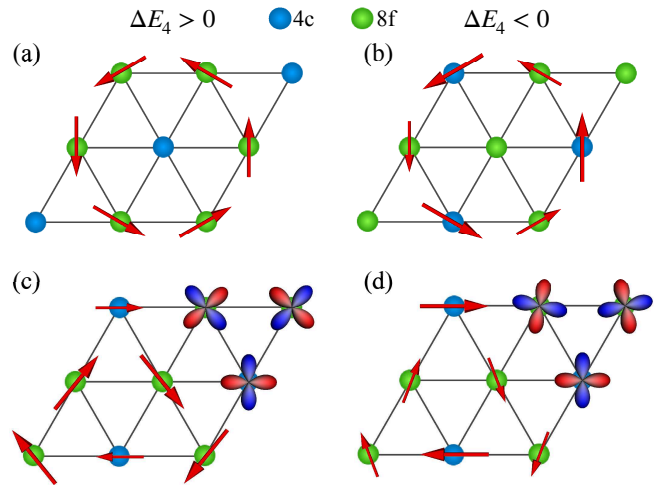}
\end{center}

\caption{Schematic configurations of the ordered structure in the presence of the CEF difference $\Delta E_4= E_4(4c)-E_4(8f)$ at $4c$ (sky blue) and $8f$ (green) sites.  Toroidal order for (a) $\Delta E_4>0$ and (b) $\Delta E_4<0$.  Triforce order for (c) $\Delta E_4>0$ and (d) $\Delta E_4<0$. }
\label{fig:CEF_4c_8f}
\end{figure}
When the CEF ground state is $\Gamma_5$ doublet at both of the $4c$ and the $8f$ sites, the difference in $E_4$ at these sites induces the modulation in the magnitudes of the magnetic moments.  Although a large $\Delta E_4 \equiv E_4(4c)-E_4(8f)$ may stabilize other orders, we focus on its perturbative effects.  Figure \ref{fig:CEF_4c_8f} shows the schematic illustrations of the order parameters in the presence of the CEF modulation.  When $\Delta E_4>0$ for the toroidal order, the magnetic moments order at the $8f$ sites, as shown in Fig.~\ref{fig:CEF_4c_8f}(a).  In the case that $\Delta E_4>0$ is much larger than the exchange interactions, the quadrupole order at the $4c$ sites is expected at low temperatures.  When $\Delta E_4<0$ for the toroidal order, the magnetic moments order at both of the $4c$ and the $8f$ sites, and their magnitudes are different, as shown in Fig.~\ref{fig:CEF_4c_8f}(b).  This state contains even-parity multipole moments when decomposed into irreps, and they have a similar symmetry to that in the triforce order;  the even parity component is $B_{2g}^-$ octupole, while that in the triforce order is $B_{1g}^-$ octupole.  The relation $A_{2u}^-\otimes B_{2g}^-=B_{1u}^+$ indicates that $B_{1u}^+$ electric octupole moments are induced.  

\begin{figure}[t]
\begin{center}
\includegraphics[width=0.47\textwidth]{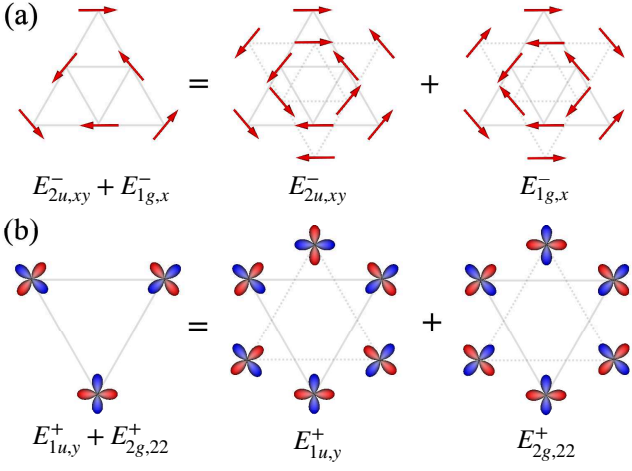}
\end{center}

\caption{Cluster multipole decomposition in the distorted triforce order: (a) magnetic dipole part and (b) electric quadrupole part.  The moments at the vertices of the dotted triangles are virtual ones.  Only site-dependent CEF- or distortion-induced components are shown. }
\label{fig:Distortion_Cluster}

\end{figure}
The triforce order for $\Delta E_4>0$ in Fig.~\ref{fig:CEF_4c_8f}(c) and $\Delta E_4<0$ in Fig.~\ref{fig:CEF_4c_8f}(d) break the $C_3$ rotational symmetry, while preserving the $x$-mirror symmetry.  We call them a canted triforce state hereafter.  The magnetic moments order at the four $8f$ and two $4c$ sites for both cases, and their magnitudes at the $8f$ sites are larger (smaller) than those at the $4c$ sites for $\Delta E_4>0$ ($\Delta E_4<0$).  
Let us discuss the symmetry of the canted triforce state.  For both cases of $\Delta E_4>0$ and $\Delta E_4<0$, the order parameter has the same symmetry.  
  Figure \ref{fig:Distortion_Cluster} shows the cluster multipole decomposition of the canted triforce state.  We show only the difference from the pure triforce state with $\Delta E_4=0$ for simplicity, i.e., ${\bm M}(\vecr)-{\bm M}(\vecr)|_{\Delta E_4=0}$ and ${\bm Q}(\vecr)-{\bm Q}(\vecr)|_{\Delta E_4=0}$.  The magnetic part in Fig.~\ref{fig:Distortion_Cluster}(a) is decomposed into $E_{2u}^-$ quadrupole and $E_{1g}^-$ dipole.  The electric part consists of $E_{1u}^+$ dipole and $E_{2g}^+$ quadrupole moments as shown in Fig.~\ref{fig:Distortion_Cluster}(b).  The presence of $E_{1g}^-$ magnetic dipole indicates that there is a finite magnetization, which has not been observed in the experiments.  The reason for the absence or smallness of the magnetization may be explained intuitively by the cancellation inside the cluster shown in Fig.~\ref{fig:Distortion_Cluster}(a).  Although the most natural moment for $E_{1g}^-$ is the magnetic dipole, the magnetization is almost canceled out in the inner and outer clusters in the right-hand side of Fig.~\ref{fig:Distortion_Cluster}(a).  Indeed, we have confirmed that the induced magnetization is small $\sim 10^{-4} \mu_{\rm B}$ ($\mu_{\rm B}$: the Bohr magneton) for $\Delta E_4=5$ K  and vanishes when the magnetic interactions are isotropic, i.e., $K^M_1=0$.  In contrast, the other components in Figs.~\ref{fig:Distortion_Cluster}(a) and \ref{fig:Distortion_Cluster}(b) do not show such cancellation.  We note that the presence of $E_{1u}^+$ electric dipole is important when we discuss the experiments of the magnetoelectric effects as discussed in Sec.~\ref{subsec:ME}.
  
\subsubsection{Macroscopic orthogonal distortion}
Lastly, we consider the effect of the macroscopic orthogonal distortion.  The space group $Cmcm$ does not possess hexagonal symmetry but orthogonal.  The distortion belongs to $E_{2g}^+$, and the orthogonal distortion-induced moments in the ordered states can be understood by the direct products of the irreps.  For the toroidal order, the distortion induces the $E_{2u}^-$ moments since $A_{2u}^- \otimes E_{2g}^+ =E_{2u}^-$.  When the site-dependent CEF is present, the modulated toroidal order with the distortion has the $E_{1g}^-$ and $E_{1u}^+$ components since $(B_{2g}^- \oplus B_{1u}^+) \otimes E_{2g}^+=E_{1g}^- \oplus E_{1u}^+$, where $B_{2g}^-$ and $B_{1u}^+$ are induced by the site-dependent CEF.  For the triforce order, they are obtained by $(A_{2u}^- \oplus B_{1g}^-) \otimes E_{2g}^+=E_{2u}^- \oplus E_{1g}^-$ for the magnetic part and $(B_{2u}^+ \oplus A_{1g}^+) \otimes E_{2g}^+=E_{1u}^+ \oplus E_{2g}^+$ for the electric part.  Note that these irreps are exactly the same as those induced by the site-dependent CEF (Fig.~\ref{fig:Distortion_Cluster}).  Thus, the macroscopic symmetry in the triforce order under the orthogonal distortion of the crystal structure is the same as that under the site-dependent CEF.  
In this sense, the orthogonal distortion induces the canted triforce order even without the site-dependent CEF levels.  For the single-$\mathcal{Q}$ state, a multi-domain structure is hardly expected, and the one with the lowest free energy realizes.  For example, the order at ${\bm k}_3$ with induced $E_{2g,22}$ moment realizes for the $\varepsilon_{22}$ type distortion.

We should comment on the degeneracy lifting of the susceptibility tensor by the orthogonal distortion.  The point group at the ordering vectors is the $C_{2v}$ ($D_{3h}$) with (without) the orthogonal distortion.  The degeneracies in the eigenmodes of the susceptibility tensor due to the $C_3$ rotational symmetry in the $D_{3h}$ are lifted by the orthogonal distortion.  In the point group $C_{2v}$, the orders with in-plane magnetic moments belong to one or both of two types of irreps: even under the $x$-mirror $B_1$ and odd $A_2$.  The pure toroidal, triforce, and single-$\mathcal{Q}$ orders are even under the $x$-mirror and belong to the $B_1$ representation, while the modulated toroidal order spanning the $4c$ and $8f$ sites is not an eigenstate of the $x$-mirror and belongs to a reducible representation.  This means that the phase transition from the paramagnetic phase to the modulated toroidal phase can occur only in accidental cases. 

We have demonstrated how the orthogonal distortion affects the symmetry of the ordered states.  Although the observed distortion in the lattice constants is tiny \cite{Tabata2021, Willwater2021}, the small but finite distortion breaks the $C_3$ rotational symmetry.  This must make the domains related by the $C_3$ symmetry inequivalent.  It is natural to consider that the component of the two-dimensional $E_{2g}^+$ irreps is fixed by the orthogonal distortion.

\subsubsection{Brief summary of Secs.~\ref{sec:neutron} and \ref{sec:realcrys}}
We now briefly summarize Secs.~\ref{sec:neutron} and \ref{sec:realcrys}, focusing on the difference between the triforce and troidal orders. First, the triforce and toroidal orders agree equally with the neutron data. When the realistic crystal structure is considered, the symmetry of the ordered phases is lowered.  The symmetry depends on the sign of the site-dependent CEF $\Delta E_4$ for the toroidal order, while not for the triforce order.  In the case that CEF at the $4c$ sites is large and unfavors the magnetic orders, the toroidal order can be stable. For the triforce order, the sign of $\Delta E_4$ does not affect the symmetry or stability of this phase as long as it is considered perturbative.
Finally, the single-$\mathcal{Q}$ order with multiple domain is unlikely in the realistic orthogonal crystal structure, since the orthogonal distortion selects one domain.

\subsection{Symmetry and magnetoelectric effects}
\label{subsec:ME}

We now carry out symmetry analyses on the current-induced magnetization (CIM) experiments and compare the experimental results with each theoretical one: triforce, toroidal, and single-$\mathcal{Q}$ phases.
This part is the most important result in this paper. 
   We describe the CIM response by the magnetoelectric (ME) coefficient $\alpha_{ij}$ defined by $M_i=\sum_j \alpha_{ij}E_j~(i,j=x,y,z)$, where $M_i$ and $E_i$ are the $i$th component of the magnetization and the electric field, respectively.
Note that $\alpha_{ij}$ is directly related to the symmetry of the order parameter below $T_{\rm N}$ \cite{Hayami2018, Watanabe2018}.
In UNi$_4$B, Saito {\it et al.}, reported that $\alpha_{yx}$ and $\alpha_{yz}$ are both finite below $T_{\rm N}$ \cite{Saito2018}.   
We will discuss possible order parameters consistent with this result.

First, we summarize the main conclusion.  
The magnetic space group under the triforce order is $Pm^{\prime}m2^{\prime}$ (No. 25.59) and consistent with the observation of the ME effects, and we consider it as the order parameter of ${\rm UNi_4B}$.  Other orders are inconsistent with the experiments: the toroidal and single-$\mathcal{Q}$ orders with $Pmm^{\prime}a$ (No. 51.292).  Although the toroidal order spanning the $8f$ and $4c$ sites with $Pm^{\prime}c2_1^{\prime}$ (No. 26.68) space group is consistent with the ME effect, we consider it is hardly realized as will be explained.  
In the following, we will discuss general symmetry arguments, focusing on the magnetic point groups and their representations, rather than the magnetic space groups, since the point group is sufficient to discuss the thermodynamic and transport phenomena.

%
%
\begin{table*}[thb]
\centering
\caption{
List of irreducible representations (irreps) and 
their direct product table for $D_{6}$ symmetry. $X_i$ and $X_{jk}$ represent the quantity $X=M$ (magnetic dipole or magnetic field), $E$ (electric dipole or electric field), $j$ (electric current), $T$ (magnetic toroidal dipole), where $i=x,y,z$, and $X=\varepsilon$ (lattice distortion or electric quadrupole) where $jk=20,22$, and $xy$ with the symmetry of $z^2$, $x^2-y^2$, and $xy$, respectively. Composite fields constructed by direct products of multiple irreps are also shown.  For highlighting which products generate magnetizations ($M_z: \ A^-_{2g}$ and $M_{x,y}: \ E^-_{1g}$) 
$A_2$'s and $E_1$'s are boxed in the table. 
}
\label{table:Rep}
	\begin{tabular}{ccccccccc}
		\hline \hline
irreps &fields, multipoles&$A_1$&$A_2$&$B_1$&$B_2$&$E_1$&$E_2$&composite fields, orders\\
		\hline
$A_1$&$\varepsilon_{20}$&$A_1$&\doublebox{$A_2$}&$B_1$&$B_2$&
\Ovalbox{$E_1$}&$E_2$&$xM_x+yM_y$, $\varepsilon_{22}\varepsilon^\prime_{22}+\varepsilon_{xy}\varepsilon^\prime_{xy}$\\
$A_2$&$M_z$, $E_z$, $j_z$, $T_z$&\doublebox{$A_2$}&$A_1$&$B_2$&$B_1$&\Ovalbox{$E_1$}&$E_2$& $xM_y-yM_x$, $\varepsilon_{22}\varepsilon^\prime_{xy}-\varepsilon_{xy}\varepsilon^\prime_{22}$\\
$B_1$& &$B_1$&$B_2$&$A_1$&\doublebox{$A_2$}&$E_2$&\Ovalbox{$E_1$}&$\varepsilon_{22}M_x-\varepsilon_{xy}M_y$, $x\varepsilon_{22}-y\varepsilon_{xy}$\\
$B_2$&&$B_2$&\doublebox{$A_2$}&$A_1$&$E_2$&\Ovalbox{$E_1$}&$E_2$&$\varepsilon_{22}M_y+\varepsilon_{xy}M_x$, $x\varepsilon_{xy}+y\varepsilon_{22}$\\
\dr{$E_1$}&$\{M_x,M_y\}$, $\{T_x,T_y\}$, &\dr{\Ovalbox{$E_1$}}&\dr{\Ovalbox{$E_1$}}&\dr{$E_2$}&\dr{$E_2$}&\dr{($A_1$, \doublebox{$A_{2}$}, $E_2$)}&\dr{($B_1$, $B_2$, $E_2$)}&$\{\varepsilon_{22}M_{x}+\varepsilon_{xy}M_{y}, \varepsilon_{22}M_{y}-\varepsilon_{xy}M_{x}\}$,\\ &$\{j_x, j_y\}$, $\{E_x, E_y\}$&&&&&&&$\{x\varepsilon_{22}+y\varepsilon_{xy}, x\varepsilon_{xy}-y\varepsilon_{22}\}$\\
\dr{$E_2$}&\dr{$\{\varepsilon_{22}, \varepsilon_{xy}\}$}&\dr{$E_2$}&\dr{$E_2$}&\dr{\Ovalbox{$E_1$}}&\dr{\Ovalbox{$E_1$}}&\dr{($B_1$, $B_2$, $E_2$)}&\dr{($A_1$, \doublebox{$A_2$}, $E_2$)}&$\{xM_x-yM_y, xM_y+yM_x\}$,\\ &&&&&&&&$\{\varepsilon_{22}\varepsilon^\prime_{22}-\varepsilon_{xy}\varepsilon^\prime_{xy}, \varepsilon_{22}\varepsilon^\prime_{xy}+\varepsilon_{xy}\varepsilon^\prime_{22}\}$\\[2mm]
\hline \hline
	\end{tabular}
\end{table*}
%
%

Table \ref{table:Rep} summarizes the irreps and their direct products for the $D_{6}$ symmetry.  The irreps for the $D_{6h}$ can be constructed from those in the $D_6$ with the inversion parity label: even ($g$) and odd ($u$) added appropriately.  
The observed $\alpha_{yx}$ implies that the order parameter possesses components of $A_{2u}\sim xM_y-yM_x$ or $E_{2u,xy}\sim xM_y+yM_x$ representations, while the finite $\alpha_{yz}$ responses indicate that there must be components of $E_{1u,x}\sim zM_y$ representation.  Note that the time-reversal parity of the order parameters can be either even ($+$) or odd ($-$) in the CIM measurements, since both electric-field-induced magnetizations by magnetic multipoles and current-induced magnetizations by electric multipoles are possible in metals \cite{Hayami2018, Watanabe2018}. For the current induced cases, one can just replace the coordinate $\{x,y\}$ by the current $\{j_x,j_y\}$: $A_{2u}\sim j_xM_y-j_yM_x$, $E_{2u,xy}\sim j_xM_y+j_yM_x$, and $E_{1u,x}\sim j_zM_y$. The choice of the time-reversal parity of the order parameter can be restricted when the candidate states are fixed from the physical ground as discussed below. 

 We here employ an assumption that the magnetic moments lie on the $xy$ ($ab$) plane, as reported by the neutron scattering experiments \cite{Mentink1994, Willwater2021}.  Under this assumption, the $E_{1u}$ part of the order parameter should be an electric $E_{1u}^+$, where $+$ represents the time-reversal parity even. 
 This is because the in-plane magnetic moments are odd under the $z$-mirror reflection $(x,y,z)\to (x,y,-z)$, while $E_{1u}$ is even under the $z$-mirror. This means that any magnetic configurations confined on the $xy$ plane are odd under the $z$-mirror operation.
 From this fact, one can conclude that the part of the order parameter with $E_{1u}^+$ representation is that of a secondary one induced by the magnetic order parameters.  In this case, the primary order parameter should contain at least one even-parity representation and one odd-parity component since their product includes an odd parity $E_{1u}^+$ representation.  As for the time-reversal parity, it is natural to assume that the finite $\alpha_{yx}$ arises from magnetic ones since if it were from non-magnetic ones, both even- and odd-parity components of the order parameters would be non-magnetic, and we consider this is unphysical in UNi$_4$B.
 
 Let us now examine possible irreps of the primary order parameters satisfying the above conditions.  Remember that, for realizing finite $\alpha_{yx}$, the order parameters must be $A_{2u}$ or $E_{2u}$.  First, consider a magnetic $A^-_{2u}$ irreps.  In Table~\ref{table:Rep}, in the horizontal row of $A_2$, there is only one $E_1$ irreps indicated by the single-line box, which represents $A_{2u}^-\otimes E_{1g}^-=E_{1u}^+$ (secondary order parameter). This means the order parameter must consist both of $A_{2u}^-$ and $E_{1g}^-$.  For the other choice, $E^-_{2u}$, one can see that there are two candidates $B_{1g}^-$ or $B_{2g}^-$ as indicated by the single-line boxes in the $E_2$ row in Table~\ref{table:Rep}. 
  
Interestingly, in-plane uniform magnetic moments should emerge in both cases.  For the first case with $\{A_{2u}^-,E_{1g}^-\}$, the part $E_{1g}^-$ 
	is classified as the same irreps as 
	the in-plane uniform magnetic moment $M_{x,y}$ as listed in Table~\ref{table:Rep}. Thus, it directly couples with $M_{x,y}$, and $M_{x,y}$ is induced in general.  For the other case with $\{E_{2u}^-,B_{1(2)g}^-\}$, the in-plane uniform magnetic moments are induced by the orthogonal distortion $\varepsilon_{22}$ with $E_{2g}^+$ irreps: $B_{1(2)g}^-\otimes E_{2g}^+=E_{1g}^-$.  Although the in-plane uniform magnetic moment $M_{x,y}$ has not been observed, it must be present from the viewpoint of the symmetry for any in-plane magnetic order parameter with orthogonal distortion.  It might be tiny due to weak couplings with the order parameters or the small distortions.  In principle, it is possible to consider that order parameters with finite magnetic moments along the $z$ direction or those not uniformly stacked in the $z$ direction.  However, as discussed in this section, their realization is not physically sound by observing the experimental data so far. 

Bearing the above symmetry argument in mind, we discuss possible candidates in our theoretical results.  The canted triforce order, which is induced by the site-dependent CEF or the orthogonal distortion, is the only candidate that is qualitatively consistent with both the neutron and the CIM results.  The triforce order contains $A_{2u}^-$, $B_{1g}^-$, and $B_{2u}^+$ irreps and additionally $E_{2u}^-$, $E_{1g}^-$, and $E_{1u}^+$ ones under the canting due to the site-dependent CEF or the orthogonal distortion, as discussed in Sec.~\ref{subsubsec:triforce}.  The presence of the $A_{2u}^-$ and $E_{1u}^+$ irreps agrees with the observed $a$- and $c$-axes CIMs, respectively.  The absence or smallness of the magnetization can be explained as a result of the cancellation shown in Fig.~\ref{fig:Distortion_Cluster}(a) with keeping the consistency with the ME effects. 
The triforce order includes several irreps in $D_{6h}$ for the highest-symmetry point at the U sites. This is because the highest-symmetry point in the triforce phase is not at the U site but at the center of the nearest-neighbor triangle with $D_{3h}$ symmetry. In the reduction $D_{6h} \downarrow D_{3h}$, $B_{2u}^+ \rightarrow A_1^{\prime +}$ and $A_{2u}^-,B_{1g}^- \rightarrow A_2^{\prime \prime -}$, where $A_1^{\prime +}$ is the totally symmetric representation. Thus, the triforce order has a single irrep $A_2^{\prime \prime -}$ other than the totally symmetric $A_1^{\prime +}$ in $D_{3h}$. When the orthogonal distortion is considered, the local symmetry at the center of the nearest-neighbor triangle is $C_{2v}$. In the reduction $D_{6h} \downarrow C_{2v}$, $B_{2u}^+, E_{2g,22}^+, E_{1u,y}^+ \rightarrow A_1^+$ and $A_{2u}^-,B_{1g}^-,E_{2u,xy}^-, E_{1g,x}^- \rightarrow B_1^-$, where $A_1^+$ is the totally symmetric representation. Again, the canted triforce order consists of a single irrep $B_1^-$ in addition to the trivial $A_1^+$ in $C_{2v}$.  In this sense, the canted triforce order is the simplest state consistent with the observed CIM.

Here, we discuss the detail of the two dimensional $E_{1u}^+$ representation, which corresponds to electric polarizations $\{E_{1u,x}^+,E_{1u,y}^+\}\sim\{x,y\}$.  As shown in Fig.~\ref{fig:Distortion_Cluster}, the induced component of $E_{1u}^+$ representation in the canted triforce order is that of $E_{1u,y}^+\sim y$.  
This is because the $B_{2u}^+\sim x\varepsilon_{xy}+y\varepsilon_{22}$ octupole moment in the triforce order couples to the distortion $\varepsilon_{22}$ with the coefficient proportional to $y$.  And we take a domain in which $\varepsilon_{22}\sim x^2-y^2$ is finite with $\varepsilon_{xy}=0$.  Although the $C_3$ rotated domains, $\sim \pm \sqrt{3}x/2-y/2$, can realize without the orthogonal distortion, polarization parallel or anti-parallel to $y$ is realized in the presence of the distortion $\varepsilon_{22}$.  The $E_{1u,y}^+$ representation has the same symmetry as $j_zM_x$ and $j_xM_z$ corresponding to $\alpha_{xz}$ and $\alpha_{zx}$.  One may consider that this is inconsistent with the experimental results with $\alpha_{yz}\neq 0$.  We emphasize that this actually agrees with the canted triforce order.  In Ref.~\onlinecite{Saito2018}, the analysis is based on the hexagonal structure.  Thus, three conventions of the axis in the $ab$ plane ($xy$ plane) exist.  Here, a trivial inversion $a,b\to -a,-b$ has not been counted.  In a single crystal with in-plane orthogonal distortions, there is one unique set of axis in the $ab$ plane.  Our results are consistent with the finite $\alpha_{yz}$ if the axis taken in Ref.~\onlinecite{Saito2018} coincide with those rotated by $\pm 60^\circ$ from ours.  

For the other symmetry-broken phases in our results, the toroidal or the single-$\mathcal{Q}$ phases are magnetic and occupy a wide region of the parameter space as the triforce phase does, as shown in the phase diagram in Fig.~\ref{fig:PhaseDiagram}.  However, the symmetry of the two phases is inconsistent with the observed CIM.  First, the toroidal order contains $A_{2u}^-$ part in its magnetic structure.  The presence of the $A_{2u}^-$ irreps is consistent with $\alpha_{yx}\ne 0$ but cannot explain $\alpha_{yz} \ne 0$.  Even when the orthogonal distortion $\varepsilon_{22}$ is taken into account, the induced moments are $A_{2u}^-\otimes E_{2g}^+=E_{2u}^-$ and are inconsistent with the experiment.  
Second, the single-$\mathcal{Q}$ order contains $A_{2u}^-$, $E_{2u,xy}^-$, and $E_{2g,22}^+$ irreps.  Again, it is impossible to construct $E_{1u}$ irreps from these irreps and the distortion $\varepsilon_{22}$ with $E_{2g,22}^+$ irreps.

We note that the modulated toroidal order on the $4c$ and $8f$ sites [Fig.~\ref{fig:CEF_4c_8f}(b)] has $B_{2g}^-$ and $B_{1u}^+$ components, when the realistic crystal structure is considered.  This leads to $E_{1u,x}^+$ with the orthogonal distortion by $B_{1u}^+\otimes E_{2g,22}^+=E_{1u,x}^+$, and is consistent with the ME experiments, which has not been recognized in the previous studies \cite{Tabata2021, Willwater2021}.  However, there are two reasons why the canted triforce order [Fig.~\ref{fig:CEF_4c_8f}(c) or \ref{fig:CEF_4c_8f}(d)] is more favorable than the modulated toroidal order.  First, the toroidal order on the $4c$ and $8f$ is hardly stable as lowering $T$.  Second, the modulated toroidal order has a finite $E_{1u}^+$ moment only when the orthogonal distortion is present, but this state is not an eigenmode of the susceptibility tensor in the presence of the orthogonal distortion and can be realized only in accidental cases, as discussed in Sec.~\ref{sec:realcrys}. In contrast, the canted triforce order in the realistic crystal structure can be stable both at high and low temperatures.  
Thus, the toroidal order on the $4c$ and the $8f$ sites does not seem to be a major candidate for ${\rm UNi_4B}$ even if it were stable at lower temperatures by an unknown mechanism. 

Lastly, we discuss the magnitudes of the ME coefficients.  The observed $\alpha_{yx}$ and $\alpha_{yz}$ are in the same order \cite{Saito2018}.  We note that this does not mean that the magnitudes of the $A_{2u}^-$ and $E_{1u}^+$ moments are similar.  The two ME coefficients $\alpha_{yx}$ and $\alpha_{yz}$ are qualitatively different; $\alpha_{yx}$ is induced by the electric field, while $\alpha_{yz}$ is induced by the electric current.  The field-induced one is owing to interband effects, while the current-induced one is owing to intraband effects.  
Although the quantitative estimation of the ME coefficients is beyond the scope of this study, we note that the magnitudes of the $A_{2u}^-$ and the $E_{1u}^+$ moments do not need to be in the same order.  Thus, the magnitude of $\Delta E_4$, which induces the $E_{1u}^+$ moment for the triforce order, cannot be estimated from the ME experiments.

\subsection{Comparison in other experiments}
\label{subsec:Comparison}
In this section, we compare our numerical data and the experimental results. Since the calculation in this paper is based on the mean-field theory and the model is rather simple to reproduce all the aspects of UNi$_4$B, we restrict ourselves to the qualitative discussions. 

\subsubsection{Thermodynamic properties}\label{subsubsec:Thermo}
We first discuss the $T$ dependence of the order parameters and the thermodynamic quantities.  Several experiments in ${\rm UNi_4B}$ have clarified that there is a clear anomaly in the specific-heat coefficient $C/T$, the susceptibility, and the resistivity at $T_{\rm N}=20$ K \cite{Mentink1994}.  It is also noted that there is a weak anomaly at $T^{\ast}=0.3$ K in the $T$ dependence of the specific heat \cite{Movshovich1999} and the ultrasound velocity \cite{Yanagisawa2021}.  So far, whether the latter is a phase transition or not is unclear. 

In our results, the triforce and the single-$\mathcal{Q}$ orders are possibly consistent with these aspects.  This is because they show a single phase transition at $T_{\rm N}$ as shown in Fig.~\ref{fig:PhaseDiagram}, while the toroidal order with disordered sites is followed by several phase transitions below $T_{\rm N}$.  As a possible explanation for the weak anomaly at $T^\ast$, we note that for the triforce state, there is shoulder-like $T$ dependence at $T=T^\star\sim 5$ K in Fig.~\ref{fig:triforce_OP_Chi}(d).  This is related to the $T$ dependence of the quadrupole moment and quadrupole susceptibility, both of which are saturated at $\sim5$ K. This characteristic temperature $T^\star$ is much higher than the observed one $T^\ast\sim 0.3$ K. When the quadrupole interactions are small, the value of $T^\star$ can be lower and it also leads to the low Curie-Weiss temperature $\theta_{\rm CW}^Q\sim -1$ K observed \cite{Yanagisawa2021}. 

However, the quadrupole interaction is essential for stabilizing the triforce order at zero temperature. 
See the discussion in Sec.~\ref{subsec:FreeEnergy} and also Appendixes \ref{sec:App_fewer_param} and \ref{subsec:OtherQuadrupole}. Thus, it is difficult to reproduce both the stability of the triforce order and an increase in the quadrupole susceptibility at low temperatures. This might be realized by considering the effects not considered here, which suppress the magnetic orders even for small quadrupole interactions.

Such suppression of the magnetic orders may be caused by magnetic fluctuations due to the frustrated interactions or the Kondo effects. Within the mean-field approximation, additional $O_{20}$ quadrupole interactions can suppress the magnetic orders. Interactions of $O_{20}$ quadrupole with $A_{1g}$ representation act as a temperature-dependent CEF and can suppress the magnetic orders (see Appendix~\ref{subsec:OtherQuadrupole}). However, the validity of such parametrization is not based on the microscopic information about UNi$_4$B, and we show the results as an example among several possibilities in Appendix~\ref{subsec:OtherQuadrupole}. The complete understanding about $T^*\sim 0.3$ K needs a more sophisticated model construction and analysis, and this is one of the future problems.

For the magnetic susceptibility, the consistency with the experiments is more subtle.  In the experiments, the susceptibility increases as $T$ decreases in the ordered state for $10~{\rm K}\lesssim T<T_{\rm N}$ \cite{Mentink1994, Saito2018}, which is consistent with the results in Fig.~\ref{fig:triforce_OP_Chi}(d). However, it decreases for $T\lesssim 10$ K \cite{Saito2018}.  The decrease in the magnetic susceptibility at low temperatures is not realized in this study.  This inconsistency will be resolved when the CEF with an orthogonal distortion is taken into account \cite{Yanagisawa2021}.

\subsubsection{Ultrasound experiments}
Let us discuss the quadrupole interactions, focusing on the ultrasound experiments.  
We emphasize that the $T$ dependence of quadrupole interactions is key to identifying the order parameters.  In Ref.~\onlinecite{Yanagisawa2021}, the sound velocity softening is observed both above and below $T_{\rm N}$.
 The softening is the consequence of the enhanced quadrupole susceptibility, and it has been analyzed by the Curie-Weiss fitting. 
  Interestingly, the Curie-Weiss temperature for the quadrupole sector $\theta_{\rm CW}^Q$ is {\it positive} ($\theta_{\rm CW}^Q=11$ K) in the paramagnetic phase $T>T_{\rm N}$, while it is {\it negative} ($\theta_{\rm CW}^Q=-1.2$ K) in the ordered phase $0.3~{\rm K}<T<10~{\rm K}$.   
  In the following, we will show that the change in $\theta_{\rm CW}^Q$ can be explained qualitatively if the ordered state is assumed to be the triforce phase, while it turns out that the quantitative agreement with the experiments at low temperatures is not achieved in our simple model.  
  
The $T$ dependence of the quadrupole susceptibility $\chi_Q$ is shown in Fig.~\ref{fig:triforce_OP_Chi}(d). The high-temperature Curie-Weiss temperature $\theta_{\rm CW}^Q$ is automatically satisfied by the constraint (\ref{eq:TCW_Q}). $\chi_Q$ shows a jump at $T=T_{\rm N}$, which might be an artifact of the mean-field theory. Below $T_{\rm N}$, it decreases once and turns to increase. The increase at low temperatures is qualitatively consistent, but the actual $T$ dependence is quantitatively different from the observed $T$ dependence of the elastic constant. Similarly to the case of the specific heat discussed before, $\chi_Q$ in Fig.~\ref{fig:triforce_OP_Chi}(d) is saturated to $\sim0.6$ below $T=T^\star\sim 5$ K. To obtain the lower $T^\star$ within the mean-field approximation, we need additional parameters as discussed in Appendix~\ref{subsec:OtherQuadrupole}. For some parameter sets, the Curie-Weiss $T$ dependence with $\theta^Q_{\rm CW}<0$ can be reproduced, but it leads to some drawbacks such as the increasing magnetic susceptibility at low temperatures.

Despite the quantitative discrepancy between the data in Fig.~\ref{fig:triforce_OP_Chi}(d) and the experiment, the triforce order gives a phenomenological explanation about the negative $\theta_{\rm CW}^Q$ in the ordered phase below $T_{\rm N}$. 
 In the triforce configuration, the magnetically disordered sites are connected by the nearest- and the third-nearest-neighbor bonds.  Suppose the quadrupole moments at these sites are nearly free while those at the magnetically ordered sites are frozen owing to the large dipole-quadrupole coupling, only the nearest-neighbor interaction appears in the Curie-Weiss form of the quadrupole susceptibility. 
 
  In Table~\ref{table:Jq}, the eigenvalues of the magnetic exchange eigenvalues are listed for $\veck=\veck_0,\ \veck_{\rm K}$, and $\veck_n$. These eigenvalues are also correct for the quadrupole ones by replacing $J_i^M$ with $J_i^Q$. The Curie-Weiss factor $T-T^Q_{\rm CW}=T+J_\Gamma^Q=T+6(J_1^Q+J_2^Q)\to T+6J^Q_1$ by discarding $J_2^Q$ in the above picture.  The triforce order appears for $J_1^Q>0$ as shown in Fig.~\ref{fig:PhaseDiagram}, which is also consistent with the Landau analysis in Sec.~\ref{subsec:LandauF}, and this indeed leads to the negative $\theta_{\rm CW}^Q=-6J_1^Q<0$.
 Such consistency is not expected for other phases.  For the single-$\mathcal{Q}$ order, the interaction will be ferroic since $J^Q_\Gamma < J^Q_{\rm K}$ is needed to realize the single-$\mathcal{Q}$ order (Fig.~\ref{fig:PhaseDiagram}) and leads to $-6(J_1^Q+J_2^Q)\to -6J_1^Q>0$.  For the toroidal order, the quadrupole interactions at the disordered sites $J_2^Q$ can be weak antiferroic.  However, it is hardly stable at low temperatures since the magnetic interactions between the disordered sites are dominant for the ordering vector at ${\bm k}_{1,2,3}$.  
 
 The validity of the above phenomenological argument strongly depends on how free the quadrupole moments are at the magnetically disordered sites.  In the mean-field data in Fig.~\ref{fig:triforce_OP_Chi}(d), the situation is applicable above $T^\star\sim 5$ K, below which the quadrupole moments are saturated.  Thus, if the $T^\star$ can be lowered by fine tuning of the parameters and/or by the higher-order many-body corrections, the observed softening in the ordered phase would be explained.  See one example in Appendix \ref{subsec:OtherQuadrupole} of such fine tuning within the mean-field approximation.  We consider that clarifying this is one of the important problems for our future studies.

\subsection{Important future experiments}
\label{subsec:Prediction}
In Secs.~\ref{sec:realcrys}, \ref{subsec:ME}, and \ref{subsec:Comparison}, we have proposed that the canted triforce order qualitatively explains the experimental data available so far.  Let us comment on the future experiments to check the triforce order scenario.

\begin{figure*}[t]
\begin{center}
\includegraphics[width=0.95\textwidth]{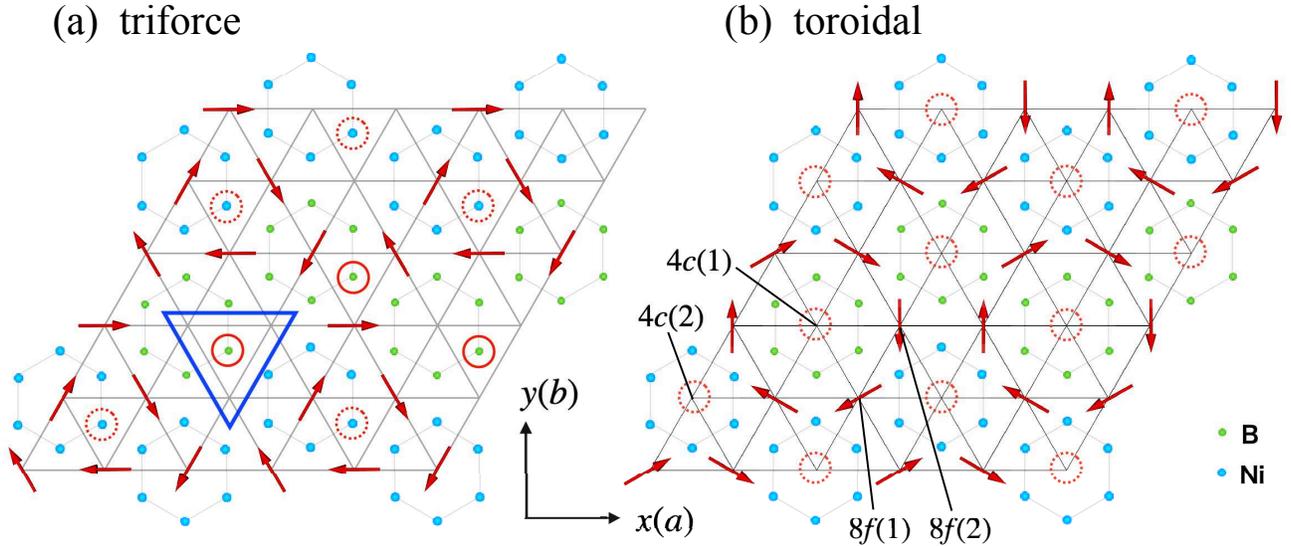}
\end{center}

\caption{Crystal and magnetic structure for the (a) triforce and (b) toroidal orders. 
At U sites forming the triangular lattice sites, arrows representing the magnetic dipole moment are drawn.  Note that sites without an arrow represent the magnetically disordered U sites.  At the center of each triangle, there are B (small, green) or Ni (large, sky blue) atoms on the same triangular plane. 
The blue triangle frame in (a) represents the paramagnetic unit cell when the difference between $8f(1)$ and $8f(2)$ [$4c(1)$ and $4c(2)$] are neglected.   
Sites where local fields vanish are enclosed by red circles.  In (a), solid (dashed) circles are B (Ni) sites, while in (b), dashed circles enclose the magnetically disordered U sites. }
\label{fig:LocalField}
\end{figure*}

The first one is $^{11}$B NQR and/or NMR experiments at low temperatures.  
The NQR and NMR can be powerful tools for clarifying local environments.  We here note that there is characteristic symmetry lowering in the B sites in the triforce order.  
Figure \ref{fig:LocalField}(a) illustrates the triforce states on the triangular plane together with B and Ni atoms in the $Cmcm$ structure.  One can see that there exist B sites where the local magnetic and quadrupole fields vanish.  For the canted triforce state with the small canting, the local fields at the B sites with the approximate $C_3$ rotational symmetry are finite but small.  Remarkably, such high-symmetry B sites do not exist for the toroidal order [Fig. \ref{fig:LocalField}(b)].  This can be useful for identifying the order parameter in the NMR/NQR experiments.

The next is the detection of the secondary quadrupole moments in resonant x-ray scattering experiments.  The triforce order has the quadrupole moments at $\p=\veck_{\rm K}$, while the toroidal order does not [see Eqs.~(\ref{eq:induced_q_toroidal}) and (\ref{eq:q_phase_triforce1})].  The presence of the quadrupole moments at $\p=\veck_{\rm K}$ can be the semi-direct evidence of the triforce orders.  Note that the K point component itself should be present in the paramagnetic phase since the crystallographically inequivalent sites form the triangular lattice in UNi$_4$B \cite{Tabata2021, Willwater2021}.  The contribution owing to this is $O_{20}\sim 2z^2-x^2-y^2$ type quadrupole with $A_{1g}^+$ irreps, while that arising from the order parameter is $O_{22}\sim x^2-y^2$ and $O_{xy}\sim xy$ types with $E_{2g}^+$ irreps.  Thus, the contribution owing to the crystal structure and the order parameter can be distinguished by the polarization or the azimuth angle dependence. In addition to this, the $O_{20}$-type quadrupole moments or charge density wave at ${\bm k}_{1,2,3}$ are expected for the triforce order, but not for the toroidal one. See also the discussion in Appendix \ref{subsec:OtherQuadrupole}.  Detection of them can be another smoking gun of the order parameter.  The resonant x-ray scattering or neutron scattering measurements can detect this $A_{1g}$ contribution.

Next, nonreciprocal transport experiments are important to understand the order parameter of ${\rm UNi_4B}$.  
The time-reversal parity of the $E_{1u}$ ($A_{2u}$) part of the order parameters, which causes the ME coefficient $\alpha_{yz}$ ($\alpha_{yx}$), can be detectable by the non-reciprocal transport experiments. 
  The magnetic $E_{1u}^-$ ($A_{2u}^-$)  contains the in-plane (out-of-plane) component of the magnetic toroidal moment (Table \ref{table:Rep}), and it causes the nonreciprocal conductivity with the current parallel to the toroidal moment at zero magnetic field \cite{Yatsushiro2021, Yatsushiro2022}.  In contrast, the electric $E_{1u}^+$ ($A_{2u}^+$), which has the same symmetry as the in-plane (out-of-plane) electric polarization, cannot cause the nonreciprocal conductivity since it is forbidden by the Onsager relation \cite{Onsager1931_I, Onsager1931_II, LandauLifshitzStatisticalPhysics, Rikken2005}.  Thus, the non-reciprocal conductivity can be direct evidence of the toroidal moments.  In the triforce order scenario, the non-reciprocal conductivity {\it for the $c$-axis current} is expected, while {\it not for the $ab$-plane currents}.  When the latter is present, the magnetic moments have components along the $c$ axis or are nonuniformly stacked along the $c$ axis, both of which have not been detected.  The determination of the time-reversal parity of $E_{1u}$ component is of significant importance, as well as the direct evidence of $A_{2u}^-$ toroidal moment.   Indeed, the nonreciprocal transport measurement has broader information beyond checking particular scenarios and is highly desired.  
 
 Finally, we comment on detections of even-parity multipole moments that can examine the triforce scenario.  The triforce state contains the $B_{1g}^-\sim (x^2-y^2)M_x-2xyM_y$ octupole moment, and this can be detected by magnetostriction experiments \cite{Patri2019}.  The anisotropies in the magnetostriction can distinguish the canted triforce order from the modulated toroidal order on the $4c$ and $8f$ sites [Fig.~\ref{fig:CEF_4c_8f}(b)].  The latter contains the $B_{2g}^-\sim (x^2-y^2)M_y+2xyM_x$ octupole moments.  For example, magnetic fields ${\bm H} \parallel x$ induce the strain $\varepsilon_{22}$ in the canted triforce order, while $\varepsilon_{xy}$ for the modulated toroidal order.  In addition to the octupole moments, small but finite uniform magnetizations are expected in any in-plane magnetic orders consistent with the ME effects.  The finite magnetization can be checked directly and indirectly by, e.g., the anomalous Hall effect.  The presence of these even-parity magnetic multipole moments implies that the odd-parity $E_{1u}^+$ moments are induced by the magnetic orders and cause finite $\alpha_{yz}$.

\subsection{Theoretical perspective}
\label{subsec:Interest}

To close Sec.~\ref{sec:discussions}, we discuss the mechanism of multiple-$\mathcal{Q}$ partial orders in this study and its possible extensions. We note that the multiple-$\mathcal Q$ order mechanism owing to the cooperation among multipole degrees of freedom can be applicable to more general systems.   

\subsubsection{Partial magnetic order in {\rm UNi$_4$B}}
\label{subsec:partial}
First, we discuss the mechanism of partial orders, focusing on UNi$_4$B with the ordering vector ${\bm k}_{1,2,3}$.  In this study, the partial magnetic orders are realized by the competing interactions of magnetic dipole and electric quadrupole moments.  Such partial magnetic order with the quadrupole moments at magnetically disordered sites was also reported in Ref. \onlinecite{Seifert2022}, where the K point version of the single-$\mathcal{Q}$ order is realized in an anisotropic $S=1$ model with a biquadratic interaction. The crucial point is the presence of the quadrupole degrees of freedom, whose importance is recently clarified by the ultrasonic experiments \cite{Yanagisawa2021}.  In contrast, the previously proposed mechanism based on the partial Kondo screening \cite{Lacroix1996,Motome2010, Ishizuka2012, Hayami2012} has difficulty when directly applied to ${\rm UNi_4B}$.  In Ref. \onlinecite{Lacroix1996}, the in-plane anisotropy of the magnetic moments is important to realize the partial orders.  However, the proposed CEF scheme in the recent ultrasound experiments suggests that the pseudo triplet CEF ground state has no in-plane anisotropy.  
In Refs. \onlinecite{Motome2010, Ishizuka2012, Hayami2012} as well as in Ref. \onlinecite{Lacroix1996}, partial orders are realized by the cooperation of Kondo singlet formation and magnetic orderings.  With one-third of the moments Kondo screened, the remaining unscreened magnetic moments lie on the honeycomb structure without frustration, and a simple N{\'e}el order is realized.  Thus, the partial screening relaxes the magnetic frustration of the triangular lattice, and the partial order is realized as a cooperative effect between the Kondo screening and the magnetic interactions. 
One may consider this mechanism is applicable to ${\rm UNi_4B}$ since the ordered sites in the presumed toroidal order in ${\rm UNi_4B} $ also form the honeycomb structure.  However, there is a crucial difference between the N{\'e}el order on the honeycomb structure and the toroidal order in ${\rm UNi_4B}$.  The former is a collinear order at the K point, while the latter is a non-collinear order at ${\bm k}_{1,2,3}$.  The K point order is favored when the nearest-neighbor antiferro interaction is dominant.  The disordered sites in the partially ordered state are connected by the next-nearest-neighbor bonds, and the interactions between the disordered sites are not necessarily large.  Thus, the energy cost in the magnetic interactions is small, and the Kondo singlet is favored at the disordered sites.
 
In contrast, the magnetic interactions and the positions of the disordered sites in the toroidal order are mismatched.  The ordering vector ${\bm k}_{1,2,3}$ implies the next-nearest-neighbor antiferro interaction is dominant.  In this case, the Kondo screening and the magnetic interaction compete since the disordered sites are connected by the next-nearest neighbor-bonds.  Furthermore, the next nearest bonds connecting the ordered moments form the $120^\circ$ structure.  This implies that the frustration is not relaxed by the partial screening.  Thus, the mechanism based on the cooperation between the Kondo screening and the magnetic order is not directly applicable to ${\rm UNi_4B}$.  

Interestingly, the partial screening mechanism can be applied to the triforce and single-$\mathcal{Q}$ orders.  Both orders can be seen as the N{\'e}el order when focused on the next-nearest-neighbor bonds between the magnetic sites.  Furthermore, the interactions between the disordered sites are of the nearest- and the third-nearest-neighbor ones, which are not necessarily large for realizing the ordering vector ${\bm k}_{1,2,3}$.  Thus, the partial screening mechanism is applicable to the two orders.  
A similar discussion is possible for the case that the partial order is owing to thermal or quantum fluctuation instead of the Kondo singlet formation.
Although we have shown that the triforce and the single-$\mathcal{Q}$ orders can be stable at zero temperature due to the quadrupole interaction, investigation of these orders in models without quadrupole degrees of freedom is an interesting future issue.  

\subsubsection{Multiple-$\mathcal{Q}$ ordering mechanism}
\label{subsec:multiple-Q}
Next, we discuss the mechanism of multiple-$\mathcal{Q}$ orders.  The triple-$\mathcal{Q}$ ordering mechanism in this study can be applied to other systems with secondary order parameters, such as the quadrupole moments in this study. In this study, we have clarified that the triple-$\mathcal{Q}$ orders become stable owing to the quadrupole interactions through the local couplings between the dipole and the quadrupole moments.  The coefficients of the inter-mode couplings between the magnetic dipoles for ${\bm p}_1$ and ${\bm p}_2$ are affected by the quadrupole interactions for the higher harmonic wave vectors ${\bm p}_1 \pm {\bm p}_2$, and it plays a role in determining the phase factors in the triple-$\mathcal{Q}$ superposition.  

The significance of the higher harmonic modes and the phase degrees of freedom in magnetic configurations is also investigated for topological magnets such as skyrmion and hedgehog lattices, particularly in itinerant magnets, where biquadratic interactions play an important role \cite{Ozawa2016, Hayami2021_Itinerant, Hayami2021_Phase, Hayami2021_Lock, Shimizu2021_Moire, Shimizu2021_Chiral, Shimizu2022_Phase}.  Since the bi-quadratic interactions for spins and the quadrupole interactions are essentially the same, the mechanism of multiple-$\mathcal{Q}$ orders assisted by the inter-mode couplings is common to these studies and the present study.  However, there is a difference between these studies for the itinerant magnets and our study: the CEF effect.  Note that the CEF excitation gap plays an important role in the partial magnetic disorder at $T=0$.  In this study, the triforce and the single-$\mathcal{Q}$ phases are such partial magnetic orders with the quadrupole order at the non-magnetic sites, even when the quadrupole interactions are weaker than the magnetic ones.  This is because the quadrupole orders gain the CEF energy.  Remember that the $E_{2g}^+$ quadrupole moments ${\bm Q}$ have the matrix elements within the CEF ground state, while those of the $E_{1g}^-$ magnetic dipole moments ${\bm M}$ span between the ground doublet and excited singlet [see Eqs.~(\ref{eq:Ope_Q}) and (\ref{eq:Ope_Jxy})].  Such partial magnetic orders are hardly stable without the CEF effect.  
Extending this CEF mechanism for partial magnetic orders to more general situations can also lead to further exotic orders.  For example, an incommensurate coplanar vortex lattice formed by magnetic moments can be realized even at zero temperature, which is forbidden for usual spin systems since the entropy is not released at the vortex cores.  The topological feature of such magnetic vortex lattices can be affected by the quadrupole interactions since they influence the phase factors of multiple-$\mathcal{Q}$ superpositions.  Investigation of such states is one of our future problems.  

We now point out possible examples that a triple-$\mathcal{Q}$ order is realized by a similar mechanism to that in this study.  In ${\rm UPd_3}$, a triple-$\mathcal{Q}$ order of $E_{1g}$ electric quadrupole moments, with one-fourth of sites disordered, is considered to be realized \cite{Walker1994, McEwen1995, McEwen1998}.  The ordering wave vectors are at M points, ${\bm k}_{{\rm M}_1}=-{\bm g}_1/2-{\bm g}_2/2$, ${\bm k}_{{\rm M}_2}={\bm g}_1/2$, and ${\bm k}_{{\rm M}_3}={\bm g}_2/2$.   The triple-$\mathcal{Q}$ order was theoretically analyzed by means of Landau theory \cite{Lingg1999}.  In Ref.~\onlinecite{Lingg1999}, the fourth-order mode-mode couplings in the Landau free energy are important to realize the triple-$\mathcal{Q}$ state.  The microscopic origin of such fourth-order mode-mode coupling is possibly similar to that in this study. The CEF scheme of ${\rm UPd_3}$ is a pseudo triplet with the ground-state doublet and the excited singlet \cite{McEwen2003}, similar to ours.  The primary order parameter is staggered along the $c$ axis, which couples to the $c$-axis uniform secondary order parameters.
  When the quadrupole interactions favor orders at the M points with uniform stacking, the triple-$\mathcal{Q}$ order is realized, while single-$\mathcal{Q}$ orders occur when the interactions prefer the orders at the $\Gamma$ point with uniform stacking.  The main difference from this study is that the single-$\mathcal{Q}$ order at the M point has no disordered site.  More detailed analysis is needed for quantitative discussions, and this is one of our future issues.  

As another example, the kagome compounds $A$V$_3$Sb$_5$ ($A$=K, Rb, Cs) \cite{Ortiz2019, Jiang2021, Tan2021, Neupert2021} show triple-$\mathcal{Q}$ charge-density-wave (CDW) orders, which have attracted considerable attention as well as their properties of superconductivity \cite{Xu2021, Neupert2021}.  In the triple-$\mathcal{Q}$ CDW states, cubic couplings in their free energy play a key role \cite{Denner2021, Miao2021}.  In $A$V$_3$Sb$_5$, whether the time-reversal symmetry is broken or preserved in the CDW ordered phase is under debate \cite{Feng2021, Li2022}.  The time-reversal broken imaginary parts of the CDW, in the chiral flux state, couple to the non-magnetic real parts via the cubic term \cite{Denner2021, Miao2021}, as in the same manner as the magnetic dipole to the electric quadrupole in this study.  Such cubic coupling is also present for purely non-chiral electric terms and seems to play a key role in the triple-$\mathcal{Q}$ order.  
Further material search for the triple-$\mathcal{Q}$ orders induced by cubic couplings between the primary and secondary order parameters is an interesting topic in the future.

\section{Summary} 
\label{sec:summary}
We have introduced a doublet-singlet ($\Gamma_5$--$\Gamma_4$) localized moment model for ${\rm UNi_4B}$ with active magnetic dipole and electric quadrupole moments, which is motivated by recent ultrasonic experiments \cite{Yanagisawa2021}.  Deriving the realistic model parameters based on the various experimental data, we have carried out the mean-field calculations.  The results clearly show that the presumed toroidal order is not stable at low temperature in the localized model and that another triple-$\mathcal{Q}$ triforce order is a promising candidate for the order parameter of ${\rm UNi_4B}$.  The two states possess a common magnetic structure factor consisting of triple-$\mathcal{Q}$ configurations but with different phase factors.  We find that the quadrupole moments play a crucial role in determining which is realized.  The phenomenological Landau analysis also leads to the two triple-$\mathcal{Q}$ states as its solutions.  Thus, it is quite natural to obtain the triforce order in the microscopic mean-field calculations. Symmetry arguments reveal that the triforce order is consistent with the ME effects, where site-dependent crystalline-electric field levels or orthogonal distortion in the realistic crystal structure is crucial for explaining the ME effects. The point is that the highest symmetry point in the ordered state is {\it not} at the U site. This is crucial to explain the observed magnetoelectric effects. 
We have also proposed several future experiments for checking the validity of the triforce order scenario.  The resonant x-ray experiment with the wavenumber at the K and ${\bm k}_{1,2,3}$ points, the NMR and/or NQR for $^{11}$B sites, and non-reciprocal transport properties can be powerful tools for identifying the order parameter in ${\rm UNi_4B}$.   

Since after the proposal of the toroidal magnetic order in the neutron scattering experiment in 1994 \cite{Mentink1994}, the toroidal order has been recognized as the order parameter of ${\rm UNi_4B}$.  The recent experiment on the ME effect offers reexamination of the validity of the presumption \cite{Saito2018}.  The observed anisotropies in the ME effects suggest that the order parameter has not only the out-of-plane $A_{2u}^-$ toroidal moments, but also an in-plane $E_{1u}^+$ component.  It leads to a remarkable fact that the order parameter with in-plane magnetic moments has both even- and odd-parity components.  Such cases with mixed parity are naturally realized when the highest-symmetry point is at the center of the nearest-neighbor triangles, as in the $120^\circ$ structure.  We note that the triforce order is such an example for the triangular systems with the ordering vectors at ${\bm k}_{1,2,3}$, as shown in Figs.~\ref{fig:Distortion_Cluster}(a) and \ref{fig:LocalField}(a).  The triforce order scenario proposed in this paper can explain both the neutron data \cite{Mentink1994, Willwater2021} and the ME effects. The quantitative discrepancy between the triforce order scenario and the experimental data still remains: the energy scale of the quadrupole sector. For analyzing the quantitative aspect of UNi$_4$B, more elaborated calculations and microscopic information are needed in the future studies.
 
We consider that our results stimulate future experimental investigations to clarify the physics of ${\rm UNi_4B}$ and shed light on analyses of ME effects and non-reciprocal transport in anisotropic correlated systems and their potential in order-parameter detection.  
Furthermore, the multiple-$\mathcal{Q}$ ordering mechanism induced by the quadrupole interactions can be applied to other systems, and it will stimulate further studies in multipole physics for $d$- and $f$-electron systems.

\section*{Acknowledgement} The authors thank H. Kusunose, S. Hayami, K. Ota, K. Izawa, H. Amitsuka, C. Tabata, T. Yanagisawa, and H. Saito for fruitful discussions. 
This work was supported by JSPS KAKENHI Grant No. JP21H01031.

\appendix
\label{sec:Appendix}

\section{Detail expressions of exchange interactions}
\label{subsec:Jq}

We show the explicit formula of the exchange interaction matrix $\hat{J}^X_\p (X=M,Q)$.  The matrix elements are decomposed into the irreducible representations and expressed in Eqs.~(\ref{eq:Hint}) and (\ref{eq:Jmat}).
The $\p$-dependent coefficient $J_\p^{X,A_{1g}}$, $J_\p^{X,E_{2g,22}}$, and $J_\p^{X,E_{2g,xy}}$ belong to the $A_{1g}$, $E_{2g,22}$, and $E_{2g,xy}$ representations in $D_{6h}$, respectively.  Their explicit forms are given by 
\begin{align}
J_\p^{X,A_{1g}} =& 2J^X_1( c_{p_1} + c_{p_2} + c_{p_2-p_1} ) \nonumber \\
&+ 2J^X_2( c_{p_1+p_2} + c_{2p_2-p_1} + c_{p_2-2p_1} ), \\
  J_\p^{X,E_{2g,22}} =& K^X_1( 2c_{p_1} - c_{p_2} - c_{p_2-p_1} ),\\
  J_\p^{X,E_{2g,xy}} =& \sqrt{3} K^X_1( c_{p_2} - c_{p_2-p_1} ), 
,\end{align}
where $c_{p_i}\equiv \cos p_i$ and $p_i=\p \cdot {\bm a_i}$ $(i=1,2)$ with the primitive translation vectors ${\bm a}_1=(1,0)$ and ${\bm a}_2=(1/2, \sqrt{3}/2)$. Note that we have assumed $K_1^Q=0$ in Eqs.~(\ref{eq:Hint_minimal}) and (\ref{eq:Hint_minimalK}).

It is useful to write down the matrix form of the exchange interactions $\hat{J}^X_{{\bm k}_{1,2,3}}$, where ${\bm k}_{1,2,3}$ is defined in Sec.~\ref{subsec:CEF}. For $n=1$ and $2$,
\begin{align}
	\hat{J}_{{\bm k}_n}^X=\begin{bmatrix}
		-3J^X_2 -\frac{3}{2}K_1^X& (-1)^n\frac{3\sqrt{3}}{2}K_1^X\\[1mm]
		(-1)^n\frac{3\sqrt{3}}{2}K_1^X & -3J^X_2+\frac{3}{2}K_1^X
	\end{bmatrix},
\end{align}
and for $n=3$,
\begin{align}
	\hat{J}_{{\bm k}_3}^X=\begin{bmatrix}
		-3J^X_2 +3K_1^X& 0\\		
		0 & -3J^X_2-3K_1^X
	\end{bmatrix}.
\end{align}
The eigenvectors are easily calculated and given by Eq.~(\ref{eq:v_n}).
For $K_1^X<0$, $\bm{v}_{n\perp}$ corresponds to the eigenvector for the smaller eigenvalue, while  $\bm{v}_{n\parallel}$ for $K_1^X>0$.

For the interactions at other important wave vectors: $\Gamma$ and K points, $J_\p$'s are given by
\begin{align}
	J^{X,A_{1g}}_{{\veck}_0} &= 6(J^X_1+J^X_2), \ 
	&J^{X,E_{2g,22,xy}}_{{\veck}_0}  = 0, \label{eq:J_Gamma}\\
	J^{X,A_{1g}}_{{\veck}_{\rm K}} &= -3(J^X_1-2J^X_2), \ 
	&J^{X,E_{2g,22,xy}}_{{\veck}_{\rm K}} = 0. \label{eq:J_K}
\end{align}
Equations (\ref{eq:J_Gamma}) and (\ref{eq:J_K}) show that the 
eigenmodes are degenerate since the exchange interaction matrices at these wave vectors are proportional to the identity matrix. 

\section{Local Landau free energy}
\label{sec:AppLocLandau}

In this appendix, we discuss the Landau free energy for a single-site crystalline-electric-field model with $\Gamma_5$--$\Gamma_4$ [Fig.~\ref{fig:CEF_Interactions}(a)].  Let the conjugate field to the dipole (quadrupole) moment be ${\bm h}$~($\tilde{\bm h}$). The
expectation values of the dipole and quadrupole are represented by $\m$ and  $\q$, respectively. The 
free energy under these conjugate fields is given by 
\begin{align}
	F({\bm h}, \tilde{\bm h}) &= -\beta^{-1}\ln Z({\bm h}, \tilde{\bm h}), \label{eq:FMQ}
\end{align}
where $\beta$ is the inverse temperature $1/T$ and $Z({\bm h}, \tilde{\bm h})$ is the partition function. 
The expectation values are calculated in a standard form by 
\begin{align}	
	\m &= -\pdv{F}{{\bm h}}=
	\begin{pmatrix}
		m_x\\
		m_y
	\end{pmatrix}
	, ~
	\q = -\pdv{F}{\tilde{\bm h}}=\begin{pmatrix}
	q_{22}\\
	-q_{xy}
	\end{pmatrix}.\label{eq:Define_mq}
\end{align}
Note that $\q$ is the expectation value of $(O_{22},-O_{xy})^{\rm T}$.  To derive the Landau free energy, we expand $Z({\bm h},\tilde{\bm h})$ up to the fourth order in terms of the conjugate fields ${\bm h}=h(\cos\Theta,\sin\Theta)^{\rm T}$ and $\tilde{\bm h}=\tilde{h}(\cos\Phi,\sin\Phi)^{\rm T}$  as
\begin{align}
	Z =& Z_0 + d_2^mh^2 + d_2^q\tilde{h}^2 + d_3h^2\tilde{h}\cos(2\Theta+\Phi) \nonumber\\
	 &+ d_4^mh^4 + d_4^q\tilde{h}^4 + d_4^{mq}h^2\tilde{h}^2. \label{eq:Z}
\end{align}
Here, the coefficients in Eq.~(\ref{eq:Z}) are 
\begin{align}	 
	Z_0 &= 2 + e^{-\beta E_4},\ 
	\frac{d_2^m}{\beta^2} = \frac{1-e^{-\beta E_4}}{\beta E_4},\ 
	\frac{d_2^q}{\beta^2} = 1, \\
	\frac{d_3}{\beta^3} &= -\frac{1}{(\beta E_4)^2}\qty( e^{-\beta E_4} - 1 + \beta E_4 ), \ \frac{d_4^q}{\beta^4} = \frac{1}{12},\\
	\frac{d_4^m}{\beta^4} &= \frac{1}{(\beta E_4)^3}\qty[ (e^{-\beta E_4}-1) + \frac{\beta E_4}{2}(1+e^{-\beta E_4}) ], \\
	\frac{d_4^{mq}}{\beta^4} &= \frac{1}{(\beta E_4)^3}\qty[ (1-e^{-\beta E_4}) - \beta E_4 + \frac{(\beta E_4)^2}{2} ].
\end{align}
Note that 
$d_3<0$, while $d_2^{m,q}$, $d_4^{m,q}$, and $d_4^{mq}$ are all positive. 

Now, substituting Eq.~(\ref{eq:Z}) into Eq.~(\ref{eq:FMQ}), and retaining up to the fourth-order terms, one finds 
\begin{align}
	F =& \tilde{F}_0 + \frac{\tilde{a}_m}{2}h^2 + \frac{\tilde{a}_q}{2}\tilde{h}^2 + \frac{\tilde{c}}{3}h^2\tilde{h}\cos(2\Theta+\Phi) \nonumber\\
	&+ \frac{\tilde{b}_m}{4}h^4 + \frac{\tilde{b}_q}{4}\tilde{h}^4 + \frac{\tilde{b}_{mq}}{4}h^2\tilde{h}^2, \\
		\tilde{F}_0 &= -\beta^{-1}\ln Z_0, \ 
		\beta \tilde{a}_m = -\frac{2d_2^m}{Z_0}, \ 
		\beta \tilde{a}_q = -\frac{2d_2^q}{Z_0}, \\
		\beta \tilde{c} &= -\frac{3d_3}{Z_0}, \ 
		\beta \tilde{b}_m = -\frac{2d_2^{m2}-4Z_0d_4^m}{Z_0^2}, \\
		\beta \tilde{b}_q &= -\frac{2d_2^{q2}-4Z_0d_4^q}{Z_0^2}, \ 
		\beta \tilde{b}_{mq} = -4\frac{d_2^md_2^q-Z_0d_4^{mq}}{Z_0^2}.
\end{align}
Note $\tilde{a}_m<0$ and $\tilde{a}_q <0 $, while others are positive. 
Using the relation (\ref{eq:Define_mq}), we obtain 
\begin{align}
	m_x =& -\tilde{a}_mh_x -\tilde{b}_mh^2h_x - \frac{1}{2}\tilde{b}_{mq}\tilde{h}^2h_x \nonumber\\
	&- \frac{2}{3}\tilde{c}(h_x\tilde{h}_{22}+h_y\tilde{h}_{xy}), \label{eq:mx_expansion}\\
	m_y =& -\tilde{a}_mh_y -\tilde{b}_mh^2h_y - \frac{1}{2}\tilde{b}_{mq}\tilde{h}^2h_y \nonumber\\
	 &- \frac{2}{3}\tilde{c}(h_x\tilde{h}_{xy}-h_y\tilde{h}_{22}), \\
	q_{22} =& -\tilde{a}_q\tilde{h}_{22} - \tilde{b}_q\tilde{h}^2\tilde{h}_{22} - \frac{1}{2}\tilde{b}_{mq}h^2\tilde{h}_{22} \nonumber\\
	&- \frac{1}{3}\tilde{c}(h_x^2-h_y^2), \\
	q_{xy} =& -\tilde{a}_q\tilde{h}_{xy} - \tilde{b}_q\tilde{h}^2\tilde{h}_{xy} - \frac{1}{2}\tilde{b}_{mq}h^2\tilde{h}_{xy} 
	- \frac{2}{3}\tilde{c}h_xh_y. \label{eq:qxy_expansion}
\end{align}
One can solve Eqs.~(\ref{eq:mx_expansion})--(\ref{eq:qxy_expansion}) in terms of ${\bm h}$ and $\tilde{\bm h}$ iteratively and obtains  
\begin{align}
	h_x =& \frac{1}{\tilde{a}_m} \Bigg[ -m_x + \frac{1}{\tilde{a}_m^3}\qty(\tilde{b}_m+\frac{2\tilde{c}^2}{9\tilde{a}_q})m^2m_x \nonumber\\
	&+ \frac{\tilde{b}_{mq}}{2\tilde{a}_q^2\tilde{a}_m}q^2m_x + \frac{2\tilde{c}}{3\tilde{a}_m\tilde{a}_q}(m_xq_{22}+m_yq_{xy}) \Bigg], \label{eq:Mx_expansion}\\
	h_y =& \frac{1}{\tilde{a}_m}  \Bigg[ -m_y +\frac{1}{\tilde{a}_m^3}\qty(\tilde{b}_m+\frac{2\tilde{c}^2}{9\tilde{a}_q})m^2m_y \nonumber\\
	&+ \frac{\tilde{b}_{mq}}{2\tilde{a}_q^2\tilde{a}_m}q^2m_y + \frac{2\tilde{c}}{3\tilde{a}_m\tilde{a}_q}(m_xq_{xy}-m_yq_{22}) \Bigg], \\
	\tilde{h}_{22} =& \frac{1}{\tilde{a}_q}  \Bigg[ -q_{22} + \frac{\tilde{b}_q}{\tilde{a}_q^3}q^2q_{22} + \frac{\tilde{b}_{mq}}{2\tilde{a}_m^2\tilde{a}_q}m^2q_{22} \nonumber\\
	&+ \frac{\tilde{c}}{3\tilde{a}_m^2}(m_x^2-m_y^2)  \Bigg], \\
	\tilde{h}_{xy} =& \frac{1}{\tilde{a}_q} \Bigg[ -q_{xy} + \frac{\tilde{b}_q}{\tilde{a}_q^3}q^2q_{xy} + \frac{\tilde{b}_{mq}}{2\tilde{a}_m^2\tilde{a}_q}m^2q_{xy} \nonumber\\
	&+ \frac{2\tilde{c}}{3\tilde{a}_m^2}m_xm_y \Bigg],\label{eq:Qxy_expansion}
\end{align}
with $\m\equiv m(\cos\theta,\sin\theta)^{\rm T}$ and $\q\equiv q(\cos\phi,\sin\phi)^{\rm T}$. 

We have succeeded in expressing the conjugate fields ${\bm h}$ and $\tilde{\bm h}$ in terms of the expectation values $\m$ and $\q$. Then, we carry out the Legendre transformation and obtain the Landau free energy $F_{\rm L}(\m, \q)$ defined as 
\begin{align}
	F_{\rm L}(\m, \q) =& F({\bm h}(\m,\q), \tilde{\bm h}(\m,\q)) \nonumber\\
	&+ {\bm h}(\m,\q)\cdot \m + \tilde{\bm h}(\m,\q)\cdot \q.
\end{align}	
Using Eqs.~(\ref{eq:Mx_expansion})--(\ref{eq:Qxy_expansion}), we find 
\begin{align}
	&{\bm h}\cdot \m + \tilde{\bm h} \cdot \q \nonumber\\
	&= -\tilde{a}_m^{-1}m^2 - \tilde{a}_q^{-1}q^2 +  \frac{1}{\tilde{a}_m^4}\qty(\tilde{b}_{m}+\frac{2\tilde{c}^2}{9\tilde{a}_q})m^4 + \frac{\tilde{b}_{q}}{\tilde{a}_q^4}q^4 \nonumber\\
	&\ \ \ + \frac{\tilde{b}_{mq}}{\tilde{a}_m^2\tilde{a}_q^2}m^2q^2 
	+ \frac{\tilde{c}}{\tilde{a}_m^2\tilde{a}_q}m^2q\cos(2\theta+\phi),\\
	&\frac{\tilde{a}_m}{2}h^2 + \frac{\tilde{a}_q}{2}\tilde{h}^2 \nonumber\\
	&=  \frac{1}{2}\qty( \frac{1}{\tilde{a}_m}m^2 +  \frac{1}{\tilde{a}_q} q^2)
	- \tilde{a}_m^{-4}\qty(\tilde{b}_m+\frac{2}{9}\frac{\tilde{c}^2}{\tilde{a}_q})m^4 - \frac{\tilde{b}_m}{\tilde{a}_q^4}q^4 \nonumber \\
	&\ \ \ - \frac{\tilde{c}}{\tilde{a}_m^{2}\tilde{a}_q}m^2q\cos(2\theta+\phi)
	- \frac{\tilde{b}_{mq}}{\tilde{a}_m^{2}\tilde{a}_q^{2}}m^2q^2,
	\end{align}
	\begin{align}
	\frac{\tilde{b}_m}{4}h^4 + \frac{\tilde{b}_m}{4}\tilde{h}^4 &= \frac{1}{4}\qty(\frac{\tilde{b}_m}{\tilde{a}_m^{4}}m^4 + \frac{\tilde{b}_q}{\tilde{a}_q^{4}}q^4 ), \\
	\frac{\tilde{b}_{mq}}{4}h^2\tilde{h}^2&= \frac{\tilde{b}_{mq}}{4\tilde{a}_m^2\tilde{a}_q^2}m^2q^2, \\
	\frac{\tilde{c}}{3}h^2\tilde{h}\cos(2\theta+\phi)&= -\frac{\tilde{c}}{3\tilde{a}_m^2\tilde{a}_q}m^2q\cos(2\theta+\phi) \nonumber\\
	&- \frac{\tilde{c}^2}{9\tilde{a}_m^4\tilde{a}_q}m^4 -\frac{4\tilde{c}^2}{9\tilde{a}_m^3\tilde{a}_q^2}m^2q^2 
.\end{align}
Finally, $F_{\rm L}(\m,\q)$ is given as 
\begin{align}
	F_{\rm L}(\m,\q) &= F_0 + \frac{a_m}{2}m^2 + \frac{a_q}{2}q^2 - \frac{c}{3}m^2q\cos(2\theta+\phi) \ \ \ \ \ \ \nonumber\\
	&\ \ \ + \frac{b_m}{4}m^4 + \frac{b_q}{4}q^4 + \frac{b_{mq}}{4}m^2q^2, \\
	F_0 &= -\beta^{-1}\ln Z_0, \ 
	a_m = \frac{-1}{\tilde{a}_m},\ 
	a_q = \frac{-1}{\tilde{a}_q},\\
	c &= \frac{\tilde{c}}{\tilde{a}_m^2\tilde{a}_q},  \ 
	b_m = \frac{1}{\tilde{a}_m^4}\qty(\tilde{b}_m-\frac{4\tilde{c}^2}{9\tilde{a}_q}), \\
	b_q &= \frac{\tilde{b}_q}{\tilde{a}_q^4},\ 
	b_{mq} = \frac{1}{\tilde{a}_m^2\tilde{a}_q^2}\qty(\tilde{b}_{mq}-\frac{16\tilde{c}^2}{9\tilde{a}_m}).
\end{align}
Note that $a_m,a_q,c,b_m,b_q,b_{mq}$ are all positive. In the main text, we do not use the terms proportional to $q^4$ and $m^2q^2$ in Sec.~\ref{subsec:FreeEnergy}, since the quadrupole moment is not the primary order parameters. However, in the microscopic mean-field analysis in Sec.~\ref{sec:results}, such terms are implicitly included and play a role in determining the stable phases e.g., for low temperatures.

\section{Minimization of $F_{4{\rm m}}^{\rm loc}+\delta F_4$}
\label{sec:App_Minimization}
We discuss $F^{\rm tot}_{4{\rm m}}=F_{4{\rm m}}^{\rm loc}+\delta F_{4{\rm m}}$, where the two terms are defined in Eqs.~(\ref{eq:F4locFin}) and (\ref{eq:dF4}). First, we minimize $F^{\rm tot}_{4{\rm m}}$ in terms of the phase degrees of freedom $\delta_n~(n=1,2,3)$. The terms including $\delta_n$ are represented as $F_{4\delta}$: 
\begin{align}
	F_{4\delta}(x,y)=-\frac{4c^2}{9}\qty(\frac{1}{a^{\rm Q}} -\frac{1}{a^{\rm Q}_{\rm K}}) m_1m_2m_3g(x,y),
\end{align}
where $x\equiv 2\delta_3-\delta_1-\delta_2$,  $y=2\delta_2-\delta_3-\delta_1$, and 
\begin{align}
	g(x,y)\equiv m_3\cos x+m_2\cos y+m_1\cos(x+y).
\end{align}
Note that  $\delta_n$ dependence of $F_{4\delta}$ arises from the two variables $x$ and $y$ through $g(x,y)$. Differentiating $g(x,y)$ by $x$ and $y$, we obtain the following stationary conditions
\begin{align}
	&\sin x+\frac{m_1}{m_3}\sin(x+y)=0,\label{eq:sinx}\\
	&\frac{m_2}{m_3}\sin y + \frac{m_1}{m_3}\sin(x+y)=0.
\end{align}
From the above two equations, one also finds 
\begin{align}
	\frac{m_2}{m_3}\sin y =\sin x.
\end{align} 
\subsection{$a^{\rm Q}<a^{\rm Q}_{\rm K}$} 
\label{sec:App_triple1}
First, we consider the case for $a^{\rm Q}<a^{\rm Q}_{\rm K}$. 
There are trivial solutions: $x,y=2\ell \pi$ with $\ell=0, \pm 1$, which leads to $g(x,y)=m_1+m_2+m_3$. We have checked numerically that the symmetric solution with $m_1=m_2=m_3$ has the lowest free energy for $a_{\rm K}^{\rm Q}<a_0^{\rm Q}$. Remember that for $a^{\rm Q}<a_{\rm K}^{\rm Q}<a_0^{\rm Q}$ the single-$\mathcal{Q}$ is stable as demonstrated in Fig.~\ref{fig:PhaseLandau}. The above results are valid only near the highest second-order transition temperature from the paramagnetic state. It is generally possible that other configurations are stabilized at low temperatures.

The $\delta_n$'s for the symmetric solution are $(\delta_1,\delta_2,\delta_3)=(\delta,\delta,\delta),(\delta+\omega_1,\delta+\omega_2,\delta+\omega_3)$, where $\delta$ is arbitrary and $\omega_n=2n\pi/3$. Note that any permutations $\delta_i\leftrightarrow \delta_j$ and simultaneous sign changes of all the $\delta_{i}$ lead to different domains with the same class of phase. These properties are also the case for other configurations.

\subsection{$a^{\rm Q}>a^{\rm Q}_{\rm K}$}
\label{sec:App_triple2}

 When $a^{\rm Q}>a^{\rm Q}_{\rm K}$, we need to search stationary solutions with $g(x,y)<0$. Since the analysis for general $m_n$ are complicated, we restrict ourselves on the limiting cases. When $m_2=m_3$, one can find a simple solution with $x=y$, which is sufficient for our discussion in Sec.~\ref{subsec:FreeEnergy}. Equation (\ref{eq:sinx}) readily reads as
\begin{align}
	\cos x=-\frac{m_3}{2m_1}\to x=\pm\frac{2\pi}{3}\ {\rm for} \ m_1=m_3.
\end{align}
When $m_1=m_2=m_3$, $x,y= 2\pi/3$ and this leads to $(\delta_1,\delta_2,\delta_3)=(\delta,\delta,\delta+2\pi/3)$, $(\delta,\delta+2\pi/3,\delta)$, and $(\delta+ 2\pi/3,\delta,\delta)$.

For checking the stability of the symmetric solution with $m_1=m_2=m_3$, we relax the condition $m_1=m_2$. We find $g(x,y)=-m_2^2/(2m_1)-m_1$, and 
\begin{align}
F_{4{\rm m}}^{\rm loc}+\delta F_{4{\rm m}} &=
\frac{3b}{2}R^4-\frac{c^2}{9}  \left(\frac{2}{a^{\rm Q}_0}+\frac{1}{a^{\rm Q}}\right)R^4 \nonumber\\
&+ \frac{2c^2}{3}  \left(\frac{1}{a^{\rm Q}_0}-\frac{1}{a^{\rm Q}_{\rm K}}\right)\left(2m^2_1m_2^2+m_2^4\right),
	\end{align}
	where $R^2\equiv m_1^2+2m_2^2$. It is trivial to find that the stationary solution is that with $m_1=m_2=m_3$. Again, this analysis is valid near the second-order transition temperature between paramagnetic and symmetry-broken phases.

\subsection{$C_3$ symmetric states}
Here, we briefly show that the above two symmetric triple-$\mathcal{Q}$ states preserve $C_3$ rotational symmetry along the $c$ axis. Let us consider the real-space magnetic configuration consisting of ${\bm m}_n~(n=1,2,3)$ :
\begin{align}
	{\bm m}(\vecr)=\sum_{n=1,2,3}{\bm m}_n\cos(\veck_n\cdot \vecr + \delta_n). \label{eq:Mr_triple}
\end{align}
Denoting a space-group operation consisting of the $C_3$ rotation and translation $\bm{T}$ in the Seitz symbol as $\{ C_3|{\bm T}\}$, we obtain
\begin{align}
	& \{ C_3|{\bm T}\}{\bm m}_n \cos(\veck_n \cdot \vecr + \delta_n)\nonumber\\
	&=	{\bm m}_{n+1} \cos(\veck_{n+1} \cdot \vecr + \veck_n \cdot {\bm T} +\delta_n).
\end{align} 
For Eq.~(\ref{eq:Mr_triple}) being preserved under $\{ C_3|{\bm T}\}$, $\delta_{n+1}-\delta_{n}= \veck_n \cdot {\bm T}$ must be satisfied. 
For simplicity, we restrict ourselves within 3 $\times$ 3 sublattice orders.
There are two such classes of the translation vectors ${\bm T}=n_1\bm{a}_1+n_2\bm{a}_2$: $(n_1,n_2)=(0,0),\ \pm(0,1),\ \pm(1,0),\ \pm(1,1)$, and $\pm(1,-1)$ within the 3 $\times$ 3 sublattice orders.

{\flushleft{Case 1:}}
\begin{align}
(n_1,n_2)= 
\begin{cases}
(0,0): & {\bm{\delta}}=\{ \delta,\delta,\delta \}, \\
(1, 1): & {\bm{\delta}}= \{\delta+\omega_1,\delta+\omega_2,\delta+\omega_3\}.
\end{cases} 
\label{eq:phase_case-I}
\end{align}
Case 2:
\begin{align}
(n_1,n_2)= 
\begin{cases}
(1,-1): & {\bm{\delta}}= \{\delta+2\pi/3,\delta,\delta \}, \\
(1, 0): & {\bm{\delta}}= \{\delta,\delta+2\pi/3,\delta \}, \\
(0, 1): & {\bm{\delta}}= \{\delta,\delta,\delta+2\pi/3 \}.
\end{cases}
\label{eq:phase_case-II}
\end{align}
Here, we have defined ${\bm{\delta}}\equiv \{ \delta_1,\delta_2,\delta_3 \}$. For $(-n_1,-n_2)$, one simply replaces $\bm{\delta}\to -\bm{\delta$} for the corresponding $(n_1,n_2)$.
Case 1 corresponds to the symmetric triple-$\mathcal{Q}$ for $a^{\rm Q}<a_{\rm K}^{\rm Q}$ discussed in Appendix.~\ref{sec:App_triple1} and case 2 is for $a^{\rm Q}>a_{\rm K}^{\rm Q}$ discussed in Appendix.~\ref{sec:App_triple2}.

\section{Phase fixing by $F^{\rm loc}_{6\rm{m}}$}
\label{sec:App_sixth}
Here, we show that the local sixth-order term $F^{\rm loc}_{6\rm{m}}$ determines the phase degree of freedom for the triple-(2) state and the single-$\mathcal{Q}$ state, which have remained arbitrary in the free energy up to the fourth order $F^{\rm tot}_{4\rm{m}}$ in Eqs.~(\ref{eq:phase_case-II}) and (\ref{eq:induced_q_single}).  
$F^{\rm loc}_{6\rm{m}}$ is given by
\begin{align}
	F^{\rm loc}_{6\rm{m}}&=\frac{d}{6N}\sum_\vecr \left[\sum_{\mu} m_\mu(\vecr)m_\mu(\vecr)\right]^3 
\label{eq:F_loc6}
,\end{align} 
where $d>0$ is a coefficient.  
For the triple-(2) state, substituting Eqs.~(\ref{eq:Mr_triple}) and (\ref{eq:phase_case-II}) into Eq.~(\ref{eq:F_loc6}), we obtain
\begin{align}
 F^{\rm triple(2)}_{6\rm{m}} &= \frac{d}{6}\left[ 8\sin^6 \delta -12\sin^4 \delta+\frac{9}{2}\sin^2\delta + \frac{9}{4} \right]m^6
 \label{eq:F6_triforce}
.\end{align}
The minima of Eq.~(\ref{eq:F6_triforce}) are at $\delta=n\pi/3$ ($n=0,\pm 1, \pm 2, 3$).  
Similarly, $F^{\rm single}_{6\rm{m}}$ for the single-$\mathcal{Q}$ order is obtained 
\begin{align}
 F^{\rm single}_{6\rm{m}} &= \frac{d}{6}\left[ 8\cos^6 \delta -12\cos^4 \delta+\frac{9}{2}\cos^2\delta + \frac{9}{4}\right]m^6
 \label{eq:F6_single}
,\end{align}
with its minima at $\delta=\pi/6+n\pi/3$ ($n=0,\pm 1, \pm 2, 3$).  
Finally, the phase factor for the triple-(1) state [Eq.~(\ref{eq:phase_case-I})] is not fixed by $F^{\rm loc}_{6\rm{m}}$ and even by the higher-order terms in the local free energy.  This is obvious since any choice of $\delta$ gives the same magnitudes of $m_\mu(\vecr)$ for the six of nine sites in the magnetic unit cell, and the other three remain disordered.  
This is an accidental degeneracy and lifted by effects beyond our model, e.g., the sixth-order anisotropic term due to the local anisotropy absent here.

\section{Results for other interaction parameters}\label{app:ResultsOther}
In the main text, we have shown the results for $J_1^M=0$, and this is because $J_1^M$ is not important for the orders at $\veck_n$. In this appendix, we will briefly show the results for finite $J_1^M$ in Appendix ~\ref{sec:App_finite_J1M} and for parameter sets without the experimental constraints in $J_1^Q$ and $J_2^Q$ in Appendix~\ref{sec:App_fewer_param}. The effects of the anisotropic interaction $K_1^M$ are also examined. In Appendix \ref{subsec:OtherQuadrupole}, as an example of various possible fine tuning of the microscopic parameters for the low-$T$ Curie-Weiss behavior observed in Ref. \onlinecite{Yanagisawa2021}, we will examine the effect of $A_{1g}$ quadrupole ($O_{20}$) interactions.  This appendix can help readers understand the more global situation/phase diagram than that for the parameters suitable to UNi$_4$B used in the main text. 

\subsection{Effect of finite $J_1^M$}
\label{sec:App_finite_J1M}
Here, we show the mean-field results for finite $J_1^M$ in order to check the phases discussed in the main text exist for finite $J_1^M$ as far as $J_1^M$ does not alter the leading instability. 

\begin{figure}[t]
\begin{center}
\includegraphics[width=0.48\textwidth]{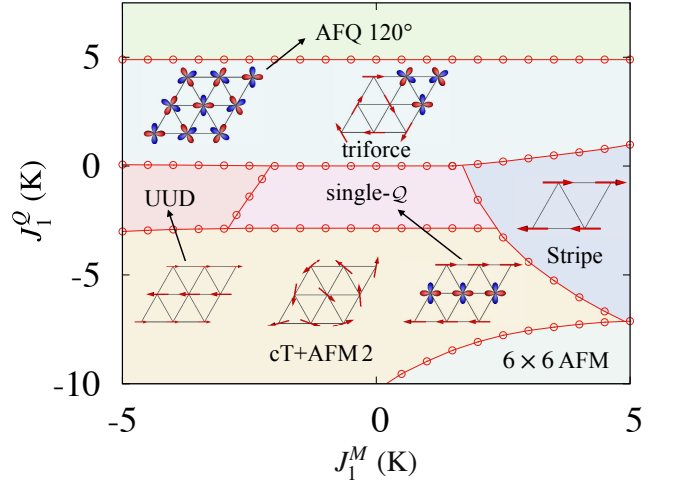}
\end{center}

\caption{$J_1^M$--$J_1^Q$ phase diagram with the constraint (\ref{eq:TCW_Q}) at $T=0$. The other parameters are the same as those in Fig. \ref{fig:PhaseDiagram}.  All the phase transitions are of the first-order.  For each phase, schematic configurations of magnetic dipole and electric quadrupole moments are illustrated. }
\label{fig:finite_J1M}
\end{figure}

Figure \ref{fig:finite_J1M} shows the $J_1^M$--$J_1^Q$ phase diagram at $T=0$. Other interaction parameters are the same as those in Fig. \ref{fig:PhaseDiagram}. 
Since the eigenvalue of $\hat{J}^M_\p$ has minima at the M points ${\bm k}_{\rm M}=(1/2,0)$ and the equivalent ones for 
$J_1^M>4$ K, there appear magnetic orders with the ordering vector at $\veck_{\rm M}$. This is clearly seen in Fig.~\ref{fig:finite_J1M} as a stripe phase for positive $J_1^M\gtrsim 2.5$ K. Other differences from Fig. \ref{fig:PhaseDiagram} include an up-up-down (UUD) phase that appears as an alternative of the single-$\mathcal{Q}$ phase for negative $J_1^M\lesssim -3$ K. The UUD phase has a finite magnetization and is stabilized for negatively large $J_1^M$.  The triforce, 120$^\circ$ AFQ, and canted AFM(cT+AFM2) orders are more stable than the single-$\mathcal{Q}$ order. For larger $J_1^M$ and $-J_1^Q$, a low-symmetry AFM order with 6$\times$6 magnetic unit cell is stabilized. From these results, one can understand that the simplified parametrization in the main text with $J_1^M=0$ contains the essential aspects of the triple-$\mathcal{Q}$ orders at $\veck_n$ for small $J_1^M$.

\subsection{Results without experimental constraints on $J_1^Q$ and $J_2^Q$}
\label{sec:App_fewer_param}
We have considered five interaction parameters $J_1^M$, $J_2^M$, $J_1^Q$, $J_2^Q$, and $K_1^M$, with $J_1^M=0$ in the main text. We have applied the experimental constraints (\ref{eq:TN})--(\ref{eq:TCW_Q}) and discussed possible orders with the ordering vector $\veck_n$ and which one can explain the observed ME effect in UNi$_4$B. Although the detailed analysis for more general parameter space is not our primary purpose in this paper, it is important to understand the stability of various phases discussed in the main text against variation in our parameters without the constraints.

\begin{figure*}[t]
\begin{center}
\includegraphics[width=0.98\textwidth]{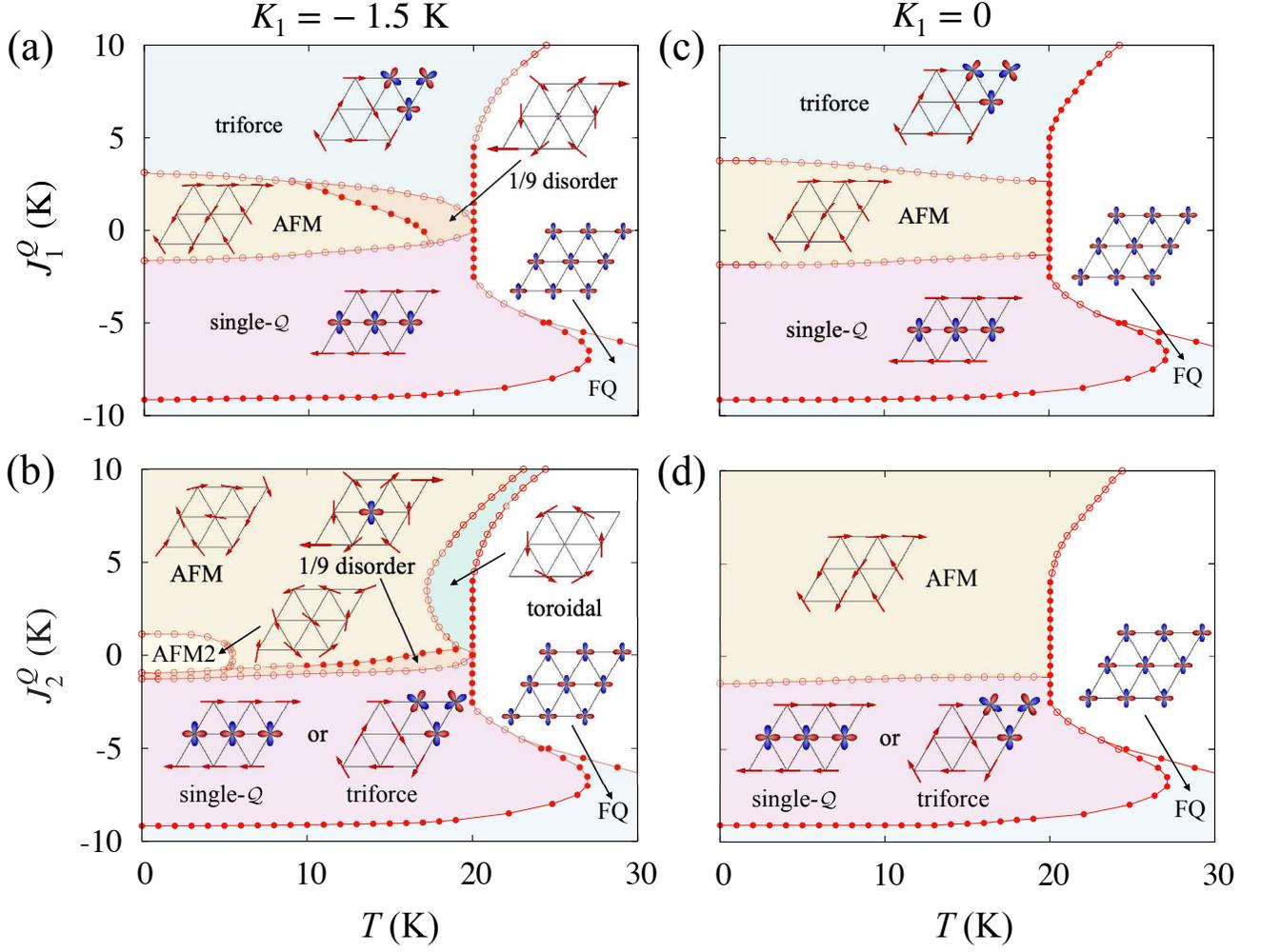}
\end{center}

\caption{$T$--$J_{1(2)}^Q$ phase diagrams without the $\theta_{\rm CW}^Q$ constraint (\ref{eq:TCW_Q}). (a) $T$--$J_1^Q$ phase diagram for $K_1^M=-1.5$, $J_2^Q=0$, and $J^M_2=11$ K. (b) $T$--$J_2^Q$ phase diagram for $K_1^M=-1.5$, $J_1^Q=0$, and $J_2=11$ K. (c) $T$--$J_1^Q$ phase diagram for $K_1^M=0$, $J_2^Q=0$, and $J_2^M=12.5$ K. (d) $T$--$J_2^Q$ phase diagram for $K_1^M=0$, $J_1^Q=0$, and $J_2^M=12.5$ K. The phase boundaries drawn by the filled circles represent second-order transitions, while the open circles mean first-order ones.   For each phase, schematic configurations of magnetic dipole and electric quadrupole moments are illustrated. States in (c) and (d) are isotropic and can be rotated globally. As a representative state, the order parameters corresponding to the infinitesimal $K_1^M<0$ are shown. }
\label{fig:fewer_param}
\end{figure*}

Figure \ref{fig:fewer_param} shows the results when the condition (\ref{eq:TCW_Q}) about $\theta_{\rm CW}^Q$ is not imposed and either $J_1^Q$ or $J_2^Q$ is finite. Figures \ref{fig:fewer_param}(a) and \ref{fig:fewer_param}(b) show the result for $K_1^M=-1.5$ K, the same as the main text, while Figs. \ref{fig:fewer_param}(c) and \ref{fig:fewer_param}(d) show the results for $K_1^M=0$ and $J_2^M=12.5$ K. 

In either case, the triforce and single-$\mathcal{Q}$ orders are stable at low temperatures when the quadrupole interactions are sufficiently large. 
The degeneracy of the triforce and single-$\mathcal{Q}$ orders for $J_2^Q<0$ and $J_1^Q=0$ is an accidental one, which is lifted by infinitesimal $J_1^Q$; triforce (single-$\mathcal{Q}$) order is favored for $J_1^Q>0$ ($J_1^Q<0$). See also the free energy expressions (\ref{eq:F_single2}) and (\ref{eq:F_triforce2}), and Table \ref{table:Jq}.

Let us focus on the regions where the quadrupole interactions are small in Fig.~\ref{fig:fewer_param}. 
For $K_1^M=-1.5$ K, a partial disordered phase with one-ninth of sites remaining disordered (1/9 disorder) is realized at high temperatures. In the quite narrow range shown in Fig. \ref{fig:fewer_param}(b), this is stable down to $T=0$. This phase has $C_2$ rotational symmetry but no $C_3$ symmetry.  Thus, quadrupole moments are induced at the magnetically-disordered sites. 
For $K_1^M=0$, a fully magnetically ordered state (denoted simply as AFM) is stable from high to low temperatures. 

We note that (i) the presence or absence of $K_1^M$ generates only the detail difference both for $J_1^Q$ and $J_2^Q$ variations as shown in Fig.~\ref{fig:fewer_param}. This is clear by the observation of similar shapes of the phases between Fig.~\ref{fig:fewer_param}(a) and (c), and (b) and (d). (ii) the triforce or single-${\mathcal Q}$ orders appear for a wide range of parameter space for relatively large $J_{1,2}^Q$. This is also consistent with the fact that the quadrupole degrees of freedom are important for their realization, as discussed in Sec.~\ref{subsec:FreeEnergy}.

\subsection{An example of fine-tuning about the low-$T$ quadrupolar Curie-Weiss temperature: $O_{20}$ quadrupole interactions}
\label{subsec:OtherQuadrupole}
We now examine the effects of quadrupole interaction on the quadrupole saturation scale $T^{\star}$ (or $T^*$ as denoted for the experimental data). As discussed in Sec.~\ref{subsubsec:Thermo}, the characteristic temperature scale $T^\star\sim 5$ K in the numerical results in the main text is much higher than $T^*\sim 0.3$ K \cite{Yanagisawa2021}. In the mean-field approximation of the localized model, it is necessary to tune the quadrupole exchange interaction at the $\Gamma$ point $J^Q_\Gamma$ small for decreasing $T^\star$. However, for the smaller quadrupole interactions, the more fragile the triforce or single-$\mathcal{Q}$ phases are. In Appendix~\ref{subsec:JQ_k0}, we will show the $J^Q_\Gamma$ dependence of the phase diagram by controlling $J_2^Q$ with fixed $J_1^Q$ for the similar parameter set used in the main text. Then, in Appendix~\ref{subsec:J_O20}, we will introduce additional $O_{20}$ quadrupole interactions 
and try to search parameter sets satisfying both the small $T^\star$ and the stable triforce phase.

\subsubsection{$J^Q_{\Gamma}$ dependence}
\label{subsec:JQ_k0}
\begin{figure*}[t]
\begin{center}
\includegraphics[width=0.85\textwidth]{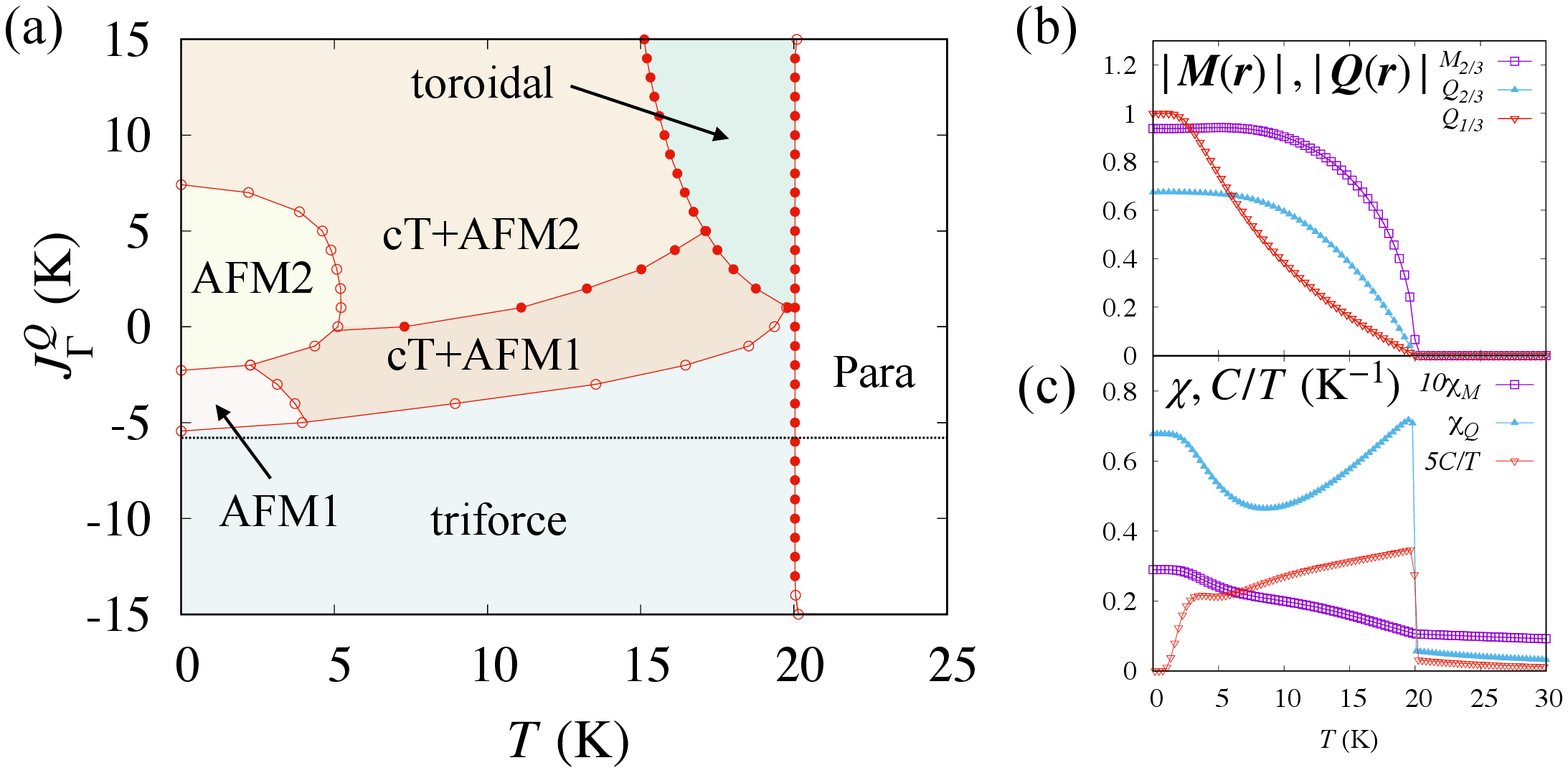}
\end{center}

\caption{Phase diagrams and temperature dependencies of order parameters and thermodynamical quantities. 
(a) $T$--$J^Q_{\Gamma}$ phase diagram. The interaction parameters are $(J_1^M,J_2^M, J_1^Q,J_2^Q,K_1^M)=(0,11,0.15,J^Q_{\Gamma}/6-0.15,-1.5)$ K.  
(b) Temperature dependence of the magnetic $M_{1/3,2/3}$ and quadrupole $Q_{1/3,2/3}$  moment in the real space. See the definition in the caption in Fig.~\ref{fig:triforce_OP_Chi}.  
(c) Temperature dependence of magnetic (quadrupole) susceptibilities $\chi_M~(\chi_Q)$ and specific heat $C$ divided by $T$, $C/T$. The interaction parameters in (b) and (c) are $(J_1^M,J_2^M,J_1^Q,J_2^Q,K_1^M)=(0,11,0.15,-1.15,-1.5)$ K, which correspond to $J_{\Gamma}^Q=-6$ K as indicated by the horizontal line in (a).}
\label{fig:Other_JQ}
\end{figure*}

First, we discuss the variation of $J^Q_{\Gamma}$.  Figure \ref{fig:Other_JQ}(a) shows $T$--$J^Q_{\Gamma}$ phase diagram for $(J_1^M,J_2^M,J_1^Q,J_2^Q,K_1^M)=(0,11,0.15,J^Q_{\Gamma}/6-0.15,-1.5)$ K, which are the same as those in Fig.~\ref{fig:triforce_OP_Chi} except for $J^Q_2$.  The triforce order is stable at $T=0$ for $J^Q_{\Gamma} \lesssim -5$ K.  The other phases at $T=0$ are magnetic without magnetically disordered sites.  Figure \ref{fig:Other_JQ}(b) shows the temperature dependence of the magnitudes of the order parameters in the real space for $J^Q_{\Gamma}=-6$ K.  In comparison with that in Fig.~\ref{fig:triforce_OP_Chi}(b) where $J^Q_{\Gamma}=-11$ K, the downward convex $T$ dependence behavior of $Q_{1/3}$ is more prominent.  Meanwhile, the quadrupole susceptibility $\chi_Q$ increases down to $T^\star\sim 3$ K, and $C/T$ has a peak at $\sim 3$ K as shown in \ref{fig:Other_JQ}(c).  However, the complete Curie-Weiss fitting of $\chi_Q$ in the ordered phase is not successful since $\chi_Q$ saturate at $\sim 3$ K.  This is because the mean fields acting on the quadrupole moments at the magnetically disordered sites from the magnetically ordered sites cannot be ignored.

\subsubsection{$O_{20}$ quadrupole interaction}
\label{subsec:J_O20}
So far, we have considered the interactions of in-plane magnetic dipole and electric quadrupole moments. Here, we discuss $O_{20}$ quadrupole moments, which affect the order parameters since the $O_{20}$ belongs to the totally symmetric representation and couples to any components of the order parameters. In particular, $O_{20}$ interactions behave as a temperature-dependent CEF energy $E_4$. They affect the relative stability between the magnetic orders and the quadrupole ones. For other quadrupole moments, such as the two-dimensional $O_{yz,zx}$ is expected to be irrelevant since they are not coupled with the primary in-plane magnetic moments. 

 We should comment that this parameter choice is not based on the microscopic information, such as the spin-wave fitting of the inelastic neutron scattering data or the first-principle calculations. The results shown below are aimed to demonstrate a possible example to describe the quantitative aspect of the quadrupole susceptibility in UNi$_4$B, and we do not rule out other unknown quantitative explanations.

\begin{figure*}[t]
\begin{center}
\includegraphics[width=0.85\textwidth]{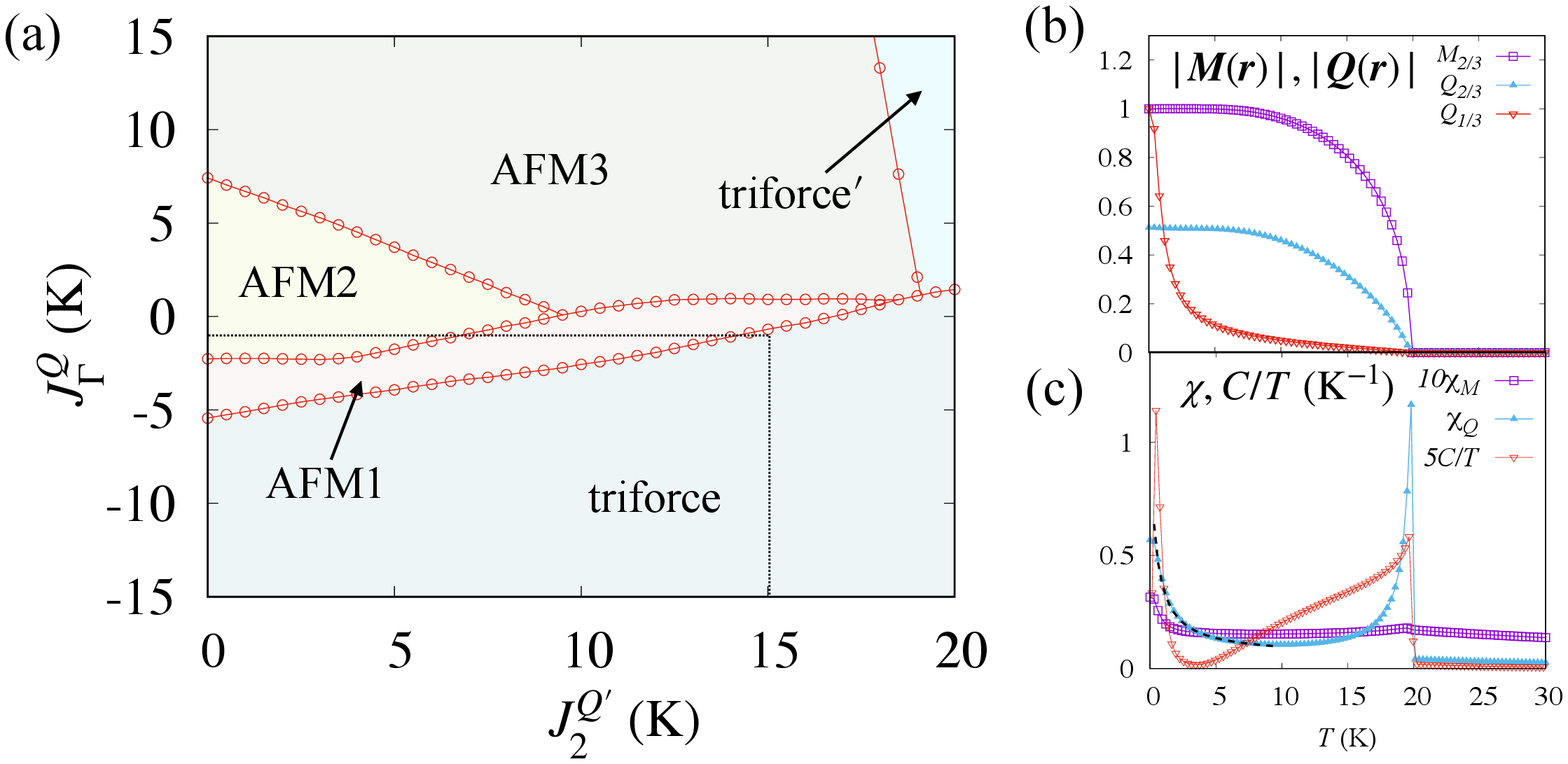}
\end{center}

\caption{(a) $J_2^{Q'}$--$J_\Gamma^Q$ phase diagram for $(J_1^M,J_2^M,J_1^Q,J_2^Q,K_1^M)=(0,11,0.15,J^Q_{\Gamma}/6-0.15,-1.5)$ K, and $J_1^{Q^\prime}=-J_2^{Q^\prime}$.  The vertical line is a guide for $J_2^{Q^\prime}=15$ K, and the horizontal line is  for $J_{\Gamma}^{Q}=-1$ K. (b) Temperature dependencies of  order parameters for $J_1^{Q^\prime}=-J_2^{Q^\prime}=-15$ and $J_\Gamma^Q=-1$ K [the crossing point of the two guided lines in (a)]. 
(c) Thermodynamical quantities: magnetic (quadrupole) susceptibilities $\chi_M~(\chi_Q)$ and specific heat $C$ divided by $T$, $C/T$ for the same parameter set in (b). In (b), $M_{1/3,2/3}$ and $Q_{1/3,2/3}$ represent the magnetic (quadrupole) moment in the real space. See the definition in the caption in Fig.~\ref{fig:triforce_OP_Chi}. The dashed line in (c) shows the Curie-Weiss fitting in $0.3~{\rm K}<T<10~{\rm K}$. }
\label{fig:O20}
\end{figure*}

The quadrupole $O_{20}$ is defined in Eq.~(\ref{eq:Ope_Jz}).  We consider up to the next-nearest-neighbor interactions $J_1^{Q^\prime}$ and $J_2^{Q^\prime}$.  To simplify the discussion, we take a constraint.  The interaction at the $\Gamma$ point $J^{Q^\prime}_{\Gamma}$ is fixed to zero since finite $J^{Q^\prime}_{\Gamma}$ modifies the transition temperature $T_{\rm N}$.  Thus, the parameter that we can vary is $J_2^{Q^\prime}=-J_1^{Q^\prime}$.  Note that the triforce and the single-$\mathcal{Q}$ orders induce the $O_{20}$ moments at ${\bm k}_{1,2,3}$ and will be stabilized by the positive $J_2^{Q^\prime}$ leading to $J^{Q^\prime}_{{\bm k}_{1,2,3}}<0$, while the toroidal order induces the $O_{20}$ moment at the K point, which is stabilized by the negative $J_2^{Q^\prime}$ leading to $J^{Q^\prime}_{{\bm k}_{\rm K}}<0$.  The wave vectors of induced $O_{20}$ moments are straightforwardly obtained by the coupling $q^{20}_{\q_1}{\bm m}_{\q_2}\cdot{\bm m}_{\q_3}$, where $q^{20}$ is the $O_{20}$-type quadrupole field and $\q_1+\q_2+\q_3$ is a reciprocal lattice vector.  We will focus on the triforce order and set $J_2^{Q^\prime}>0$.  The impact of the $O_{20}$ interactions on the stability of the single-$\mathcal{Q}$ order is considered to be similar to that of the triforce order.  For the toroidal order, it cannot be realized at $T=0$ since it has disordered sites.  We have discussed the stability of the toroidal order under a more realistic situation reflecting the crystal structure in Sec.~\ref{sec:realcrys}

Figure \ref{fig:O20}(a) shows $J^{Q^\prime}_2$--$J^Q_{\Gamma}$ phase diagram at $T=0$.  The interaction parameters are $(J_1^M,J_2^M,J_1^Q,J_2^Q,K_1^M)=(0,11,0.15,J^Q_{\Gamma}/6-0.15,-1.5)$ K, and $J_1^{Q^\prime}=-J_2^{Q'}$. The triforce order is stable even at $J^Q_{\Gamma}=0$ for $J_2^{Q^\prime}\sim 20$ K.  The triforce$^\prime$ phase in large  $J^Q_{\Gamma}$ and $J^{Q^\prime}_2$ is similar to the triforce order but with a different quadrupole configuration at the magnetically disordered sites.  The other phases are magnetic ones, with all the sites magnetically ordered.  Figure \ref{fig:O20}(b) shows the temperature dependence of the magnitudes of the order parameters in the real space for $J_2^{Q^\prime}=15$ K and $J^Q_{\Gamma}=-1$ K.  As a consequence of the small in-plane quadrupole interactions, the downward convex behavior is more prominent.  Interestingly, the quadrupole susceptibility increases down to $T^\star\sim 0.3$ K, and $C/T$ has a peak also at $\sim0.3$ K as shown in Fig.~\ref{fig:O20}(c).  The quadrupole susceptibility is well fitted by the Curie-Weiss low in $0.3~{\rm K}<T<10~{\rm K}$ with a weak antiferro quadrupole interaction $\sim0.5$ K.  

The above parameter tuning leads to the small quadrupole energy scale $T^\star$.  However, the drawback of the small $J^{Q}_\Gamma$ is that the Curie-Weiss constraint (\ref{eq:TCW_Q}) is not satisfied, and the value of $J_2^{Q'}$ is rather large.  Such large $O_{20}$ coupling is not impossible but seems to be difficult to naively expect.  The examination of such a possibility and also explorations of other mechanisms to lower $T^*$ are important 
future problems for the triforce scenario.  The issue about the Curie-Weiss temperature $\theta_{\rm CW}^Q$ of the quadrupole susceptibility $\chi_Q$ at the paramagnetic phase includes some ambiguity since there is a jump at $T_{\rm N}$ in $\chi_Q$ in the mean-field approximation, which might influence the estimation of $\theta_{\rm CW}^Q$.  To clarify this, one needs more elaborate calculations beyond the mean-field theory.

\bibliography{Ref_UNi4B.bib}

\begin{thebibliography}{113}%
\makeatletter
\providecommand \@ifxundefined [1]{%
 \@ifx{#1\undefined}
}%
\providecommand \@ifnum [1]{%
 \ifnum #1\expandafter \@firstoftwo
 \else \expandafter \@secondoftwo
 \fi
}%
\providecommand \@ifx [1]{%
 \ifx #1\expandafter \@firstoftwo
 \else \expandafter \@secondoftwo
 \fi
}%
\providecommand \natexlab [1]{#1}%
\providecommand \enquote  [1]{``#1''}%
\providecommand \bibnamefont  [1]{#1}%
\providecommand \bibfnamefont [1]{#1}%
\providecommand \citenamefont [1]{#1}%
\providecommand \href@noop [0]{\@secondoftwo}%
\providecommand \href [0]{\begingroup \@sanitize@url \@href}%
\providecommand \@href[1]{\@@startlink{#1}\@@href}%
\providecommand \@@href[1]{\endgroup#1\@@endlink}%
\providecommand \@sanitize@url [0]{\catcode `\\12\catcode `\$12\catcode
  `\&12\catcode `\#12\catcode `\^12\catcode `\_12\catcode `\%12\relax}%
\providecommand \@@startlink[1]{}%
\providecommand \@@endlink[0]{}%
\providecommand \url  [0]{\begingroup\@sanitize@url \@url }%
\providecommand \@url [1]{\endgroup\@href {#1}{\urlprefix }}%
\providecommand \urlprefix  [0]{URL }%
\providecommand \Eprint [0]{\href }%
\providecommand \doibase [0]{https://doi.org/}%
\providecommand \selectlanguage [0]{\@gobble}%
\providecommand \bibinfo  [0]{\@secondoftwo}%
\providecommand \bibfield  [0]{\@secondoftwo}%
\providecommand \translation [1]{[#1]}%
\providecommand \BibitemOpen [0]{}%
\providecommand \bibitemStop [0]{}%
\providecommand \bibitemNoStop [0]{.\EOS\space}%
\providecommand \EOS [0]{\spacefactor3000\relax}%
\providecommand \BibitemShut  [1]{\csname bibitem#1\endcsname}%
\let\auto@bib@innerbib\@empty
\bibitem [{\citenamefont {Landau}\ \emph {et~al.}(1960)\citenamefont {Landau},
  \citenamefont {Lifshitz},\ and\ \citenamefont
  {Pitaevskii}}]{LandauLifshitzContinuousMedia}%
  \BibitemOpen
  \bibfield  {author} {\bibinfo {author} {\bibfnamefont {L.~D.}\ \bibnamefont
  {Landau}}, \bibinfo {author} {\bibfnamefont {E.~M.}\ \bibnamefont
  {Lifshitz}},\ and\ \bibinfo {author} {\bibfnamefont {L.~P.}\ \bibnamefont
  {Pitaevskii}},\ }\href@noop {} {\emph {\bibinfo {title} {Electrodynamics of
  continuous media}}}\ (\bibinfo  {publisher} {Pergamon press, Oxford},\
  \bibinfo {year} {1960})\BibitemShut {NoStop}%
\bibitem [{Dzy()}]{Dzyaloshinskii1959}%
  \BibitemOpen
  \href@noop {} {}\bibinfo {note} {I. E. Dzyaloshinskii, Zh. Eksp. Teor. Fiz.
  {\bf 37}, 881 (1959) [Sov. Phys. JETP {\bf 10}, 628 (1960)].}\BibitemShut
  {Stop}%
\bibitem [{Ast()}]{Astrov1960}%
  \BibitemOpen
  \href@noop {} {}\bibinfo {note} {D. N. Astrov, Zh. Eksp. Teor. Fiz. {\bf 38},
  984 (1960) [Sov. Phys. JETP {\bf 11}, 708 (1960)].}\BibitemShut {Stop}%
\bibitem [{\citenamefont {Nagaosa}\ \emph {et~al.}(2010)\citenamefont
  {Nagaosa}, \citenamefont {Sinova}, \citenamefont {Onoda}, \citenamefont
  {MacDonald},\ and\ \citenamefont {Ong}}]{Nagaosa2010}%
  \BibitemOpen
  \bibfield  {author} {\bibinfo {author} {\bibfnamefont {N.}~\bibnamefont
  {Nagaosa}}, \bibinfo {author} {\bibfnamefont {J.}~\bibnamefont {Sinova}},
  \bibinfo {author} {\bibfnamefont {S.}~\bibnamefont {Onoda}}, \bibinfo
  {author} {\bibfnamefont {A.~H.}\ \bibnamefont {MacDonald}},\ and\ \bibinfo
  {author} {\bibfnamefont {N.~P.}\ \bibnamefont {Ong}},\ }\href
  {https://doi.org/10.1103/RevModPhys.82.1539} {\bibfield  {journal} {\bibinfo
  {journal} {Rev. Mod. Phys.}\ }\textbf {\bibinfo {volume} {82}},\ \bibinfo
  {pages} {1539} (\bibinfo {year} {2010})}\BibitemShut {NoStop}%
\bibitem [{\citenamefont {Tokura}\ and\ \citenamefont
  {Nagaosa}(2018)}]{Tokura2018}%
  \BibitemOpen
  \bibfield  {author} {\bibinfo {author} {\bibfnamefont {Y.}~\bibnamefont
  {Tokura}}\ and\ \bibinfo {author} {\bibfnamefont {N.}~\bibnamefont
  {Nagaosa}},\ }\href {https://doi.org/10.1038/s41467-018-05759-4} {\bibfield
  {journal} {\bibinfo  {journal} {Nat. Commun.}\ }\textbf {\bibinfo {volume}
  {9}},\ \bibinfo {pages} {3740} (\bibinfo {year} {2018})}\BibitemShut
  {NoStop}%
\bibitem [{\citenamefont {Fert}\ \emph {et~al.}(2017)\citenamefont {Fert},
  \citenamefont {Reyren},\ and\ \citenamefont {Cros}}]{Fert2017}%
  \BibitemOpen
  \bibfield  {author} {\bibinfo {author} {\bibfnamefont {A.}~\bibnamefont
  {Fert}}, \bibinfo {author} {\bibfnamefont {N.}~\bibnamefont {Reyren}},\ and\
  \bibinfo {author} {\bibfnamefont {V.}~\bibnamefont {Cros}},\ }\href
  {https://doi.org/10.1038/natrevmats.2017.31} {\bibfield  {journal} {\bibinfo
  {journal} {Nature Reviews Materials}\ }\textbf {\bibinfo {volume} {2}},\
  \bibinfo {pages} {1} (\bibinfo {year} {2017})}\BibitemShut {NoStop}%
\bibitem [{\citenamefont {Nagaosa}\ and\ \citenamefont
  {Tokura}(2013)}]{Nagaosa2013}%
  \BibitemOpen
  \bibfield  {author} {\bibinfo {author} {\bibfnamefont {N.}~\bibnamefont
  {Nagaosa}}\ and\ \bibinfo {author} {\bibfnamefont {Y.}~\bibnamefont
  {Tokura}},\ }\href {https://doi.org/10.1038/nnano.2013.243} {\bibfield
  {journal} {\bibinfo  {journal} {Nat. Nanotechnol.}\ }\textbf {\bibinfo
  {volume} {8}},\ \bibinfo {pages} {899} (\bibinfo {year} {2013})}\BibitemShut
  {NoStop}%
\bibitem [{\citenamefont {R{\"o}ssler}\ \emph {et~al.}(2006)\citenamefont
  {R{\"o}ssler}, \citenamefont {Bogdanov},\ and\ \citenamefont
  {Pfleiderer}}]{Rossler2006}%
  \BibitemOpen
  \bibfield  {author} {\bibinfo {author} {\bibfnamefont {U.~K.}\ \bibnamefont
  {R{\"o}ssler}}, \bibinfo {author} {\bibfnamefont {A.~N.}\ \bibnamefont
  {Bogdanov}},\ and\ \bibinfo {author} {\bibfnamefont {C.}~\bibnamefont
  {Pfleiderer}},\ }\href {https://doi.org/10.1038/nature05056} {\bibfield
  {journal} {\bibinfo  {journal} {Nature}\ }\textbf {\bibinfo {volume} {442}},\
  \bibinfo {pages} {797} (\bibinfo {year} {2006})}\BibitemShut {NoStop}%
\bibitem [{\citenamefont {M{\"u}hlbauer}\ \emph {et~al.}(2009)\citenamefont
  {M{\"u}hlbauer}, \citenamefont {Binz}, \citenamefont {Jonietz}, \citenamefont
  {Pfleiderer}, \citenamefont {Rosch}, \citenamefont {Neubauer}, \citenamefont
  {Georgii},\ and\ \citenamefont {B{\"o}ni}}]{Muhlbauer2009}%
  \BibitemOpen
  \bibfield  {author} {\bibinfo {author} {\bibfnamefont {S.}~\bibnamefont
  {M{\"u}hlbauer}}, \bibinfo {author} {\bibfnamefont {B.}~\bibnamefont {Binz}},
  \bibinfo {author} {\bibfnamefont {F.}~\bibnamefont {Jonietz}}, \bibinfo
  {author} {\bibfnamefont {C.}~\bibnamefont {Pfleiderer}}, \bibinfo {author}
  {\bibfnamefont {A.}~\bibnamefont {Rosch}}, \bibinfo {author} {\bibfnamefont
  {A.}~\bibnamefont {Neubauer}}, \bibinfo {author} {\bibfnamefont
  {R.}~\bibnamefont {Georgii}},\ and\ \bibinfo {author} {\bibfnamefont
  {P.}~\bibnamefont {B{\"o}ni}},\ }\href
  {https://doi.org/10.1126/science.1166767} {\bibfield  {journal} {\bibinfo
  {journal} {Science}\ }\textbf {\bibinfo {volume} {323}},\ \bibinfo {pages}
  {915} (\bibinfo {year} {2009})}\BibitemShut {NoStop}%
\bibitem [{\citenamefont {Nayak}\ \emph {et~al.}(2017)\citenamefont {Nayak},
  \citenamefont {Kumar}, \citenamefont {Ma}, \citenamefont {Werner},
  \citenamefont {Pippel}, \citenamefont {Sahoo}, \citenamefont {Damay},
  \citenamefont {R{\"o}{\ss}ler}, \citenamefont {Felser},\ and\ \citenamefont
  {Parkin}}]{Nayak2017}%
  \BibitemOpen
  \bibfield  {author} {\bibinfo {author} {\bibfnamefont {A.~K.}\ \bibnamefont
  {Nayak}}, \bibinfo {author} {\bibfnamefont {V.}~\bibnamefont {Kumar}},
  \bibinfo {author} {\bibfnamefont {T.}~\bibnamefont {Ma}}, \bibinfo {author}
  {\bibfnamefont {P.}~\bibnamefont {Werner}}, \bibinfo {author} {\bibfnamefont
  {E.}~\bibnamefont {Pippel}}, \bibinfo {author} {\bibfnamefont
  {R.}~\bibnamefont {Sahoo}}, \bibinfo {author} {\bibfnamefont
  {F.}~\bibnamefont {Damay}}, \bibinfo {author} {\bibfnamefont {U.~K.}\
  \bibnamefont {R{\"o}{\ss}ler}}, \bibinfo {author} {\bibfnamefont
  {C.}~\bibnamefont {Felser}},\ and\ \bibinfo {author} {\bibfnamefont
  {S.~S.~P.}\ \bibnamefont {Parkin}},\ }\href
  {https://doi.org/10.1038/nature23466} {\bibfield  {journal} {\bibinfo
  {journal} {Nature}\ }\textbf {\bibinfo {volume} {548}},\ \bibinfo {pages}
  {561} (\bibinfo {year} {2017})}\BibitemShut {NoStop}%
\bibitem [{\citenamefont {K{\'e}zsm{\'a}rki}\ \emph {et~al.}(2015)\citenamefont
  {K{\'e}zsm{\'a}rki}, \citenamefont {Bord{\'a}cs}, \citenamefont {Milde},
  \citenamefont {Neuber}, \citenamefont {Eng}, \citenamefont {White},
  \citenamefont {R{\o}nnow}, \citenamefont {Dewhurst}, \citenamefont
  {Mochizuki}, \citenamefont {Yanai}, \citenamefont {Nakamura}, \citenamefont
  {Ehlers}, \citenamefont {Tsurkan},\ and\ \citenamefont
  {Loidl}}]{Kezsmarki2015}%
  \BibitemOpen
  \bibfield  {author} {\bibinfo {author} {\bibfnamefont {I.}~\bibnamefont
  {K{\'e}zsm{\'a}rki}}, \bibinfo {author} {\bibfnamefont {S.}~\bibnamefont
  {Bord{\'a}cs}}, \bibinfo {author} {\bibfnamefont {P.}~\bibnamefont {Milde}},
  \bibinfo {author} {\bibfnamefont {E.}~\bibnamefont {Neuber}}, \bibinfo
  {author} {\bibfnamefont {L.~M.}\ \bibnamefont {Eng}}, \bibinfo {author}
  {\bibfnamefont {J.~S.}\ \bibnamefont {White}}, \bibinfo {author}
  {\bibfnamefont {H.~M.}\ \bibnamefont {R{\o}nnow}}, \bibinfo {author}
  {\bibfnamefont {C.~D.}\ \bibnamefont {Dewhurst}}, \bibinfo {author}
  {\bibfnamefont {M.}~\bibnamefont {Mochizuki}}, \bibinfo {author}
  {\bibfnamefont {K.}~\bibnamefont {Yanai}}, \bibinfo {author} {\bibfnamefont
  {H.}~\bibnamefont {Nakamura}}, \bibinfo {author} {\bibfnamefont
  {D.}~\bibnamefont {Ehlers}}, \bibinfo {author} {\bibfnamefont
  {V.}~\bibnamefont {Tsurkan}},\ and\ \bibinfo {author} {\bibfnamefont
  {A.}~\bibnamefont {Loidl}},\ }\href {https://doi.org/10.1038/nmat4402}
  {\bibfield  {journal} {\bibinfo  {journal} {Nat. Mater.}\ }\textbf {\bibinfo
  {volume} {14}},\ \bibinfo {pages} {1116} (\bibinfo {year}
  {2015})}\BibitemShut {NoStop}%
\bibitem [{\citenamefont {Tokunaga}\ \emph {et~al.}(2015)\citenamefont
  {Tokunaga}, \citenamefont {Yu}, \citenamefont {White}, \citenamefont
  {R{\o}nnow}, \citenamefont {Morikawa}, \citenamefont {Taguchi},\ and\
  \citenamefont {Tokura}}]{Tokunaga2015}%
  \BibitemOpen
  \bibfield  {author} {\bibinfo {author} {\bibfnamefont {Y.}~\bibnamefont
  {Tokunaga}}, \bibinfo {author} {\bibfnamefont {X.~Z.}\ \bibnamefont {Yu}},
  \bibinfo {author} {\bibfnamefont {J.~S.}\ \bibnamefont {White}}, \bibinfo
  {author} {\bibfnamefont {H.~M.}\ \bibnamefont {R{\o}nnow}}, \bibinfo {author}
  {\bibfnamefont {D.}~\bibnamefont {Morikawa}}, \bibinfo {author}
  {\bibfnamefont {Y.}~\bibnamefont {Taguchi}},\ and\ \bibinfo {author}
  {\bibfnamefont {Y.}~\bibnamefont {Tokura}},\ }\href
  {https://doi.org/10.1038/ncomms8638} {\bibfield  {journal} {\bibinfo
  {journal} {Nat. Commun.}\ }\textbf {\bibinfo {volume} {6}},\ \bibinfo {pages}
  {7638} (\bibinfo {year} {2015})}\BibitemShut {NoStop}%
\bibitem [{\citenamefont {Seki}\ \emph {et~al.}(2012)\citenamefont {Seki},
  \citenamefont {Yu}, \citenamefont {Ishiwata},\ and\ \citenamefont
  {Tokura}}]{Seki2012}%
  \BibitemOpen
  \bibfield  {author} {\bibinfo {author} {\bibfnamefont {S.}~\bibnamefont
  {Seki}}, \bibinfo {author} {\bibfnamefont {X.~Z.}\ \bibnamefont {Yu}},
  \bibinfo {author} {\bibfnamefont {S.}~\bibnamefont {Ishiwata}},\ and\
  \bibinfo {author} {\bibfnamefont {Y.}~\bibnamefont {Tokura}},\ }\href
  {https://doi.org/10.1126/science.1214143} {\bibfield  {journal} {\bibinfo
  {journal} {Science}\ }\textbf {\bibinfo {volume} {336}},\ \bibinfo {pages}
  {198} (\bibinfo {year} {2012})}\BibitemShut {NoStop}%
\bibitem [{\citenamefont {Yu}\ \emph {et~al.}(2010)\citenamefont {Yu},
  \citenamefont {Onose}, \citenamefont {Kanazawa}, \citenamefont {Park},
  \citenamefont {Han}, \citenamefont {Matsui}, \citenamefont {Nagaosa},\ and\
  \citenamefont {Tokura}}]{Yu2010}%
  \BibitemOpen
  \bibfield  {author} {\bibinfo {author} {\bibfnamefont {X.~Z.}\ \bibnamefont
  {Yu}}, \bibinfo {author} {\bibfnamefont {Y.}~\bibnamefont {Onose}}, \bibinfo
  {author} {\bibfnamefont {N.}~\bibnamefont {Kanazawa}}, \bibinfo {author}
  {\bibfnamefont {J.~H.}\ \bibnamefont {Park}}, \bibinfo {author}
  {\bibfnamefont {J.~H.}\ \bibnamefont {Han}}, \bibinfo {author} {\bibfnamefont
  {Y.}~\bibnamefont {Matsui}}, \bibinfo {author} {\bibfnamefont
  {N.}~\bibnamefont {Nagaosa}},\ and\ \bibinfo {author} {\bibfnamefont
  {Y.}~\bibnamefont {Tokura}},\ }\href {https://doi.org/10.1038/nature09124}
  {\bibfield  {journal} {\bibinfo  {journal} {Nature}\ }\textbf {\bibinfo
  {volume} {465}},\ \bibinfo {pages} {901} (\bibinfo {year}
  {2010})}\BibitemShut {NoStop}%
\bibitem [{\citenamefont {Hayami}\ and\ \citenamefont
  {Motome}(2021)}]{Hayami2021_Itinerant}%
  \BibitemOpen
  \bibfield  {author} {\bibinfo {author} {\bibfnamefont {S.}~\bibnamefont
  {Hayami}}\ and\ \bibinfo {author} {\bibfnamefont {Y.}~\bibnamefont
  {Motome}},\ }\href {https://doi.org/10.1088/1361-648X/ac1a30} {\bibfield
  {journal} {\bibinfo  {journal} {J. Phys. Condens. Matter}\ }\textbf {\bibinfo
  {volume} {33}},\ \bibinfo {pages} {443001} (\bibinfo {year}
  {2021})}\BibitemShut {NoStop}%
\bibitem [{\citenamefont {Kanazawa}\ \emph {et~al.}(2016)\citenamefont
  {Kanazawa}, \citenamefont {Nii}, \citenamefont {Zhang}, \citenamefont
  {Mishchenko}, \citenamefont {De~Filippis}, \citenamefont {Kagawa},
  \citenamefont {Iwasa}, \citenamefont {Nagaosa},\ and\ \citenamefont
  {Tokura}}]{Kanazawa2016}%
  \BibitemOpen
  \bibfield  {author} {\bibinfo {author} {\bibfnamefont {N.}~\bibnamefont
  {Kanazawa}}, \bibinfo {author} {\bibfnamefont {Y.}~\bibnamefont {Nii}},
  \bibinfo {author} {\bibfnamefont {X.-X.}\ \bibnamefont {Zhang}}, \bibinfo
  {author} {\bibfnamefont {A.~S.}\ \bibnamefont {Mishchenko}}, \bibinfo
  {author} {\bibfnamefont {G.}~\bibnamefont {De~Filippis}}, \bibinfo {author}
  {\bibfnamefont {F.}~\bibnamefont {Kagawa}}, \bibinfo {author} {\bibfnamefont
  {Y.}~\bibnamefont {Iwasa}}, \bibinfo {author} {\bibfnamefont
  {N.}~\bibnamefont {Nagaosa}},\ and\ \bibinfo {author} {\bibfnamefont
  {Y.}~\bibnamefont {Tokura}},\ }\href {https://doi.org/10.1038/ncomms11622}
  {\bibfield  {journal} {\bibinfo  {journal} {Nat. Commun.}\ }\textbf {\bibinfo
  {volume} {7}},\ \bibinfo {pages} {11622} (\bibinfo {year}
  {2016})}\BibitemShut {NoStop}%
\bibitem [{\citenamefont {Fujishiro}\ \emph {et~al.}(2019)\citenamefont
  {Fujishiro}, \citenamefont {Kanazawa}, \citenamefont {Nakajima},
  \citenamefont {Yu}, \citenamefont {Ohishi}, \citenamefont {Kawamura},
  \citenamefont {Kakurai}, \citenamefont {Arima}, \citenamefont {Mitamura},
  \citenamefont {Miyake}, \citenamefont {Akiba}, \citenamefont {Tokunaga},
  \citenamefont {Matsuo}, \citenamefont {Kindo}, \citenamefont {Koretsune},
  \citenamefont {Arita},\ and\ \citenamefont {Tokura}}]{Fujishiro2019}%
  \BibitemOpen
  \bibfield  {author} {\bibinfo {author} {\bibfnamefont {Y.}~\bibnamefont
  {Fujishiro}}, \bibinfo {author} {\bibfnamefont {N.}~\bibnamefont {Kanazawa}},
  \bibinfo {author} {\bibfnamefont {T.}~\bibnamefont {Nakajima}}, \bibinfo
  {author} {\bibfnamefont {X.~Z.}\ \bibnamefont {Yu}}, \bibinfo {author}
  {\bibfnamefont {K.}~\bibnamefont {Ohishi}}, \bibinfo {author} {\bibfnamefont
  {Y.}~\bibnamefont {Kawamura}}, \bibinfo {author} {\bibfnamefont
  {K.}~\bibnamefont {Kakurai}}, \bibinfo {author} {\bibfnamefont
  {T.}~\bibnamefont {Arima}}, \bibinfo {author} {\bibfnamefont
  {H.}~\bibnamefont {Mitamura}}, \bibinfo {author} {\bibfnamefont
  {A.}~\bibnamefont {Miyake}}, \bibinfo {author} {\bibfnamefont
  {K.}~\bibnamefont {Akiba}}, \bibinfo {author} {\bibfnamefont
  {M.}~\bibnamefont {Tokunaga}}, \bibinfo {author} {\bibfnamefont
  {A.}~\bibnamefont {Matsuo}}, \bibinfo {author} {\bibfnamefont
  {K.}~\bibnamefont {Kindo}}, \bibinfo {author} {\bibfnamefont
  {T.}~\bibnamefont {Koretsune}}, \bibinfo {author} {\bibfnamefont
  {R.}~\bibnamefont {Arita}},\ and\ \bibinfo {author} {\bibfnamefont
  {Y.}~\bibnamefont {Tokura}},\ }\href
  {https://doi.org/10.1038/s41467-019-08985-6} {\bibfield  {journal} {\bibinfo
  {journal} {Nat. Commun.}\ }\textbf {\bibinfo {volume} {10}},\ \bibinfo
  {pages} {1059} (\bibinfo {year} {2019})}\BibitemShut {NoStop}%
\bibitem [{\citenamefont {Okumura}\ \emph {et~al.}(2020)\citenamefont
  {Okumura}, \citenamefont {Hayami}, \citenamefont {Kato},\ and\ \citenamefont
  {Motome}}]{Okumura2020}%
  \BibitemOpen
  \bibfield  {author} {\bibinfo {author} {\bibfnamefont {S.}~\bibnamefont
  {Okumura}}, \bibinfo {author} {\bibfnamefont {S.}~\bibnamefont {Hayami}},
  \bibinfo {author} {\bibfnamefont {Y.}~\bibnamefont {Kato}},\ and\ \bibinfo
  {author} {\bibfnamefont {Y.}~\bibnamefont {Motome}},\ }\href
  {https://doi.org/10.1103/PhysRevB.101.144416} {\bibfield  {journal} {\bibinfo
   {journal} {Phys. Rev. B}\ }\textbf {\bibinfo {volume} {101}},\ \bibinfo
  {pages} {144416} (\bibinfo {year} {2020})}\BibitemShut {NoStop}%
\bibitem [{\citenamefont {Aoyama}\ and\ \citenamefont
  {Kawamura}(2021)}]{Aoyama2021}%
  \BibitemOpen
  \bibfield  {author} {\bibinfo {author} {\bibfnamefont {K.}~\bibnamefont
  {Aoyama}}\ and\ \bibinfo {author} {\bibfnamefont {H.}~\bibnamefont
  {Kawamura}},\ }\href {https://doi.org/10.1103/PhysRevB.103.014406} {\bibfield
   {journal} {\bibinfo  {journal} {Phys. Rev. B}\ }\textbf {\bibinfo {volume}
  {103}},\ \bibinfo {pages} {014406} (\bibinfo {year} {2021})}\BibitemShut
  {NoStop}%
\bibitem [{\citenamefont {Nagaosa}\ and\ \citenamefont
  {Tokura}(2012)}]{Nagaosa2012_Emergent}%
  \BibitemOpen
  \bibfield  {author} {\bibinfo {author} {\bibfnamefont {N.}~\bibnamefont
  {Nagaosa}}\ and\ \bibinfo {author} {\bibfnamefont {Y.}~\bibnamefont
  {Tokura}},\ }\href {https://doi.org/10.1088/0031-8949/2012/T146/014020}
  {\bibfield  {journal} {\bibinfo  {journal} {Phys. Scr.}\ }\textbf {\bibinfo
  {volume} {2012}},\ \bibinfo {pages} {014020} (\bibinfo {year}
  {2012})}\BibitemShut {NoStop}%
\bibitem [{\citenamefont {Nagaosa}\ \emph {et~al.}(2012)\citenamefont
  {Nagaosa}, \citenamefont {Yu},\ and\ \citenamefont
  {Tokura}}]{Nagaosa2012_Gauge}%
  \BibitemOpen
  \bibfield  {author} {\bibinfo {author} {\bibfnamefont {N.}~\bibnamefont
  {Nagaosa}}, \bibinfo {author} {\bibfnamefont {X.~Z.}\ \bibnamefont {Yu}},\
  and\ \bibinfo {author} {\bibfnamefont {Y.}~\bibnamefont {Tokura}},\ }\href
  {https://doi.org/10.1098/rsta.2011.0405} {\bibfield  {journal} {\bibinfo
  {journal} {Philos. Trans. A Math. Phys. Eng. Sci.}\ }\textbf {\bibinfo
  {volume} {370}},\ \bibinfo {pages} {5806} (\bibinfo {year}
  {2012})}\BibitemShut {NoStop}%
\bibitem [{\citenamefont {Xiao}\ \emph {et~al.}(2010)\citenamefont {Xiao},
  \citenamefont {Chang},\ and\ \citenamefont {Niu}}]{Xiao2010}%
  \BibitemOpen
  \bibfield  {author} {\bibinfo {author} {\bibfnamefont {D.}~\bibnamefont
  {Xiao}}, \bibinfo {author} {\bibfnamefont {M.-C.}\ \bibnamefont {Chang}},\
  and\ \bibinfo {author} {\bibfnamefont {Q.}~\bibnamefont {Niu}},\ }\href
  {https://doi.org/10.1103/RevModPhys.82.1959} {\bibfield  {journal} {\bibinfo
  {journal} {Rev. Mod. Phys.}\ }\textbf {\bibinfo {volume} {82}},\ \bibinfo
  {pages} {1959} (\bibinfo {year} {2010})}\BibitemShut {NoStop}%
\bibitem [{\citenamefont {Berry}(1984)}]{Berry1984}%
  \BibitemOpen
  \bibfield  {author} {\bibinfo {author} {\bibfnamefont {M.~V.}\ \bibnamefont
  {Berry}},\ }\href {https://doi.org/10.1098/rspa.1984.0023} {\bibfield
  {journal} {\bibinfo  {journal} {Proc. R. Soc. Lond. A Math. Phys. Sci.}\
  }\textbf {\bibinfo {volume} {392}},\ \bibinfo {pages} {45} (\bibinfo {year}
  {1984})}\BibitemShut {NoStop}%
\bibitem [{\citenamefont {Volovik}(1987)}]{Volovik1987}%
  \BibitemOpen
  \bibfield  {author} {\bibinfo {author} {\bibfnamefont {G.~E.}\ \bibnamefont
  {Volovik}},\ }\href {https://doi.org/10.1088/0022-3719/20/7/003} {\bibfield
  {journal} {\bibinfo  {journal} {J. Phys. C: Solid State Phys.}\ }\textbf
  {\bibinfo {volume} {20}},\ \bibinfo {pages} {L83} (\bibinfo {year}
  {1987})}\BibitemShut {NoStop}%
\bibitem [{\citenamefont {Ye}\ \emph {et~al.}(1999)\citenamefont {Ye},
  \citenamefont {Kim}, \citenamefont {Millis}, \citenamefont {Shraiman},
  \citenamefont {Majumdar},\ and\ \citenamefont {Te{\v s}anovi{\'c}}}]{Ye1999}%
  \BibitemOpen
  \bibfield  {author} {\bibinfo {author} {\bibfnamefont {J.}~\bibnamefont
  {Ye}}, \bibinfo {author} {\bibfnamefont {Y.~B.}\ \bibnamefont {Kim}},
  \bibinfo {author} {\bibfnamefont {A.~J.}\ \bibnamefont {Millis}}, \bibinfo
  {author} {\bibfnamefont {B.~I.}\ \bibnamefont {Shraiman}}, \bibinfo {author}
  {\bibfnamefont {P.}~\bibnamefont {Majumdar}},\ and\ \bibinfo {author}
  {\bibfnamefont {Z.}~\bibnamefont {Te{\v s}anovi{\'c}}},\ }\href
  {https://doi.org/10.1103/PhysRevLett.83.3737} {\bibfield  {journal} {\bibinfo
   {journal} {Phys. Rev. Lett.}\ }\textbf {\bibinfo {volume} {83}},\ \bibinfo
  {pages} {3737} (\bibinfo {year} {1999})}\BibitemShut {NoStop}%
\bibitem [{\citenamefont {Bruno}\ \emph {et~al.}(2004)\citenamefont {Bruno},
  \citenamefont {Dugaev},\ and\ \citenamefont {Taillefumier}}]{Bruno2004}%
  \BibitemOpen
  \bibfield  {author} {\bibinfo {author} {\bibfnamefont {P.}~\bibnamefont
  {Bruno}}, \bibinfo {author} {\bibfnamefont {V.~K.}\ \bibnamefont {Dugaev}},\
  and\ \bibinfo {author} {\bibfnamefont {M.}~\bibnamefont {Taillefumier}},\
  }\href {https://doi.org/10.1103/PhysRevLett.93.096806} {\bibfield  {journal}
  {\bibinfo  {journal} {Phys. Rev. Lett.}\ }\textbf {\bibinfo {volume} {93}},\
  \bibinfo {pages} {096806} (\bibinfo {year} {2004})}\BibitemShut {NoStop}%
\bibitem [{\citenamefont {Onoda}\ \emph {et~al.}(2004)\citenamefont {Onoda},
  \citenamefont {Tatara},\ and\ \citenamefont {Nagaosa}}]{Onoda2004}%
  \BibitemOpen
  \bibfield  {author} {\bibinfo {author} {\bibfnamefont {M.}~\bibnamefont
  {Onoda}}, \bibinfo {author} {\bibfnamefont {G.}~\bibnamefont {Tatara}},\ and\
  \bibinfo {author} {\bibfnamefont {N.}~\bibnamefont {Nagaosa}},\ }\href
  {https://doi.org/10.1143/JPSJ.73.2624} {\bibfield  {journal} {\bibinfo
  {journal} {J. Phys. Soc. Jpn.}\ }\textbf {\bibinfo {volume} {73}},\ \bibinfo
  {pages} {2624} (\bibinfo {year} {2004})}\BibitemShut {NoStop}%
\bibitem [{\citenamefont {Binz}\ and\ \citenamefont
  {Vishwanath}(2008)}]{Binz2008}%
  \BibitemOpen
  \bibfield  {author} {\bibinfo {author} {\bibfnamefont {B.}~\bibnamefont
  {Binz}}\ and\ \bibinfo {author} {\bibfnamefont {A.}~\bibnamefont
  {Vishwanath}},\ }\href {https://doi.org/10.1016/j.physb.2007.10.136}
  {\bibfield  {journal} {\bibinfo  {journal} {Physica B Condens. Matter}\
  }\textbf {\bibinfo {volume} {403}},\ \bibinfo {pages} {1336} (\bibinfo {year}
  {2008})}\BibitemShut {NoStop}%
\bibitem [{\citenamefont {Nakazawa}\ and\ \citenamefont
  {Kohno}(2019)}]{Nakazawa2019}%
  \BibitemOpen
  \bibfield  {author} {\bibinfo {author} {\bibfnamefont {K.}~\bibnamefont
  {Nakazawa}}\ and\ \bibinfo {author} {\bibfnamefont {H.}~\bibnamefont
  {Kohno}},\ }\href {https://doi.org/10.1103/PhysRevB.99.174425} {\bibfield
  {journal} {\bibinfo  {journal} {Phys. Rev. B}\ }\textbf {\bibinfo {volume}
  {99}},\ \bibinfo {pages} {174425} (\bibinfo {year} {2019})}\BibitemShut
  {NoStop}%
\bibitem [{\citenamefont {{\v S}mejkal}\ \emph {et~al.}(2022)\citenamefont {{\v
  S}mejkal}, \citenamefont {MacDonald}, \citenamefont {Sinova}, \citenamefont
  {Nakatsuji},\ and\ \citenamefont {Jungwirth}}]{Smejkal2022}%
  \BibitemOpen
  \bibfield  {author} {\bibinfo {author} {\bibfnamefont {L.}~\bibnamefont {{\v
  S}mejkal}}, \bibinfo {author} {\bibfnamefont {A.~H.}\ \bibnamefont
  {MacDonald}}, \bibinfo {author} {\bibfnamefont {J.}~\bibnamefont {Sinova}},
  \bibinfo {author} {\bibfnamefont {S.}~\bibnamefont {Nakatsuji}},\ and\
  \bibinfo {author} {\bibfnamefont {T.}~\bibnamefont {Jungwirth}},\ }\href
  {https://doi.org/10.1038/s41578-022-00430-3} {\bibfield  {journal} {\bibinfo
  {journal} {Nature Reviews Materials}\ ,\ \bibinfo {pages} {1}} (\bibinfo
  {year} {2022})}\BibitemShut {NoStop}%
\bibitem [{\citenamefont {Khanh}\ \emph {et~al.}(2022)\citenamefont {Khanh},
  \citenamefont {Nakajima}, \citenamefont {Hayami}, \citenamefont {Gao},
  \citenamefont {Yamasaki}, \citenamefont {Sagayama}, \citenamefont {Nakao},
  \citenamefont {Takagi}, \citenamefont {Motome}, \citenamefont {Tokura},
  \citenamefont {Arima},\ and\ \citenamefont {Seki}}]{Khanh2022}%
  \BibitemOpen
  \bibfield  {author} {\bibinfo {author} {\bibfnamefont {N.~D.}\ \bibnamefont
  {Khanh}}, \bibinfo {author} {\bibfnamefont {T.}~\bibnamefont {Nakajima}},
  \bibinfo {author} {\bibfnamefont {S.}~\bibnamefont {Hayami}}, \bibinfo
  {author} {\bibfnamefont {S.}~\bibnamefont {Gao}}, \bibinfo {author}
  {\bibfnamefont {Y.}~\bibnamefont {Yamasaki}}, \bibinfo {author}
  {\bibfnamefont {H.}~\bibnamefont {Sagayama}}, \bibinfo {author}
  {\bibfnamefont {H.}~\bibnamefont {Nakao}}, \bibinfo {author} {\bibfnamefont
  {R.}~\bibnamefont {Takagi}}, \bibinfo {author} {\bibfnamefont
  {Y.}~\bibnamefont {Motome}}, \bibinfo {author} {\bibfnamefont
  {Y.}~\bibnamefont {Tokura}}, \bibinfo {author} {\bibfnamefont
  {T.}~\bibnamefont {Arima}},\ and\ \bibinfo {author} {\bibfnamefont
  {S.}~\bibnamefont {Seki}},\ }\href {https://doi.org/10.1002/advs.202105452}
  {\bibfield  {journal} {\bibinfo  {journal} {Adv. Sci.}\ }\textbf {\bibinfo
  {volume} {9}},\ \bibinfo {pages} {2105452} (\bibinfo {year}
  {2022})}\BibitemShut {NoStop}%
\bibitem [{\citenamefont {Shimizu}\ \emph
  {et~al.}(2021{\natexlab{a}})\citenamefont {Shimizu}, \citenamefont {Okumura},
  \citenamefont {Kato},\ and\ \citenamefont {Motome}}]{Shimizu2021_Moire}%
  \BibitemOpen
  \bibfield  {author} {\bibinfo {author} {\bibfnamefont {K.}~\bibnamefont
  {Shimizu}}, \bibinfo {author} {\bibfnamefont {S.}~\bibnamefont {Okumura}},
  \bibinfo {author} {\bibfnamefont {Y.}~\bibnamefont {Kato}},\ and\ \bibinfo
  {author} {\bibfnamefont {Y.}~\bibnamefont {Motome}},\ }\href
  {https://doi.org/10.1103/PhysRevB.103.184421} {\bibfield  {journal} {\bibinfo
   {journal} {Phys. Rev. B}\ }\textbf {\bibinfo {volume} {103}},\ \bibinfo
  {pages} {184421} (\bibinfo {year} {2021}{\natexlab{a}})}\BibitemShut
  {NoStop}%
\bibitem [{\citenamefont {Shimizu}\ \emph
  {et~al.}(2021{\natexlab{b}})\citenamefont {Shimizu}, \citenamefont {Okumura},
  \citenamefont {Kato},\ and\ \citenamefont {Motome}}]{Shimizu2021_Chiral}%
  \BibitemOpen
  \bibfield  {author} {\bibinfo {author} {\bibfnamefont {K.}~\bibnamefont
  {Shimizu}}, \bibinfo {author} {\bibfnamefont {S.}~\bibnamefont {Okumura}},
  \bibinfo {author} {\bibfnamefont {Y.}~\bibnamefont {Kato}},\ and\ \bibinfo
  {author} {\bibfnamefont {Y.}~\bibnamefont {Motome}},\ }\href
  {https://doi.org/10.1103/PhysRevB.103.054427} {\bibfield  {journal} {\bibinfo
   {journal} {Phys. Rev. B}\ }\textbf {\bibinfo {volume} {103}},\ \bibinfo
  {pages} {054427} (\bibinfo {year} {2021}{\natexlab{b}})}\BibitemShut
  {NoStop}%
\bibitem [{\citenamefont {Hayami}\ \emph {et~al.}(2021)\citenamefont {Hayami},
  \citenamefont {Okubo},\ and\ \citenamefont {Motome}}]{Hayami2021_Phase}%
  \BibitemOpen
  \bibfield  {author} {\bibinfo {author} {\bibfnamefont {S.}~\bibnamefont
  {Hayami}}, \bibinfo {author} {\bibfnamefont {T.}~\bibnamefont {Okubo}},\ and\
  \bibinfo {author} {\bibfnamefont {Y.}~\bibnamefont {Motome}},\ }\href
  {https://doi.org/10.1038/s41467-021-27083-0} {\bibfield  {journal} {\bibinfo
  {journal} {Nat. Commun.}\ }\textbf {\bibinfo {volume} {12}},\ \bibinfo
  {pages} {1} (\bibinfo {year} {2021})}\BibitemShut {NoStop}%
\bibitem [{\citenamefont {Hayami}\ and\ \citenamefont
  {Yambe}(2021)}]{Hayami2021_Lock}%
  \BibitemOpen
  \bibfield  {author} {\bibinfo {author} {\bibfnamefont {S.}~\bibnamefont
  {Hayami}}\ and\ \bibinfo {author} {\bibfnamefont {R.}~\bibnamefont {Yambe}},\
  }\href {https://doi.org/10.1103/PhysRevResearch.3.043158} {\bibfield
  {journal} {\bibinfo  {journal} {Phys. Rev. Research}\ }\textbf {\bibinfo
  {volume} {3}},\ \bibinfo {pages} {043158} (\bibinfo {year}
  {2021})}\BibitemShut {NoStop}%
\bibitem [{\citenamefont {Mannix}\ \emph {et~al.}(1999)\citenamefont {Mannix},
  \citenamefont {Lander}, \citenamefont {Rebizant}, \citenamefont {Caciuffo},
  \citenamefont {Bernhoeft}, \citenamefont {Lidstr{\"o}m},\ and\ \citenamefont
  {Vettier}}]{Mannix1999}%
  \BibitemOpen
  \bibfield  {author} {\bibinfo {author} {\bibfnamefont {D.}~\bibnamefont
  {Mannix}}, \bibinfo {author} {\bibfnamefont {G.~H.}\ \bibnamefont {Lander}},
  \bibinfo {author} {\bibfnamefont {J.}~\bibnamefont {Rebizant}}, \bibinfo
  {author} {\bibfnamefont {R.}~\bibnamefont {Caciuffo}}, \bibinfo {author}
  {\bibfnamefont {N.}~\bibnamefont {Bernhoeft}}, \bibinfo {author}
  {\bibfnamefont {E.}~\bibnamefont {Lidstr{\"o}m}},\ and\ \bibinfo {author}
  {\bibfnamefont {C.}~\bibnamefont {Vettier}},\ }\href
  {https://doi.org/10.1103/physrevb.60.15187} {\bibfield  {journal} {\bibinfo
  {journal} {Phys. Rev. B}\ }\textbf {\bibinfo {volume} {60}},\ \bibinfo
  {pages} {15187} (\bibinfo {year} {1999})}\BibitemShut {NoStop}%
\bibitem [{\citenamefont {Caciuffo}\ \emph {et~al.}(2003)\citenamefont
  {Caciuffo}, \citenamefont {Paix{\~a}o}, \citenamefont {Detlefs},
  \citenamefont {Longfield}, \citenamefont {Santini}, \citenamefont
  {Bernhoeft}, \citenamefont {Rebizant},\ and\ \citenamefont
  {Lander}}]{Caciuffo2003}%
  \BibitemOpen
  \bibfield  {author} {\bibinfo {author} {\bibfnamefont {R.}~\bibnamefont
  {Caciuffo}}, \bibinfo {author} {\bibfnamefont {J.~A.}\ \bibnamefont
  {Paix{\~a}o}}, \bibinfo {author} {\bibfnamefont {C.}~\bibnamefont {Detlefs}},
  \bibinfo {author} {\bibfnamefont {M.~J.}\ \bibnamefont {Longfield}}, \bibinfo
  {author} {\bibfnamefont {P.}~\bibnamefont {Santini}}, \bibinfo {author}
  {\bibfnamefont {N.}~\bibnamefont {Bernhoeft}}, \bibinfo {author}
  {\bibfnamefont {J.}~\bibnamefont {Rebizant}},\ and\ \bibinfo {author}
  {\bibfnamefont {G.~H.}\ \bibnamefont {Lander}},\ }\href
  {https://doi.org/10.1088/0953-8984/15/28/370} {\bibfield  {journal} {\bibinfo
   {journal} {J. Phys. Condens. Matter}\ }\textbf {\bibinfo {volume} {15}},\
  \bibinfo {pages} {S2287} (\bibinfo {year} {2003})}\BibitemShut {NoStop}%
\bibitem [{\citenamefont {Tokunaga}\ \emph {et~al.}(2005)\citenamefont
  {Tokunaga}, \citenamefont {Homma}, \citenamefont {Kambe}, \citenamefont
  {Aoki}, \citenamefont {Sakai}, \citenamefont {Yamamoto}, \citenamefont
  {Nakamura}, \citenamefont {Shiokawa}, \citenamefont {Walstedt},\ and\
  \citenamefont {Yasuoka}}]{Tokunaga2005}%
  \BibitemOpen
  \bibfield  {author} {\bibinfo {author} {\bibfnamefont {Y.}~\bibnamefont
  {Tokunaga}}, \bibinfo {author} {\bibfnamefont {Y.}~\bibnamefont {Homma}},
  \bibinfo {author} {\bibfnamefont {S.}~\bibnamefont {Kambe}}, \bibinfo
  {author} {\bibfnamefont {D.}~\bibnamefont {Aoki}}, \bibinfo {author}
  {\bibfnamefont {H.}~\bibnamefont {Sakai}}, \bibinfo {author} {\bibfnamefont
  {E.}~\bibnamefont {Yamamoto}}, \bibinfo {author} {\bibfnamefont
  {A.}~\bibnamefont {Nakamura}}, \bibinfo {author} {\bibfnamefont
  {Y.}~\bibnamefont {Shiokawa}}, \bibinfo {author} {\bibfnamefont {R.~E.}\
  \bibnamefont {Walstedt}},\ and\ \bibinfo {author} {\bibfnamefont
  {H.}~\bibnamefont {Yasuoka}},\ }\href
  {https://doi.org/10.1103/PhysRevLett.94.137209} {\bibfield  {journal}
  {\bibinfo  {journal} {Phys. Rev. Lett.}\ }\textbf {\bibinfo {volume} {94}},\
  \bibinfo {pages} {137209} (\bibinfo {year} {2005})}\BibitemShut {NoStop}%
\bibitem [{\citenamefont {Tokunaga}\ \emph {et~al.}(2006)\citenamefont
  {Tokunaga}, \citenamefont {Aoki}, \citenamefont {Homma}, \citenamefont
  {Kambe}, \citenamefont {Sakai}, \citenamefont {Ikeda}, \citenamefont
  {Fujimoto}, \citenamefont {Walstedt}, \citenamefont {Yasuoka}, \citenamefont
  {Yamamoto}, \citenamefont {Nakamura},\ and\ \citenamefont
  {Shiokawa}}]{Tokunaga2006}%
  \BibitemOpen
  \bibfield  {author} {\bibinfo {author} {\bibfnamefont {Y.}~\bibnamefont
  {Tokunaga}}, \bibinfo {author} {\bibfnamefont {D.}~\bibnamefont {Aoki}},
  \bibinfo {author} {\bibfnamefont {Y.}~\bibnamefont {Homma}}, \bibinfo
  {author} {\bibfnamefont {S.}~\bibnamefont {Kambe}}, \bibinfo {author}
  {\bibfnamefont {H.}~\bibnamefont {Sakai}}, \bibinfo {author} {\bibfnamefont
  {S.}~\bibnamefont {Ikeda}}, \bibinfo {author} {\bibfnamefont
  {T.}~\bibnamefont {Fujimoto}}, \bibinfo {author} {\bibfnamefont {R.~E.}\
  \bibnamefont {Walstedt}}, \bibinfo {author} {\bibfnamefont {H.}~\bibnamefont
  {Yasuoka}}, \bibinfo {author} {\bibfnamefont {E.}~\bibnamefont {Yamamoto}},
  \bibinfo {author} {\bibfnamefont {A.}~\bibnamefont {Nakamura}},\ and\
  \bibinfo {author} {\bibfnamefont {Y.}~\bibnamefont {Shiokawa}},\ }\href
  {https://doi.org/10.1103/PhysRevLett.97.257601} {\bibfield  {journal}
  {\bibinfo  {journal} {Phys. Rev. Lett.}\ }\textbf {\bibinfo {volume} {97}},\
  \bibinfo {pages} {257601} (\bibinfo {year} {2006})}\BibitemShut {NoStop}%
\bibitem [{\citenamefont {Magnani}\ \emph {et~al.}(2008)\citenamefont
  {Magnani}, \citenamefont {Carretta}, \citenamefont {Caciuffo}, \citenamefont
  {Santini}, \citenamefont {Amoretti}, \citenamefont {Hiess}, \citenamefont
  {Rebizant},\ and\ \citenamefont {Lander}}]{Magnani2008}%
  \BibitemOpen
  \bibfield  {author} {\bibinfo {author} {\bibfnamefont {N.}~\bibnamefont
  {Magnani}}, \bibinfo {author} {\bibfnamefont {S.}~\bibnamefont {Carretta}},
  \bibinfo {author} {\bibfnamefont {R.}~\bibnamefont {Caciuffo}}, \bibinfo
  {author} {\bibfnamefont {P.}~\bibnamefont {Santini}}, \bibinfo {author}
  {\bibfnamefont {G.}~\bibnamefont {Amoretti}}, \bibinfo {author}
  {\bibfnamefont {A.}~\bibnamefont {Hiess}}, \bibinfo {author} {\bibfnamefont
  {J.}~\bibnamefont {Rebizant}},\ and\ \bibinfo {author} {\bibfnamefont
  {G.~H.}\ \bibnamefont {Lander}},\ }\href
  {https://doi.org/10.1103/PhysRevB.78.104425} {\bibfield  {journal} {\bibinfo
  {journal} {Phys. Rev. B}\ }\textbf {\bibinfo {volume} {78}},\ \bibinfo
  {pages} {104425} (\bibinfo {year} {2008})}\BibitemShut {NoStop}%
\bibitem [{\citenamefont {Kubo}\ and\ \citenamefont {Hotta}(2005)}]{Kubo2005}%
  \BibitemOpen
  \bibfield  {author} {\bibinfo {author} {\bibfnamefont {K.}~\bibnamefont
  {Kubo}}\ and\ \bibinfo {author} {\bibfnamefont {T.}~\bibnamefont {Hotta}},\
  }\href {https://doi.org/10.1103/PhysRevB.71.140404} {\bibfield  {journal}
  {\bibinfo  {journal} {Phys. Rev. B}\ }\textbf {\bibinfo {volume} {71}},\
  \bibinfo {pages} {140404} (\bibinfo {year} {2005})}\BibitemShut {NoStop}%
\bibitem [{\citenamefont {Walker}\ \emph {et~al.}(1994)\citenamefont {Walker},
  \citenamefont {Kappler}, \citenamefont {McEwen}, \citenamefont
  {Steigenberger},\ and\ \citenamefont {Clausen}}]{Walker1994}%
  \BibitemOpen
  \bibfield  {author} {\bibinfo {author} {\bibfnamefont {M.~B.}\ \bibnamefont
  {Walker}}, \bibinfo {author} {\bibfnamefont {C.}~\bibnamefont {Kappler}},
  \bibinfo {author} {\bibfnamefont {K.~A.}\ \bibnamefont {McEwen}}, \bibinfo
  {author} {\bibfnamefont {U.}~\bibnamefont {Steigenberger}},\ and\ \bibinfo
  {author} {\bibfnamefont {K.~N.}\ \bibnamefont {Clausen}},\ }\href
  {https://doi.org/10.1088/0953-8984/6/36/018} {\bibfield  {journal} {\bibinfo
  {journal} {J. Phys. Condens. Matter}\ }\textbf {\bibinfo {volume} {6}},\
  \bibinfo {pages} {7365} (\bibinfo {year} {1994})}\BibitemShut {NoStop}%
\bibitem [{\citenamefont {McEwen}\ \emph {et~al.}(1995)\citenamefont {McEwen},
  \citenamefont {Steigenberger}, \citenamefont {Clausen}, \citenamefont {Bi},
  \citenamefont {Walker},\ and\ \citenamefont {Kappler}}]{McEwen1995}%
  \BibitemOpen
  \bibfield  {author} {\bibinfo {author} {\bibfnamefont {K.~A.}\ \bibnamefont
  {McEwen}}, \bibinfo {author} {\bibfnamefont {U.}~\bibnamefont
  {Steigenberger}}, \bibinfo {author} {\bibfnamefont {K.~N.}\ \bibnamefont
  {Clausen}}, \bibinfo {author} {\bibfnamefont {Y.~J.}\ \bibnamefont {Bi}},
  \bibinfo {author} {\bibfnamefont {M.~B.}\ \bibnamefont {Walker}},\ and\
  \bibinfo {author} {\bibfnamefont {C.}~\bibnamefont {Kappler}},\ }\href
  {https://doi.org/10.1016/0921-4526(95)00081-J} {\bibfield  {journal}
  {\bibinfo  {journal} {Physica B Condens. Matter}\ }\textbf {\bibinfo {volume}
  {213-214}},\ \bibinfo {pages} {128} (\bibinfo {year} {1995})}\BibitemShut
  {NoStop}%
\bibitem [{\citenamefont {McEwen}\ \emph {et~al.}(1998)\citenamefont {McEwen},
  \citenamefont {Steigenberger}, \citenamefont {Clausen}, \citenamefont
  {Kulda}, \citenamefont {Park},\ and\ \citenamefont {Walker}}]{McEwen1998}%
  \BibitemOpen
  \bibfield  {author} {\bibinfo {author} {\bibfnamefont {K.~A.}\ \bibnamefont
  {McEwen}}, \bibinfo {author} {\bibfnamefont {U.}~\bibnamefont
  {Steigenberger}}, \bibinfo {author} {\bibfnamefont {K.~N.}\ \bibnamefont
  {Clausen}}, \bibinfo {author} {\bibfnamefont {J.}~\bibnamefont {Kulda}},
  \bibinfo {author} {\bibfnamefont {J.-G.}\ \bibnamefont {Park}},\ and\
  \bibinfo {author} {\bibfnamefont {M.~B.}\ \bibnamefont {Walker}},\ }\href
  {https://doi.org/10.1016/S0304-8853(97)00993-1} {\bibfield  {journal}
  {\bibinfo  {journal} {J. Magn. Magn. Mater.}\ }\textbf {\bibinfo {volume}
  {177-181}},\ \bibinfo {pages} {37} (\bibinfo {year} {1998})}\BibitemShut
  {NoStop}%
\bibitem [{\citenamefont {Hanzawa}\ and\ \citenamefont
  {Yamada}(2019)}]{Hanzawa2019}%
  \BibitemOpen
  \bibfield  {author} {\bibinfo {author} {\bibfnamefont {K.}~\bibnamefont
  {Hanzawa}}\ and\ \bibinfo {author} {\bibfnamefont {T.}~\bibnamefont
  {Yamada}},\ }\href {https://doi.org/10.7566/JPSJ.88.124713} {\bibfield
  {journal} {\bibinfo  {journal} {J. Phys. Soc. Jpn.}\ }\textbf {\bibinfo
  {volume} {88}},\ \bibinfo {pages} {124713} (\bibinfo {year}
  {2019})}\BibitemShut {NoStop}%
\bibitem [{\citenamefont {Tsunetsugu}\ \emph {et~al.}(2021)\citenamefont
  {Tsunetsugu}, \citenamefont {Ishitobi},\ and\ \citenamefont
  {Hattori}}]{Tsunetsugu2021}%
  \BibitemOpen
  \bibfield  {author} {\bibinfo {author} {\bibfnamefont {H.}~\bibnamefont
  {Tsunetsugu}}, \bibinfo {author} {\bibfnamefont {T.}~\bibnamefont
  {Ishitobi}},\ and\ \bibinfo {author} {\bibfnamefont {K.}~\bibnamefont
  {Hattori}},\ }\href {https://doi.org/10.7566/JPSJ.90.043701} {\bibfield
  {journal} {\bibinfo  {journal} {J. Phys. Soc. Jpn.}\ }\textbf {\bibinfo
  {volume} {90}},\ \bibinfo {pages} {043701} (\bibinfo {year}
  {2021})}\BibitemShut {NoStop}%
\bibitem [{\citenamefont {Ishitobi}\ and\ \citenamefont
  {Hattori}(2021)}]{Ishitobi2021}%
  \BibitemOpen
  \bibfield  {author} {\bibinfo {author} {\bibfnamefont {T.}~\bibnamefont
  {Ishitobi}}\ and\ \bibinfo {author} {\bibfnamefont {K.}~\bibnamefont
  {Hattori}},\ }\href {https://doi.org/10.1103/PhysRevB.104.L241110} {\bibfield
   {journal} {\bibinfo  {journal} {Phys. Rev. B}\ }\textbf {\bibinfo {volume}
  {104}},\ \bibinfo {pages} {L241110} (\bibinfo {year} {2021})}\BibitemShut
  {NoStop}%
\bibitem [{\citenamefont {Hattori}\ \emph {et~al.}()\citenamefont {Hattori},
  \citenamefont {Ishitobi},\ and\ \citenamefont {Tsunetsugu}}]{Hattori2022}%
  \BibitemOpen
  \bibfield  {author} {\bibinfo {author} {\bibfnamefont {K.}~\bibnamefont
  {Hattori}}, \bibinfo {author} {\bibfnamefont {T.}~\bibnamefont {Ishitobi}},\
  and\ \bibinfo {author} {\bibfnamefont {H.}~\bibnamefont {Tsunetsugu}},\
  }\href {http://arxiv.org/abs/2212.12920} {\ }\Eprint
  {https://arxiv.org/abs/2212.12920} {arXiv:2212.12920} \BibitemShut {NoStop}%
\bibitem [{\citenamefont {Yamauchi}\ \emph {et~al.}(1999)\citenamefont
  {Yamauchi}, \citenamefont {Onodera}, \citenamefont {Ohashi}, \citenamefont
  {Ohoyama}, \citenamefont {Onimaru}, \citenamefont {Kosaka},\ and\
  \citenamefont {Yamaguchi}}]{Yamauchi1999}%
  \BibitemOpen
  \bibfield  {author} {\bibinfo {author} {\bibfnamefont {H.}~\bibnamefont
  {Yamauchi}}, \bibinfo {author} {\bibfnamefont {H.}~\bibnamefont {Onodera}},
  \bibinfo {author} {\bibfnamefont {M.}~\bibnamefont {Ohashi}}, \bibinfo
  {author} {\bibfnamefont {K.}~\bibnamefont {Ohoyama}}, \bibinfo {author}
  {\bibfnamefont {T.}~\bibnamefont {Onimaru}}, \bibinfo {author} {\bibfnamefont
  {M.}~\bibnamefont {Kosaka}},\ and\ \bibinfo {author} {\bibfnamefont
  {Y.}~\bibnamefont {Yamaguchi}},\ }\href
  {https://doi.org/10.1016/S0022-3697(99)00088-8} {\bibfield  {journal}
  {\bibinfo  {journal} {J. Phys. Chem. Solids}\ }\textbf {\bibinfo {volume}
  {60}},\ \bibinfo {pages} {1217} (\bibinfo {year} {1999})}\BibitemShut
  {NoStop}%
\bibitem [{\citenamefont {Tanaka}\ \emph {et~al.}(1999)\citenamefont {Tanaka},
  \citenamefont {Inami}, \citenamefont {Nakamura}, \citenamefont {Yamauchi},
  \citenamefont {Onodera}, \citenamefont {Ohoyama},\ and\ \citenamefont
  {Yamaguchi}}]{Tanaka1999}%
  \BibitemOpen
  \bibfield  {author} {\bibinfo {author} {\bibfnamefont {Y.}~\bibnamefont
  {Tanaka}}, \bibinfo {author} {\bibfnamefont {T.}~\bibnamefont {Inami}},
  \bibinfo {author} {\bibfnamefont {T.}~\bibnamefont {Nakamura}}, \bibinfo
  {author} {\bibfnamefont {H.}~\bibnamefont {Yamauchi}}, \bibinfo {author}
  {\bibfnamefont {H.}~\bibnamefont {Onodera}}, \bibinfo {author} {\bibfnamefont
  {K.}~\bibnamefont {Ohoyama}},\ and\ \bibinfo {author} {\bibfnamefont
  {Y.}~\bibnamefont {Yamaguchi}},\ }\href
  {https://doi.org/10.1088/0953-8984/11/44/102} {\bibfield  {journal} {\bibinfo
   {journal} {J. Phys. Condens. Matter}\ }\textbf {\bibinfo {volume} {11}},\
  \bibinfo {pages} {L505} (\bibinfo {year} {1999})}\BibitemShut {NoStop}%
\bibitem [{\citenamefont {Yamauchi}\ \emph {et~al.}(2003)\citenamefont
  {Yamauchi}, \citenamefont {Ohoyama}, \citenamefont {Katano}, \citenamefont
  {Matsuda}, \citenamefont {Indoh}, \citenamefont {Onodera},\ and\
  \citenamefont {Yamaguchi}}]{Yamauchi2003}%
  \BibitemOpen
  \bibfield  {author} {\bibinfo {author} {\bibfnamefont {H.}~\bibnamefont
  {Yamauchi}}, \bibinfo {author} {\bibfnamefont {K.}~\bibnamefont {Ohoyama}},
  \bibinfo {author} {\bibfnamefont {S.}~\bibnamefont {Katano}}, \bibinfo
  {author} {\bibfnamefont {M.}~\bibnamefont {Matsuda}}, \bibinfo {author}
  {\bibfnamefont {K.}~\bibnamefont {Indoh}}, \bibinfo {author} {\bibfnamefont
  {H.}~\bibnamefont {Onodera}},\ and\ \bibinfo {author} {\bibfnamefont
  {Y.}~\bibnamefont {Yamaguchi}},\ }\href
  {https://iopscience.iop.org/article/10.1088/0953-8984/15/28/337/meta}
  {\bibfield  {journal} {\bibinfo  {journal} {J. Phys. Condens. Matter}\
  }\textbf {\bibinfo {volume} {15}},\ \bibinfo {pages} {S2137} (\bibinfo {year}
  {2003})}\BibitemShut {NoStop}%
\bibitem [{\citenamefont {Chen}\ \emph {et~al.}(2010)\citenamefont {Chen},
  \citenamefont {Pereira},\ and\ \citenamefont {Balents}}]{Chen2010}%
  \BibitemOpen
  \bibfield  {author} {\bibinfo {author} {\bibfnamefont {G.}~\bibnamefont
  {Chen}}, \bibinfo {author} {\bibfnamefont {R.}~\bibnamefont {Pereira}},\ and\
  \bibinfo {author} {\bibfnamefont {L.}~\bibnamefont {Balents}},\ }\href
  {https://doi.org/10.1103/PhysRevB.82.174440} {\bibfield  {journal} {\bibinfo
  {journal} {Phys. Rev. B}\ }\textbf {\bibinfo {volume} {82}},\ \bibinfo
  {pages} {174440} (\bibinfo {year} {2010})}\BibitemShut {NoStop}%
\bibitem [{\citenamefont {Hirai}\ \emph {et~al.}(2020)\citenamefont {Hirai},
  \citenamefont {Sagayama}, \citenamefont {Gao}, \citenamefont {Ohsumi},
  \citenamefont {Chen}, \citenamefont {Arima},\ and\ \citenamefont
  {Hiroi}}]{Hirai2020}%
  \BibitemOpen
  \bibfield  {author} {\bibinfo {author} {\bibfnamefont {D.}~\bibnamefont
  {Hirai}}, \bibinfo {author} {\bibfnamefont {H.}~\bibnamefont {Sagayama}},
  \bibinfo {author} {\bibfnamefont {S.}~\bibnamefont {Gao}}, \bibinfo {author}
  {\bibfnamefont {H.}~\bibnamefont {Ohsumi}}, \bibinfo {author} {\bibfnamefont
  {G.}~\bibnamefont {Chen}}, \bibinfo {author} {\bibfnamefont {T.}~\bibnamefont
  {Arima}},\ and\ \bibinfo {author} {\bibfnamefont {Z.}~\bibnamefont {Hiroi}},\
  }\href {https://doi.org/10.1103/PhysRevResearch.2.022063} {\bibfield
  {journal} {\bibinfo  {journal} {Phys. Rev. Research}\ }\textbf {\bibinfo
  {volume} {2}},\ \bibinfo {pages} {022063} (\bibinfo {year}
  {2020})}\BibitemShut {NoStop}%
\bibitem [{\citenamefont {L{\'\i}bero}\ and\ \citenamefont
  {Cox}(1993)}]{Libero1993}%
  \BibitemOpen
  \bibfield  {author} {\bibinfo {author} {\bibfnamefont {V.~L.}\ \bibnamefont
  {L{\'\i}bero}}\ and\ \bibinfo {author} {\bibfnamefont {D.~L.}\ \bibnamefont
  {Cox}},\ }\href {https://doi.org/10.1103/PhysRevB.48.3783} {\bibfield
  {journal} {\bibinfo  {journal} {Phys. Rev. B}\ }\textbf {\bibinfo {volume}
  {48}},\ \bibinfo {pages} {3783} (\bibinfo {year} {1993})}\BibitemShut
  {NoStop}%
\bibitem [{\citenamefont {de~Lima}\ and\ \citenamefont
  {L{\'\i}bero}(2000)}]{De_Lima2000}%
  \BibitemOpen
  \bibfield  {author} {\bibinfo {author} {\bibfnamefont {N.~A.}\ \bibnamefont
  {de~Lima}}\ and\ \bibinfo {author} {\bibfnamefont {V.~L.}\ \bibnamefont
  {L{\'\i}bero}},\ }\href {https://doi.org/10.1103/PhysRevB.61.3425} {\bibfield
   {journal} {\bibinfo  {journal} {Phys. Rev. B}\ }\textbf {\bibinfo {volume}
  {61}},\ \bibinfo {pages} {3425} (\bibinfo {year} {2000})}\BibitemShut
  {NoStop}%
\bibitem [{\citenamefont {Allen}(1968)}]{Allen1968}%
  \BibitemOpen
  \bibfield  {author} {\bibinfo {author} {\bibfnamefont {S.~J.}\ \bibnamefont
  {Allen}},\ }\href {https://doi.org/10.1103/PhysRev.167.492} {\bibfield
  {journal} {\bibinfo  {journal} {Phys. Rev.}\ }\textbf {\bibinfo {volume}
  {167}},\ \bibinfo {pages} {492} (\bibinfo {year} {1968})}\BibitemShut
  {NoStop}%
\bibitem [{\citenamefont {Ederer}\ and\ \citenamefont
  {Spaldin}(2007)}]{Ederer2007}%
  \BibitemOpen
  \bibfield  {author} {\bibinfo {author} {\bibfnamefont {C.}~\bibnamefont
  {Ederer}}\ and\ \bibinfo {author} {\bibfnamefont {N.~A.}\ \bibnamefont
  {Spaldin}},\ }\href {https://doi.org/10.1103/PhysRevB.76.214404} {\bibfield
  {journal} {\bibinfo  {journal} {Phys. Rev. B}\ }\textbf {\bibinfo {volume}
  {76}},\ \bibinfo {pages} {214404} (\bibinfo {year} {2007})}\BibitemShut
  {NoStop}%
\bibitem [{\citenamefont {Suzuki}\ \emph {et~al.}(2017)\citenamefont {Suzuki},
  \citenamefont {Koretsune}, \citenamefont {Ochi},\ and\ \citenamefont
  {Arita}}]{Suzuki2017}%
  \BibitemOpen
  \bibfield  {author} {\bibinfo {author} {\bibfnamefont {M.~T.}\ \bibnamefont
  {Suzuki}}, \bibinfo {author} {\bibfnamefont {T.}~\bibnamefont {Koretsune}},
  \bibinfo {author} {\bibfnamefont {M.}~\bibnamefont {Ochi}},\ and\ \bibinfo
  {author} {\bibfnamefont {R.}~\bibnamefont {Arita}},\ }\href
  {https://doi.org/10.1103/PhysRevB.95.094406} {\bibfield  {journal} {\bibinfo
  {journal} {Phys. Rev. B}\ }\textbf {\bibinfo {volume} {95}},\ \bibinfo
  {pages} {1} (\bibinfo {year} {2017})}\BibitemShut {NoStop}%
\bibitem [{\citenamefont {Suzuki}\ \emph {et~al.}(2019)\citenamefont {Suzuki},
  \citenamefont {Nomoto}, \citenamefont {Arita}, \citenamefont {Yanagi},
  \citenamefont {Hayami},\ and\ \citenamefont {Kusunose}}]{Suzuki2019}%
  \BibitemOpen
  \bibfield  {author} {\bibinfo {author} {\bibfnamefont {M.~T.}\ \bibnamefont
  {Suzuki}}, \bibinfo {author} {\bibfnamefont {T.}~\bibnamefont {Nomoto}},
  \bibinfo {author} {\bibfnamefont {R.}~\bibnamefont {Arita}}, \bibinfo
  {author} {\bibfnamefont {Y.}~\bibnamefont {Yanagi}}, \bibinfo {author}
  {\bibfnamefont {S.}~\bibnamefont {Hayami}},\ and\ \bibinfo {author}
  {\bibfnamefont {H.}~\bibnamefont {Kusunose}},\ }\href
  {https://doi.org/10.1103/PhysRevB.99.174407} {\bibfield  {journal} {\bibinfo
  {journal} {Phys. Rev. B}\ }\textbf {\bibinfo {volume} {99}},\ \bibinfo
  {pages} {174407} (\bibinfo {year} {2019})}\BibitemShut {NoStop}%
\bibitem [{\citenamefont {Yanagi}\ \emph {et~al.}(2023)\citenamefont {Yanagi},
  \citenamefont {Kusunose}, \citenamefont {Nomoto}, \citenamefont {Arita},\
  and\ \citenamefont {Suzuki}}]{Yanagi2023}%
  \BibitemOpen
  \bibfield  {author} {\bibinfo {author} {\bibfnamefont {Y.}~\bibnamefont
  {Yanagi}}, \bibinfo {author} {\bibfnamefont {H.}~\bibnamefont {Kusunose}},
  \bibinfo {author} {\bibfnamefont {T.}~\bibnamefont {Nomoto}}, \bibinfo
  {author} {\bibfnamefont {R.}~\bibnamefont {Arita}},\ and\ \bibinfo {author}
  {\bibfnamefont {M.-T.}\ \bibnamefont {Suzuki}},\ }\href
  {https://doi.org/10.1103/PhysRevB.107.014407} {\bibfield  {journal} {\bibinfo
   {journal} {Phys. Rev. B}\ }\textbf {\bibinfo {volume} {107}},\ \bibinfo
  {pages} {014407} (\bibinfo {year} {2023})}\BibitemShut {NoStop}%
\bibitem [{\citenamefont {Shannon}\ \emph {et~al.}(2006)\citenamefont
  {Shannon}, \citenamefont {Momoi},\ and\ \citenamefont
  {Sindzingre}}]{Shannon2006}%
  \BibitemOpen
  \bibfield  {author} {\bibinfo {author} {\bibfnamefont {N.}~\bibnamefont
  {Shannon}}, \bibinfo {author} {\bibfnamefont {T.}~\bibnamefont {Momoi}},\
  and\ \bibinfo {author} {\bibfnamefont {P.}~\bibnamefont {Sindzingre}},\
  }\href {https://doi.org/10.1103/PhysRevLett.96.027213} {\bibfield  {journal}
  {\bibinfo  {journal} {Phys. Rev. Lett.}\ }\textbf {\bibinfo {volume} {96}},\
  \bibinfo {pages} {027213} (\bibinfo {year} {2006})}\BibitemShut {NoStop}%
\bibitem [{\citenamefont {Hikihara}\ \emph {et~al.}(2008)\citenamefont
  {Hikihara}, \citenamefont {Kecke}, \citenamefont {Momoi},\ and\ \citenamefont
  {Furusaki}}]{Hikihara2008}%
  \BibitemOpen
  \bibfield  {author} {\bibinfo {author} {\bibfnamefont {T.}~\bibnamefont
  {Hikihara}}, \bibinfo {author} {\bibfnamefont {L.}~\bibnamefont {Kecke}},
  \bibinfo {author} {\bibfnamefont {T.}~\bibnamefont {Momoi}},\ and\ \bibinfo
  {author} {\bibfnamefont {A.}~\bibnamefont {Furusaki}},\ }\href
  {https://doi.org/10.1103/PhysRevB.78.144404} {\bibfield  {journal} {\bibinfo
  {journal} {Phys. Rev. B}\ }\textbf {\bibinfo {volume} {78}},\ \bibinfo
  {pages} {144404} (\bibinfo {year} {2008})}\BibitemShut {NoStop}%
\bibitem [{\citenamefont {Hayami}\ \emph {et~al.}(2019)\citenamefont {Hayami},
  \citenamefont {Yanagi}, \citenamefont {Kusunose},\ and\ \citenamefont
  {Motome}}]{Hayami2019}%
  \BibitemOpen
  \bibfield  {author} {\bibinfo {author} {\bibfnamefont {S.}~\bibnamefont
  {Hayami}}, \bibinfo {author} {\bibfnamefont {Y.}~\bibnamefont {Yanagi}},
  \bibinfo {author} {\bibfnamefont {H.}~\bibnamefont {Kusunose}},\ and\
  \bibinfo {author} {\bibfnamefont {Y.}~\bibnamefont {Motome}},\ }\href
  {https://doi.org/10.1103/PhysRevLett.122.147602} {\bibfield  {journal}
  {\bibinfo  {journal} {Phys. Rev. Lett.}\ }\textbf {\bibinfo {volume} {122}},\
  \bibinfo {pages} {1} (\bibinfo {year} {2019})}\BibitemShut {NoStop}%
\bibitem [{\citenamefont {Hayami}\ \emph {et~al.}(2020)\citenamefont {Hayami},
  \citenamefont {Yanagi},\ and\ \citenamefont {Kusunose}}]{Hayami2020}%
  \BibitemOpen
  \bibfield  {author} {\bibinfo {author} {\bibfnamefont {S.}~\bibnamefont
  {Hayami}}, \bibinfo {author} {\bibfnamefont {Y.}~\bibnamefont {Yanagi}},\
  and\ \bibinfo {author} {\bibfnamefont {H.}~\bibnamefont {Kusunose}},\ }\href
  {https://doi.org/10.1103/PhysRevB.102.144441} {\bibfield  {journal} {\bibinfo
   {journal} {Phys. Rev. B}\ }\textbf {\bibinfo {volume} {102}},\ \bibinfo
  {pages} {144441} (\bibinfo {year} {2020})}\BibitemShut {NoStop}%
\bibitem [{\citenamefont {Hayami}\ \emph {et~al.}(2018)\citenamefont {Hayami},
  \citenamefont {Yatsushiro}, \citenamefont {Yanagi},\ and\ \citenamefont
  {Kusunose}}]{Hayami2018}%
  \BibitemOpen
  \bibfield  {author} {\bibinfo {author} {\bibfnamefont {S.}~\bibnamefont
  {Hayami}}, \bibinfo {author} {\bibfnamefont {M.}~\bibnamefont {Yatsushiro}},
  \bibinfo {author} {\bibfnamefont {Y.}~\bibnamefont {Yanagi}},\ and\ \bibinfo
  {author} {\bibfnamefont {H.}~\bibnamefont {Kusunose}},\ }\href
  {https://doi.org/10.1103/PhysRevB.98.165110} {\bibfield  {journal} {\bibinfo
  {journal} {Phys. Rev. B}\ }\textbf {\bibinfo {volume} {98}},\ \bibinfo
  {pages} {165110} (\bibinfo {year} {2018})}\BibitemShut {NoStop}%
\bibitem [{\citenamefont {Watanabe}\ and\ \citenamefont
  {Yanase}(2018)}]{Watanabe2018}%
  \BibitemOpen
  \bibfield  {author} {\bibinfo {author} {\bibfnamefont {H.}~\bibnamefont
  {Watanabe}}\ and\ \bibinfo {author} {\bibfnamefont {Y.}~\bibnamefont
  {Yanase}},\ }\href {https://doi.org/10.1103/PhysRevB.98.245129} {\bibfield
  {journal} {\bibinfo  {journal} {Phys. Rev. B}\ }\textbf {\bibinfo {volume}
  {98}},\ \bibinfo {pages} {245129} (\bibinfo {year} {2018})}\BibitemShut
  {NoStop}%
\bibitem [{\citenamefont {Yatsushiro}\ \emph {et~al.}(2021)\citenamefont
  {Yatsushiro}, \citenamefont {Kusunose},\ and\ \citenamefont
  {Hayami}}]{Yatsushiro2021}%
  \BibitemOpen
  \bibfield  {author} {\bibinfo {author} {\bibfnamefont {M.}~\bibnamefont
  {Yatsushiro}}, \bibinfo {author} {\bibfnamefont {H.}~\bibnamefont
  {Kusunose}},\ and\ \bibinfo {author} {\bibfnamefont {S.}~\bibnamefont
  {Hayami}},\ }\href {https://doi.org/10.1103/PhysRevB.104.054412} {\bibfield
  {journal} {\bibinfo  {journal} {Phys. Rev. B}\ }\textbf {\bibinfo {volume}
  {104}},\ \bibinfo {pages} {054412} (\bibinfo {year} {2021})}\BibitemShut
  {NoStop}%
\bibitem [{\citenamefont {Oiwa}\ and\ \citenamefont
  {Kusunose}(2022)}]{Oiwa2022}%
  \BibitemOpen
  \bibfield  {author} {\bibinfo {author} {\bibfnamefont {R.}~\bibnamefont
  {Oiwa}}\ and\ \bibinfo {author} {\bibfnamefont {H.}~\bibnamefont
  {Kusunose}},\ }\href {https://doi.org/10.7566/JPSJ.91.014701} {\bibfield
  {journal} {\bibinfo  {journal} {J. Phys. Soc. Jpn.}\ }\textbf {\bibinfo
  {volume} {91}},\ \bibinfo {pages} {014701} (\bibinfo {year}
  {2022})}\BibitemShut {NoStop}%
\bibitem [{\citenamefont {Khomskii}(2006)}]{Khomskii2006}%
  \BibitemOpen
  \bibfield  {author} {\bibinfo {author} {\bibfnamefont {D.}~\bibnamefont
  {Khomskii}},\ }\href {https://doi.org/10.1016/j.jmmm.2006.01.238} {\bibfield
  {journal} {\bibinfo  {journal} {J. Magn. Magn. Mater.}\ }\textbf {\bibinfo
  {volume} {306}},\ \bibinfo {pages} {1} (\bibinfo {year} {2006})}\BibitemShut
  {NoStop}%
\bibitem [{\citenamefont {Khomskii}(2009)}]{Khomskii2009}%
  \BibitemOpen
  \bibfield  {author} {\bibinfo {author} {\bibfnamefont {D.}~\bibnamefont
  {Khomskii}},\ }\href {https://doi.org/10.1103/Physics.2.20} {\bibfield
  {journal} {\bibinfo  {journal} {Physics (College. Park. Md).}\ }\textbf
  {\bibinfo {volume} {2}},\ \bibinfo {pages} {20} (\bibinfo {year}
  {2009})}\BibitemShut {NoStop}%
\bibitem [{\citenamefont {Edelstein}(1990)}]{Edelstein1990}%
  \BibitemOpen
  \bibfield  {author} {\bibinfo {author} {\bibfnamefont {V.}~\bibnamefont
  {Edelstein}},\ }\href
  {https://doi.org/https://doi.org/10.1016/0038-1098(90)90963-C} {\bibfield
  {journal} {\bibinfo  {journal} {Solid State Commun.}\ }\textbf {\bibinfo
  {volume} {73}},\ \bibinfo {pages} {233 } (\bibinfo {year}
  {1990})}\BibitemShut {NoStop}%
\bibitem [{\citenamefont {Fujimoto}(2007)}]{Fujimoto2007}%
  \BibitemOpen
  \bibfield  {author} {\bibinfo {author} {\bibfnamefont {S.}~\bibnamefont
  {Fujimoto}},\ }\href {https://doi.org/10.1143/JPSJ.76.034712} {\bibfield
  {journal} {\bibinfo  {journal} {J. Phys. Soc. Jpn.}\ }\textbf {\bibinfo
  {volume} {76}},\ \bibinfo {pages} {034712} (\bibinfo {year}
  {2007})}\BibitemShut {NoStop}%
\bibitem [{\citenamefont {Yanase}(2014)}]{Yanase2014}%
  \BibitemOpen
  \bibfield  {author} {\bibinfo {author} {\bibfnamefont {Y.}~\bibnamefont
  {Yanase}},\ }\href {https://doi.org/10.7566/JPSJ.83.014703} {\bibfield
  {journal} {\bibinfo  {journal} {J. Phys. Soc. Jpn.}\ }\textbf {\bibinfo
  {volume} {83}},\ \bibinfo {pages} {014703} (\bibinfo {year}
  {2014})}\BibitemShut {NoStop}%
\bibitem [{\citenamefont {Hayami}\ \emph
  {et~al.}(2014{\natexlab{a}})\citenamefont {Hayami}, \citenamefont
  {Kusunose},\ and\ \citenamefont {Motome}}]{Hayami2014_Parity}%
  \BibitemOpen
  \bibfield  {author} {\bibinfo {author} {\bibfnamefont {S.}~\bibnamefont
  {Hayami}}, \bibinfo {author} {\bibfnamefont {H.}~\bibnamefont {Kusunose}},\
  and\ \bibinfo {author} {\bibfnamefont {Y.}~\bibnamefont {Motome}},\ }\href
  {https://doi.org/10.1103/PhysRevB.90.081115} {\bibfield  {journal} {\bibinfo
  {journal} {Phys. Rev. B}\ }\textbf {\bibinfo {volume} {90}},\ \bibinfo
  {pages} {081115} (\bibinfo {year} {2014}{\natexlab{a}})}\BibitemShut
  {NoStop}%
\bibitem [{\citenamefont {Th{\"o}le}\ and\ \citenamefont
  {Spaldin}(2018)}]{Thole2018}%
  \BibitemOpen
  \bibfield  {author} {\bibinfo {author} {\bibfnamefont {F.}~\bibnamefont
  {Th{\"o}le}}\ and\ \bibinfo {author} {\bibfnamefont {N.~A.}\ \bibnamefont
  {Spaldin}},\ }\href {https://doi.org/10.1098/rsta.2017.0450} {\bibfield
  {journal} {\bibinfo  {journal} {Phil. Trans. R. Soc. A.}\ }\textbf {\bibinfo
  {volume} {376}},\ \bibinfo {pages} {20170450} (\bibinfo {year}
  {2018})}\BibitemShut {NoStop}%
\bibitem [{\citenamefont {Saito}\ \emph {et~al.}(2018)\citenamefont {Saito},
  \citenamefont {Uenishi}, \citenamefont {Miura}, \citenamefont {Tabata},
  \citenamefont {Hidaka}, \citenamefont {Yanagisawa},\ and\ \citenamefont
  {Amitsuka}}]{Saito2018}%
  \BibitemOpen
  \bibfield  {author} {\bibinfo {author} {\bibfnamefont {H.}~\bibnamefont
  {Saito}}, \bibinfo {author} {\bibfnamefont {K.}~\bibnamefont {Uenishi}},
  \bibinfo {author} {\bibfnamefont {N.}~\bibnamefont {Miura}}, \bibinfo
  {author} {\bibfnamefont {C.}~\bibnamefont {Tabata}}, \bibinfo {author}
  {\bibfnamefont {H.}~\bibnamefont {Hidaka}}, \bibinfo {author} {\bibfnamefont
  {T.}~\bibnamefont {Yanagisawa}},\ and\ \bibinfo {author} {\bibfnamefont
  {H.}~\bibnamefont {Amitsuka}},\ }\href
  {https://doi.org/10.7566/JPSJ.87.033702} {\bibfield  {journal} {\bibinfo
  {journal} {J. Phys. Soc. Jpn.}\ }\textbf {\bibinfo {volume} {87}},\ \bibinfo
  {pages} {033702} (\bibinfo {year} {2018})}\BibitemShut {NoStop}%
\bibitem [{\citenamefont {Hayami}\ \emph
  {et~al.}(2014{\natexlab{b}})\citenamefont {Hayami}, \citenamefont
  {Kusunose},\ and\ \citenamefont {Motome}}]{Hayami2014_Toroidal}%
  \BibitemOpen
  \bibfield  {author} {\bibinfo {author} {\bibfnamefont {S.}~\bibnamefont
  {Hayami}}, \bibinfo {author} {\bibfnamefont {H.}~\bibnamefont {Kusunose}},\
  and\ \bibinfo {author} {\bibfnamefont {Y.}~\bibnamefont {Motome}},\ }\href
  {https://doi.org/10.1103/PhysRevB.90.024432} {\bibfield  {journal} {\bibinfo
  {journal} {Phys. Rev. B}\ }\textbf {\bibinfo {volume} {90}},\ \bibinfo
  {pages} {024432} (\bibinfo {year} {2014}{\natexlab{b}})}\BibitemShut
  {NoStop}%
\bibitem [{\citenamefont {Furukawa}\ \emph {et~al.}(2017)\citenamefont
  {Furukawa}, \citenamefont {Shimokawa}, \citenamefont {Kobayashi},\ and\
  \citenamefont {Itou}}]{Furukawa2017}%
  \BibitemOpen
  \bibfield  {author} {\bibinfo {author} {\bibfnamefont {T.}~\bibnamefont
  {Furukawa}}, \bibinfo {author} {\bibfnamefont {Y.}~\bibnamefont {Shimokawa}},
  \bibinfo {author} {\bibfnamefont {K.}~\bibnamefont {Kobayashi}},\ and\
  \bibinfo {author} {\bibfnamefont {T.}~\bibnamefont {Itou}},\ }\href
  {https://doi.org/10.1038/s41467-017-01093-3} {\bibfield  {journal} {\bibinfo
  {journal} {Nat. Commun.}\ }\textbf {\bibinfo {volume} {8}},\ \bibinfo {pages}
  {954} (\bibinfo {year} {2017})}\BibitemShut {NoStop}%
\bibitem [{\citenamefont {Mentink}\ \emph {et~al.}(1994)\citenamefont
  {Mentink}, \citenamefont {Drost}, \citenamefont {Nieuwenhuys}, \citenamefont
  {Frikkee}, \citenamefont {Menovsky},\ and\ \citenamefont
  {Mydosh}}]{Mentink1994}%
  \BibitemOpen
  \bibfield  {author} {\bibinfo {author} {\bibfnamefont {S.~A.~M.}\
  \bibnamefont {Mentink}}, \bibinfo {author} {\bibfnamefont {A.}~\bibnamefont
  {Drost}}, \bibinfo {author} {\bibfnamefont {G.~J.}\ \bibnamefont
  {Nieuwenhuys}}, \bibinfo {author} {\bibfnamefont {E.}~\bibnamefont
  {Frikkee}}, \bibinfo {author} {\bibfnamefont {A.~A.}\ \bibnamefont
  {Menovsky}},\ and\ \bibinfo {author} {\bibfnamefont {J.~A.}\ \bibnamefont
  {Mydosh}},\ }\href {https://doi.org/10.1103/PhysRevLett.73.1031} {\bibfield
  {journal} {\bibinfo  {journal} {Phys. Rev. Lett.}\ }\textbf {\bibinfo
  {volume} {73}},\ \bibinfo {pages} {1031} (\bibinfo {year}
  {1994})}\BibitemShut {NoStop}%
\bibitem [{\citenamefont {Haga}\ \emph {et~al.}(2008)\citenamefont {Haga},
  \citenamefont {Oyamada}, \citenamefont {Matsuda}, \citenamefont {Ikeda},\
  and\ \citenamefont {^^c5^^8cunki}}]{Haga2008}%
  \BibitemOpen
  \bibfield  {author} {\bibinfo {author} {\bibfnamefont {Y.}~\bibnamefont
  {Haga}}, \bibinfo {author} {\bibfnamefont {A.}~\bibnamefont {Oyamada}},
  \bibinfo {author} {\bibfnamefont {T.}~\bibnamefont {Matsuda}}, \bibinfo
  {author} {\bibfnamefont {S.}~\bibnamefont {Ikeda}},\ and\ \bibinfo {author}
  {\bibfnamefont {Y.}~\bibnamefont {^^c5^^8cunki}},\ }\href
  {https://doi.org/10.1016/j.physb.2007.10.232} {\bibfield  {journal} {\bibinfo
   {journal} {Physica B Condens. Matter}\ }\textbf {\bibinfo {volume} {403}},\
  \bibinfo {pages} {900} (\bibinfo {year} {2008})}\BibitemShut {NoStop}%
\bibitem [{\citenamefont {Tabata}\ \emph {et~al.}(2021)\citenamefont {Tabata},
  \citenamefont {Sagayama}, \citenamefont {Saito}, \citenamefont {Nakao},\ and\
  \citenamefont {Amitsuka}}]{Tabata2021}%
  \BibitemOpen
  \bibfield  {author} {\bibinfo {author} {\bibfnamefont {C.}~\bibnamefont
  {Tabata}}, \bibinfo {author} {\bibfnamefont {H.}~\bibnamefont {Sagayama}},
  \bibinfo {author} {\bibfnamefont {H.}~\bibnamefont {Saito}}, \bibinfo
  {author} {\bibfnamefont {H.}~\bibnamefont {Nakao}},\ and\ \bibinfo {author}
  {\bibfnamefont {H.}~\bibnamefont {Amitsuka}},\ }\href
  {https://doi.org/10.7566/JPSJ.90.064601} {\bibfield  {journal} {\bibinfo
  {journal} {J. Phys. Soc. Jpn.}\ }\textbf {\bibinfo {volume} {90}},\ \bibinfo
  {pages} {064601} (\bibinfo {year} {2021})}\BibitemShut {NoStop}%
\bibitem [{\citenamefont {Willwater}\ \emph {et~al.}(2021)\citenamefont
  {Willwater}, \citenamefont {S{\"{u}}llow}, \citenamefont {Reehuis},
  \citenamefont {Feyerherm}, \citenamefont {Amitsuka}, \citenamefont
  {Ouladdiaf}, \citenamefont {Suard}, \citenamefont {Klicpera}, \citenamefont
  {Vali{\v{s}}ka}, \citenamefont {Posp{\'{i}}{\v{s}}il},\ and\ \citenamefont
  {Sechovsk{\'{y}}}}]{Willwater2021}%
  \BibitemOpen
  \bibfield  {author} {\bibinfo {author} {\bibfnamefont {J.}~\bibnamefont
  {Willwater}}, \bibinfo {author} {\bibfnamefont {S.}~\bibnamefont
  {S{\"{u}}llow}}, \bibinfo {author} {\bibfnamefont {M.}~\bibnamefont
  {Reehuis}}, \bibinfo {author} {\bibfnamefont {R.}~\bibnamefont {Feyerherm}},
  \bibinfo {author} {\bibfnamefont {H.}~\bibnamefont {Amitsuka}}, \bibinfo
  {author} {\bibfnamefont {B.}~\bibnamefont {Ouladdiaf}}, \bibinfo {author}
  {\bibfnamefont {E.}~\bibnamefont {Suard}}, \bibinfo {author} {\bibfnamefont
  {M.}~\bibnamefont {Klicpera}}, \bibinfo {author} {\bibfnamefont
  {M.}~\bibnamefont {Vali{\v{s}}ka}}, \bibinfo {author} {\bibfnamefont
  {J.}~\bibnamefont {Posp{\'{i}}{\v{s}}il}},\ and\ \bibinfo {author}
  {\bibfnamefont {V.}~\bibnamefont {Sechovsk{\'{y}}}},\ }\href
  {https://doi.org/10.1103/PhysRevB.103.184426} {\bibfield  {journal} {\bibinfo
   {journal} {Phys. Rev. B}\ }\textbf {\bibinfo {volume} {103}},\ \bibinfo
  {pages} {184426} (\bibinfo {year} {2021})}\BibitemShut {NoStop}%
\bibitem [{\citenamefont {Yanagisawa}\ \emph {et~al.}(2021)\citenamefont
  {Yanagisawa}, \citenamefont {Matsumori}, \citenamefont {Saito}, \citenamefont
  {Hidaka}, \citenamefont {Amitsuka}, \citenamefont {Nakamura}, \citenamefont
  {Awaji}, \citenamefont {Gorbunov}, \citenamefont {Zherlitsyn}, \citenamefont
  {Wosnitza}, \citenamefont {Uhl{\'{i}}^^c5^^99ov{\'{a}}}, \citenamefont
  {Vali{\v{s}}ka},\ and\ \citenamefont {Sechovsk{\'{y}}}}]{Yanagisawa2021}%
  \BibitemOpen
  \bibfield  {author} {\bibinfo {author} {\bibfnamefont {T.}~\bibnamefont
  {Yanagisawa}}, \bibinfo {author} {\bibfnamefont {H.}~\bibnamefont
  {Matsumori}}, \bibinfo {author} {\bibfnamefont {H.}~\bibnamefont {Saito}},
  \bibinfo {author} {\bibfnamefont {H.}~\bibnamefont {Hidaka}}, \bibinfo
  {author} {\bibfnamefont {H.}~\bibnamefont {Amitsuka}}, \bibinfo {author}
  {\bibfnamefont {S.}~\bibnamefont {Nakamura}}, \bibinfo {author}
  {\bibfnamefont {S.}~\bibnamefont {Awaji}}, \bibinfo {author} {\bibfnamefont
  {D.~I.}\ \bibnamefont {Gorbunov}}, \bibinfo {author} {\bibfnamefont
  {S.}~\bibnamefont {Zherlitsyn}}, \bibinfo {author} {\bibfnamefont
  {J.}~\bibnamefont {Wosnitza}}, \bibinfo {author} {\bibfnamefont
  {K.}~\bibnamefont {Uhl{\'{i}}^^c5^^99ov{\'{a}}}}, \bibinfo {author}
  {\bibfnamefont {M.}~\bibnamefont {Vali{\v{s}}ka}},\ and\ \bibinfo {author}
  {\bibfnamefont {V.}~\bibnamefont {Sechovsk{\'{y}}}},\ }\href
  {https://doi.org/10.1103/PhysRevLett.126.157201} {\bibfield  {journal}
  {\bibinfo  {journal} {Phys. Rev. Lett.}\ }\textbf {\bibinfo {volume} {126}},\
  \bibinfo {pages} {157201} (\bibinfo {year} {2021})}\BibitemShut {NoStop}%
\bibitem [{\citenamefont {Movshovich}\ \emph {et~al.}(1999)\citenamefont
  {Movshovich}, \citenamefont {Jaime}, \citenamefont {Mentink}, \citenamefont
  {Menovsky},\ and\ \citenamefont {Mydosh}}]{Movshovich1999}%
  \BibitemOpen
  \bibfield  {author} {\bibinfo {author} {\bibfnamefont {R.}~\bibnamefont
  {Movshovich}}, \bibinfo {author} {\bibfnamefont {M.}~\bibnamefont {Jaime}},
  \bibinfo {author} {\bibfnamefont {S.}~\bibnamefont {Mentink}}, \bibinfo
  {author} {\bibfnamefont {A.~A.}\ \bibnamefont {Menovsky}},\ and\ \bibinfo
  {author} {\bibfnamefont {J.~A.}\ \bibnamefont {Mydosh}},\ }\href
  {https://doi.org/10.1103/PhysRevLett.83.2065} {\bibfield  {journal} {\bibinfo
   {journal} {Phys. Rev. Lett.}\ }\textbf {\bibinfo {volume} {83}},\ \bibinfo
  {pages} {2065} (\bibinfo {year} {1999})}\BibitemShut {NoStop}%
\bibitem [{\citenamefont {Lacroix}\ \emph {et~al.}(1996)\citenamefont
  {Lacroix}, \citenamefont {Canals},\ and\ \citenamefont
  {N{\'{u}}{\~{n}}ez-Regueiro}}]{Lacroix1996}%
  \BibitemOpen
  \bibfield  {author} {\bibinfo {author} {\bibfnamefont {C.}~\bibnamefont
  {Lacroix}}, \bibinfo {author} {\bibfnamefont {B.}~\bibnamefont {Canals}},\
  and\ \bibinfo {author} {\bibfnamefont {M.~D.}\ \bibnamefont
  {N{\'{u}}{\~{n}}ez-Regueiro}},\ }\href
  {https://doi.org/10.1103/PhysRevLett.77.5126} {\bibfield  {journal} {\bibinfo
   {journal} {Phys. Rev. Lett.}\ }\textbf {\bibinfo {volume} {77}},\ \bibinfo
  {pages} {5126} (\bibinfo {year} {1996})}\BibitemShut {NoStop}%
\bibitem [{\citenamefont {Fox}\ \emph {et~al.}(2020)\citenamefont {Fox},
  \citenamefont {Sah}, \citenamefont {Sawhney}, \citenamefont {Stoner},\ and\
  \citenamefont {Zhao}}]{Fox2020}%
  \BibitemOpen
  \bibfield  {author} {\bibinfo {author} {\bibfnamefont {J.}~\bibnamefont
  {Fox}}, \bibinfo {author} {\bibfnamefont {A.}~\bibnamefont {Sah}}, \bibinfo
  {author} {\bibfnamefont {M.}~\bibnamefont {Sawhney}}, \bibinfo {author}
  {\bibfnamefont {D.}~\bibnamefont {Stoner}},\ and\ \bibinfo {author}
  {\bibfnamefont {Y.}~\bibnamefont {Zhao}},\ }\href
  {https://doi.org/10.1017/S0305004119000173} {\bibfield  {journal} {\bibinfo
  {journal} {Math. Proc. Cambridge Philos. Soc.}\ }\textbf {\bibinfo {volume}
  {169}},\ \bibinfo {pages} {209} (\bibinfo {year} {2020})}\BibitemShut
  {NoStop}%
\bibitem [{\citenamefont {Lee}\ \emph {et~al.}(1984)\citenamefont {Lee},
  \citenamefont {Joannopoulos}, \citenamefont {Negele},\ and\ \citenamefont
  {Landau}}]{Lee1984}%
  \BibitemOpen
  \bibfield  {author} {\bibinfo {author} {\bibfnamefont {D.~H.}\ \bibnamefont
  {Lee}}, \bibinfo {author} {\bibfnamefont {J.~D.}\ \bibnamefont
  {Joannopoulos}}, \bibinfo {author} {\bibfnamefont {J.~W.}\ \bibnamefont
  {Negele}},\ and\ \bibinfo {author} {\bibfnamefont {D.~P.}\ \bibnamefont
  {Landau}},\ }\href {https://doi.org/10.1103/PhysRevLett.52.433} {\bibfield
  {journal} {\bibinfo  {journal} {Phys. Rev. Lett.}\ }\textbf {\bibinfo
  {volume} {52}},\ \bibinfo {pages} {433} (\bibinfo {year} {1984})}\BibitemShut
  {NoStop}%
\bibitem [{\citenamefont {Ramirez}(1994)}]{Ramirez1994}%
  \BibitemOpen
  \bibfield  {author} {\bibinfo {author} {\bibfnamefont {A.~P.}\ \bibnamefont
  {Ramirez}},\ }\href {https://doi.org/10.1146/annurev.ms.24.080194.002321}
  {\bibfield  {journal} {\bibinfo  {journal} {Annu. Rev. Mater. Sci.}\ }\textbf
  {\bibinfo {volume} {24}},\ \bibinfo {pages} {453} (\bibinfo {year}
  {1994})}\BibitemShut {NoStop}%
\bibitem [{\citenamefont {Kubo}\ and\ \citenamefont
  {Kuramoto}(2004)}]{Kubo2004}%
  \BibitemOpen
  \bibfield  {author} {\bibinfo {author} {\bibfnamefont {K.}~\bibnamefont
  {Kubo}}\ and\ \bibinfo {author} {\bibfnamefont {Y.}~\bibnamefont
  {Kuramoto}},\ }\href {https://doi.org/10.1143/JPSJ.73.216} {\bibfield
  {journal} {\bibinfo  {journal} {J. Phys. Soc. Jpn.}\ }\textbf {\bibinfo
  {volume} {73}},\ \bibinfo {pages} {216} (\bibinfo {year} {2004})}\BibitemShut
  {NoStop}%
\bibitem [{\citenamefont {Hattori}\ and\ \citenamefont
  {Tsunetsugu}(2014)}]{Hattori2014}%
  \BibitemOpen
  \bibfield  {author} {\bibinfo {author} {\bibfnamefont {K.}~\bibnamefont
  {Hattori}}\ and\ \bibinfo {author} {\bibfnamefont {H.}~\bibnamefont
  {Tsunetsugu}},\ }\href {https://doi.org/10.7566/JPSJ.83.034709} {\bibfield
  {journal} {\bibinfo  {journal} {J. Phys. Soc. Jpn.}\ }\textbf {\bibinfo
  {volume} {83}},\ \bibinfo {pages} {034709} (\bibinfo {year}
  {2014})}\BibitemShut {NoStop}%
\bibitem [{\citenamefont {Yatsushiro}\ \emph {et~al.}(2022)\citenamefont
  {Yatsushiro}, \citenamefont {Oiwa}, \citenamefont {Kusunose},\ and\
  \citenamefont {Hayami}}]{Yatsushiro2022}%
  \BibitemOpen
  \bibfield  {author} {\bibinfo {author} {\bibfnamefont {M.}~\bibnamefont
  {Yatsushiro}}, \bibinfo {author} {\bibfnamefont {R.}~\bibnamefont {Oiwa}},
  \bibinfo {author} {\bibfnamefont {H.}~\bibnamefont {Kusunose}},\ and\
  \bibinfo {author} {\bibfnamefont {S.}~\bibnamefont {Hayami}},\ }\href
  {https://doi.org/10.1103/PhysRevB.105.155157} {\bibfield  {journal} {\bibinfo
   {journal} {Phys. Rev. B}\ }\textbf {\bibinfo {volume} {105}},\ \bibinfo
  {pages} {155157} (\bibinfo {year} {2022})}\BibitemShut {NoStop}%
\bibitem [{\citenamefont {Onsager}(1931{\natexlab{a}})}]{Onsager1931_I}%
  \BibitemOpen
  \bibfield  {author} {\bibinfo {author} {\bibfnamefont {L.}~\bibnamefont
  {Onsager}},\ }\href {https://doi.org/10.1103/PhysRev.37.405} {\bibfield
  {journal} {\bibinfo  {journal} {Phys. Rev.}\ }\textbf {\bibinfo {volume}
  {37}},\ \bibinfo {pages} {405} (\bibinfo {year}
  {1931}{\natexlab{a}})}\BibitemShut {NoStop}%
\bibitem [{\citenamefont {Onsager}(1931{\natexlab{b}})}]{Onsager1931_II}%
  \BibitemOpen
  \bibfield  {author} {\bibinfo {author} {\bibfnamefont {L.}~\bibnamefont
  {Onsager}},\ }\href {https://doi.org/10.1103/PhysRev.38.2265} {\bibfield
  {journal} {\bibinfo  {journal} {Phys. Rev.}\ }\textbf {\bibinfo {volume}
  {38}},\ \bibinfo {pages} {2265} (\bibinfo {year}
  {1931}{\natexlab{b}})}\BibitemShut {NoStop}%
\bibitem [{\citenamefont {Landau}\ and\ \citenamefont
  {Lifshitz}(1980)}]{LandauLifshitzStatisticalPhysics}%
  \BibitemOpen
  \bibfield  {author} {\bibinfo {author} {\bibfnamefont {L.~D.}\ \bibnamefont
  {Landau}}\ and\ \bibinfo {author} {\bibfnamefont {E.~M.}\ \bibnamefont
  {Lifshitz}},\ }\href@noop {} {\emph {\bibinfo {title} {Statistical
  Physics}}}\ (\bibinfo  {publisher} {Pergamon press, Oxford},\ \bibinfo {year}
  {1980})\BibitemShut {NoStop}%
\bibitem [{\citenamefont {Rikken}\ and\ \citenamefont
  {Wyder}(2005)}]{Rikken2005}%
  \BibitemOpen
  \bibfield  {author} {\bibinfo {author} {\bibfnamefont {G.~L. J.~A.}\
  \bibnamefont {Rikken}}\ and\ \bibinfo {author} {\bibfnamefont
  {P.}~\bibnamefont {Wyder}},\ }\href
  {https://doi.org/10.1103/PhysRevLett.94.016601} {\bibfield  {journal}
  {\bibinfo  {journal} {Phys. Rev. Lett.}\ }\textbf {\bibinfo {volume} {94}},\
  \bibinfo {pages} {016601} (\bibinfo {year} {2005})}\BibitemShut {NoStop}%
\bibitem [{\citenamefont {Patri}\ \emph {et~al.}(2019)\citenamefont {Patri},
  \citenamefont {Sakai}, \citenamefont {Lee}, \citenamefont {Paramekanti},
  \citenamefont {Nakatsuji},\ and\ \citenamefont {Kim}}]{Patri2019}%
  \BibitemOpen
  \bibfield  {author} {\bibinfo {author} {\bibfnamefont {A.~S.}\ \bibnamefont
  {Patri}}, \bibinfo {author} {\bibfnamefont {A.}~\bibnamefont {Sakai}},
  \bibinfo {author} {\bibfnamefont {S.}~\bibnamefont {Lee}}, \bibinfo {author}
  {\bibfnamefont {A.}~\bibnamefont {Paramekanti}}, \bibinfo {author}
  {\bibfnamefont {S.}~\bibnamefont {Nakatsuji}},\ and\ \bibinfo {author}
  {\bibfnamefont {Y.~B.}\ \bibnamefont {Kim}},\ }\href
  {https://doi.org/10.1038/s41467-019-11913-3} {\bibfield  {journal} {\bibinfo
  {journal} {Nat. Commun.}\ }\textbf {\bibinfo {volume} {10}},\ \bibinfo
  {pages} {4092} (\bibinfo {year} {2019})}\BibitemShut {NoStop}%
\bibitem [{\citenamefont {Seifert}\ and\ \citenamefont
  {Savary}(2022)}]{Seifert2022}%
  \BibitemOpen
  \bibfield  {author} {\bibinfo {author} {\bibfnamefont {U.~F.~P.}\
  \bibnamefont {Seifert}}\ and\ \bibinfo {author} {\bibfnamefont
  {L.}~\bibnamefont {Savary}},\ }\href
  {https://doi.org/10.1103/PhysRevB.106.195147} {\bibfield  {journal} {\bibinfo
   {journal} {Phys. Rev. B}\ }\textbf {\bibinfo {volume} {106}},\ \bibinfo
  {pages} {195147} (\bibinfo {year} {2022})}\BibitemShut {NoStop}%
\bibitem [{\citenamefont {Motome}\ \emph {et~al.}(2010)\citenamefont {Motome},
  \citenamefont {Nakamikawa}, \citenamefont {Yamaji},\ and\ \citenamefont
  {Udagawa}}]{Motome2010}%
  \BibitemOpen
  \bibfield  {author} {\bibinfo {author} {\bibfnamefont {Y.}~\bibnamefont
  {Motome}}, \bibinfo {author} {\bibfnamefont {K.}~\bibnamefont {Nakamikawa}},
  \bibinfo {author} {\bibfnamefont {Y.}~\bibnamefont {Yamaji}},\ and\ \bibinfo
  {author} {\bibfnamefont {M.}~\bibnamefont {Udagawa}},\ }\href
  {https://doi.org/10.1103/PhysRevLett.105.036403} {\bibfield  {journal}
  {\bibinfo  {journal} {Phys. Rev. Lett.}\ }\textbf {\bibinfo {volume} {105}},\
  \bibinfo {pages} {036403} (\bibinfo {year} {2010})}\BibitemShut {NoStop}%
\bibitem [{\citenamefont {Ishizuka}\ and\ \citenamefont
  {Motome}(2012)}]{Ishizuka2012}%
  \BibitemOpen
  \bibfield  {author} {\bibinfo {author} {\bibfnamefont {H.}~\bibnamefont
  {Ishizuka}}\ and\ \bibinfo {author} {\bibfnamefont {Y.}~\bibnamefont
  {Motome}},\ }\href {https://doi.org/10.1103/PhysRevLett.108.257205}
  {\bibfield  {journal} {\bibinfo  {journal} {Phys. Rev. Lett.}\ }\textbf
  {\bibinfo {volume} {108}},\ \bibinfo {pages} {257205} (\bibinfo {year}
  {2012})}\BibitemShut {NoStop}%
\bibitem [{\citenamefont {Hayami}\ \emph {et~al.}(2012)\citenamefont {Hayami},
  \citenamefont {Udagawa},\ and\ \citenamefont {Motome}}]{Hayami2012}%
  \BibitemOpen
  \bibfield  {author} {\bibinfo {author} {\bibfnamefont {S.}~\bibnamefont
  {Hayami}}, \bibinfo {author} {\bibfnamefont {M.}~\bibnamefont {Udagawa}},\
  and\ \bibinfo {author} {\bibfnamefont {Y.}~\bibnamefont {Motome}},\ }\href
  {https://doi.org/10.1143/JPSJ.81.103707} {\bibfield  {journal} {\bibinfo
  {journal} {J. Phys. Soc. Jpn.}\ }\textbf {\bibinfo {volume} {81}},\ \bibinfo
  {pages} {10} (\bibinfo {year} {2012})}\BibitemShut {NoStop}%
\bibitem [{\citenamefont {Ozawa}\ \emph {et~al.}(2016)\citenamefont {Ozawa},
  \citenamefont {Hayami}, \citenamefont {Barros}, \citenamefont {Chern},
  \citenamefont {Motome},\ and\ \citenamefont {Batista}}]{Ozawa2016}%
  \BibitemOpen
  \bibfield  {author} {\bibinfo {author} {\bibfnamefont {R.}~\bibnamefont
  {Ozawa}}, \bibinfo {author} {\bibfnamefont {S.}~\bibnamefont {Hayami}},
  \bibinfo {author} {\bibfnamefont {K.}~\bibnamefont {Barros}}, \bibinfo
  {author} {\bibfnamefont {G.-W.}\ \bibnamefont {Chern}}, \bibinfo {author}
  {\bibfnamefont {Y.}~\bibnamefont {Motome}},\ and\ \bibinfo {author}
  {\bibfnamefont {C.~D.}\ \bibnamefont {Batista}},\ }\href
  {https://doi.org/10.7566/JPSJ.85.103703} {\bibfield  {journal} {\bibinfo
  {journal} {J. Phys. Soc. Jpn.}\ }\textbf {\bibinfo {volume} {85}},\ \bibinfo
  {pages} {103703} (\bibinfo {year} {2016})}\BibitemShut {NoStop}%
\bibitem [{\citenamefont {Shimizu}\ \emph {et~al.}(2022)\citenamefont
  {Shimizu}, \citenamefont {Okumura}, \citenamefont {Kato},\ and\ \citenamefont
  {Motome}}]{Shimizu2022_Phase}%
  \BibitemOpen
  \bibfield  {author} {\bibinfo {author} {\bibfnamefont {K.}~\bibnamefont
  {Shimizu}}, \bibinfo {author} {\bibfnamefont {S.}~\bibnamefont {Okumura}},
  \bibinfo {author} {\bibfnamefont {Y.}~\bibnamefont {Kato}},\ and\ \bibinfo
  {author} {\bibfnamefont {Y.}~\bibnamefont {Motome}},\ }\href
  {https://doi.org/10.1103/PhysRevB.105.224405} {\bibfield  {journal} {\bibinfo
   {journal} {Phys. Rev. B}\ }\textbf {\bibinfo {volume} {105}},\ \bibinfo
  {pages} {224405} (\bibinfo {year} {2022})}\BibitemShut {NoStop}%
\bibitem [{\citenamefont {Lingg}\ \emph {et~al.}(1999)\citenamefont {Lingg},
  \citenamefont {Maurer}, \citenamefont {M{\"u}ller},\ and\ \citenamefont
  {McEwen}}]{Lingg1999}%
  \BibitemOpen
  \bibfield  {author} {\bibinfo {author} {\bibfnamefont {N.}~\bibnamefont
  {Lingg}}, \bibinfo {author} {\bibfnamefont {D.}~\bibnamefont {Maurer}},
  \bibinfo {author} {\bibfnamefont {V.}~\bibnamefont {M{\"u}ller}},\ and\
  \bibinfo {author} {\bibfnamefont {K.~A.}\ \bibnamefont {McEwen}},\ }\href
  {https://doi.org/10.1103/PhysRevB.60.R8430} {\bibfield  {journal} {\bibinfo
  {journal} {Phys. Rev. B}\ }\textbf {\bibinfo {volume} {60}},\ \bibinfo
  {pages} {R8430} (\bibinfo {year} {1999})}\BibitemShut {NoStop}%
\bibitem [{\citenamefont {McEwen}\ \emph {et~al.}(2003)\citenamefont {McEwen},
  \citenamefont {Park}, \citenamefont {Gipson},\ and\ \citenamefont
  {Gehring}}]{McEwen2003}%
  \BibitemOpen
  \bibfield  {author} {\bibinfo {author} {\bibfnamefont {K.~A.}\ \bibnamefont
  {McEwen}}, \bibinfo {author} {\bibfnamefont {J.-G.}\ \bibnamefont {Park}},
  \bibinfo {author} {\bibfnamefont {A.~J.}\ \bibnamefont {Gipson}},\ and\
  \bibinfo {author} {\bibfnamefont {G.~A.}\ \bibnamefont {Gehring}},\ }\href
  {https://doi.org/10.1088/0953-8984/15/28/304} {\bibfield  {journal} {\bibinfo
   {journal} {J. Phys. Condens. Matter}\ }\textbf {\bibinfo {volume} {15}},\
  \bibinfo {pages} {S1923} (\bibinfo {year} {2003})}\BibitemShut {NoStop}%
\bibitem [{\citenamefont {Ortiz}\ \emph {et~al.}(2019)\citenamefont {Ortiz},
  \citenamefont {Gomes}, \citenamefont {Morey}, \citenamefont {Winiarski},
  \citenamefont {Bordelon}, \citenamefont {Mangum}, \citenamefont {Oswald},
  \citenamefont {Rodriguez-Rivera}, \citenamefont {Neilson}, \citenamefont
  {Wilson}, \citenamefont {Ertekin}, \citenamefont {McQueen},\ and\
  \citenamefont {Toberer}}]{Ortiz2019}%
  \BibitemOpen
  \bibfield  {author} {\bibinfo {author} {\bibfnamefont {B.~R.}\ \bibnamefont
  {Ortiz}}, \bibinfo {author} {\bibfnamefont {L.~C.}\ \bibnamefont {Gomes}},
  \bibinfo {author} {\bibfnamefont {J.~R.}\ \bibnamefont {Morey}}, \bibinfo
  {author} {\bibfnamefont {M.}~\bibnamefont {Winiarski}}, \bibinfo {author}
  {\bibfnamefont {M.}~\bibnamefont {Bordelon}}, \bibinfo {author}
  {\bibfnamefont {J.~S.}\ \bibnamefont {Mangum}}, \bibinfo {author}
  {\bibfnamefont {I.~W.~H.}\ \bibnamefont {Oswald}}, \bibinfo {author}
  {\bibfnamefont {J.~A.}\ \bibnamefont {Rodriguez-Rivera}}, \bibinfo {author}
  {\bibfnamefont {J.~R.}\ \bibnamefont {Neilson}}, \bibinfo {author}
  {\bibfnamefont {S.~D.}\ \bibnamefont {Wilson}}, \bibinfo {author}
  {\bibfnamefont {E.}~\bibnamefont {Ertekin}}, \bibinfo {author} {\bibfnamefont
  {T.~M.}\ \bibnamefont {McQueen}},\ and\ \bibinfo {author} {\bibfnamefont
  {E.~S.}\ \bibnamefont {Toberer}},\ }\href
  {https://doi.org/10.1103/PhysRevMaterials.3.094407} {\bibfield  {journal}
  {\bibinfo  {journal} {Phys. Rev. Materials}\ }\textbf {\bibinfo {volume}
  {3}},\ \bibinfo {pages} {094407} (\bibinfo {year} {2019})}\BibitemShut
  {NoStop}%
\bibitem [{\citenamefont {Jiang}\ \emph {et~al.}(2021)\citenamefont {Jiang},
  \citenamefont {Yin}, \citenamefont {Denner}, \citenamefont {Shumiya},
  \citenamefont {Ortiz}, \citenamefont {Xu}, \citenamefont {Guguchia},
  \citenamefont {He}, \citenamefont {Hossain}, \citenamefont {Liu},
  \citenamefont {Ruff}, \citenamefont {Kautzsch}, \citenamefont {Zhang},
  \citenamefont {Chang}, \citenamefont {Belopolski}, \citenamefont {Zhang},
  \citenamefont {Cochran}, \citenamefont {Multer}, \citenamefont {Litskevich},
  \citenamefont {Cheng}, \citenamefont {Yang}, \citenamefont {Wang},
  \citenamefont {Thomale}, \citenamefont {Neupert}, \citenamefont {Wilson},\
  and\ \citenamefont {Hasan}}]{Jiang2021}%
  \BibitemOpen
  \bibfield  {author} {\bibinfo {author} {\bibfnamefont {Y.-X.}\ \bibnamefont
  {Jiang}}, \bibinfo {author} {\bibfnamefont {J.-X.}\ \bibnamefont {Yin}},
  \bibinfo {author} {\bibfnamefont {M.~M.}\ \bibnamefont {Denner}}, \bibinfo
  {author} {\bibfnamefont {N.}~\bibnamefont {Shumiya}}, \bibinfo {author}
  {\bibfnamefont {B.~R.}\ \bibnamefont {Ortiz}}, \bibinfo {author}
  {\bibfnamefont {G.}~\bibnamefont {Xu}}, \bibinfo {author} {\bibfnamefont
  {Z.}~\bibnamefont {Guguchia}}, \bibinfo {author} {\bibfnamefont
  {J.}~\bibnamefont {He}}, \bibinfo {author} {\bibfnamefont {M.~S.}\
  \bibnamefont {Hossain}}, \bibinfo {author} {\bibfnamefont {X.}~\bibnamefont
  {Liu}}, \bibinfo {author} {\bibfnamefont {J.}~\bibnamefont {Ruff}}, \bibinfo
  {author} {\bibfnamefont {L.}~\bibnamefont {Kautzsch}}, \bibinfo {author}
  {\bibfnamefont {S.~S.}\ \bibnamefont {Zhang}}, \bibinfo {author}
  {\bibfnamefont {G.}~\bibnamefont {Chang}}, \bibinfo {author} {\bibfnamefont
  {I.}~\bibnamefont {Belopolski}}, \bibinfo {author} {\bibfnamefont
  {Q.}~\bibnamefont {Zhang}}, \bibinfo {author} {\bibfnamefont {T.~A.}\
  \bibnamefont {Cochran}}, \bibinfo {author} {\bibfnamefont {D.}~\bibnamefont
  {Multer}}, \bibinfo {author} {\bibfnamefont {M.}~\bibnamefont {Litskevich}},
  \bibinfo {author} {\bibfnamefont {Z.-J.}\ \bibnamefont {Cheng}}, \bibinfo
  {author} {\bibfnamefont {X.~P.}\ \bibnamefont {Yang}}, \bibinfo {author}
  {\bibfnamefont {Z.}~\bibnamefont {Wang}}, \bibinfo {author} {\bibfnamefont
  {R.}~\bibnamefont {Thomale}}, \bibinfo {author} {\bibfnamefont
  {T.}~\bibnamefont {Neupert}}, \bibinfo {author} {\bibfnamefont {S.~D.}\
  \bibnamefont {Wilson}},\ and\ \bibinfo {author} {\bibfnamefont {M.~Z.}\
  \bibnamefont {Hasan}},\ }\href {https://doi.org/10.1038/s41563-021-01034-y}
  {\bibfield  {journal} {\bibinfo  {journal} {Nat. Mater.}\ }\textbf {\bibinfo
  {volume} {20}},\ \bibinfo {pages} {1353} (\bibinfo {year}
  {2021})}\BibitemShut {NoStop}%
\bibitem [{\citenamefont {Tan}\ \emph {et~al.}(2021)\citenamefont {Tan},
  \citenamefont {Liu}, \citenamefont {Wang},\ and\ \citenamefont
  {Yan}}]{Tan2021}%
  \BibitemOpen
  \bibfield  {author} {\bibinfo {author} {\bibfnamefont {H.}~\bibnamefont
  {Tan}}, \bibinfo {author} {\bibfnamefont {Y.}~\bibnamefont {Liu}}, \bibinfo
  {author} {\bibfnamefont {Z.}~\bibnamefont {Wang}},\ and\ \bibinfo {author}
  {\bibfnamefont {B.}~\bibnamefont {Yan}},\ }\href
  {https://doi.org/10.1103/PhysRevLett.127.046401} {\bibfield  {journal}
  {\bibinfo  {journal} {Phys. Rev. Lett.}\ }\textbf {\bibinfo {volume} {127}},\
  \bibinfo {pages} {046401} (\bibinfo {year} {2021})}\BibitemShut {NoStop}%
\bibitem [{\citenamefont {Neupert}\ \emph {et~al.}(2021)\citenamefont
  {Neupert}, \citenamefont {Denner}, \citenamefont {Yin}, \citenamefont
  {Thomale},\ and\ \citenamefont {Hasan}}]{Neupert2021}%
  \BibitemOpen
  \bibfield  {author} {\bibinfo {author} {\bibfnamefont {T.}~\bibnamefont
  {Neupert}}, \bibinfo {author} {\bibfnamefont {M.~M.}\ \bibnamefont {Denner}},
  \bibinfo {author} {\bibfnamefont {J.-X.}\ \bibnamefont {Yin}}, \bibinfo
  {author} {\bibfnamefont {R.}~\bibnamefont {Thomale}},\ and\ \bibinfo {author}
  {\bibfnamefont {M.~Z.}\ \bibnamefont {Hasan}},\ }\href
  {https://doi.org/10.1038/s41567-021-01404-y} {\bibfield  {journal} {\bibinfo
  {journal} {Nat. Phys.}\ }\textbf {\bibinfo {volume} {18}},\ \bibinfo {pages}
  {137} (\bibinfo {year} {2021})}\BibitemShut {NoStop}%
\bibitem [{\citenamefont {Xu}\ \emph {et~al.}(2021)\citenamefont {Xu},
  \citenamefont {Yan}, \citenamefont {Yin}, \citenamefont {Xia}, \citenamefont
  {Fang}, \citenamefont {Chen}, \citenamefont {Li}, \citenamefont {Yang},
  \citenamefont {Guo},\ and\ \citenamefont {Feng}}]{Xu2021}%
  \BibitemOpen
  \bibfield  {author} {\bibinfo {author} {\bibfnamefont {H.-S.}\ \bibnamefont
  {Xu}}, \bibinfo {author} {\bibfnamefont {Y.-J.}\ \bibnamefont {Yan}},
  \bibinfo {author} {\bibfnamefont {R.}~\bibnamefont {Yin}}, \bibinfo {author}
  {\bibfnamefont {W.}~\bibnamefont {Xia}}, \bibinfo {author} {\bibfnamefont
  {S.}~\bibnamefont {Fang}}, \bibinfo {author} {\bibfnamefont {Z.}~\bibnamefont
  {Chen}}, \bibinfo {author} {\bibfnamefont {Y.}~\bibnamefont {Li}}, \bibinfo
  {author} {\bibfnamefont {W.}~\bibnamefont {Yang}}, \bibinfo {author}
  {\bibfnamefont {Y.}~\bibnamefont {Guo}},\ and\ \bibinfo {author}
  {\bibfnamefont {D.-L.}\ \bibnamefont {Feng}},\ }\href
  {https://doi.org/10.1103/PhysRevLett.127.187004} {\bibfield  {journal}
  {\bibinfo  {journal} {Phys. Rev. Lett.}\ }\textbf {\bibinfo {volume} {127}},\
  \bibinfo {pages} {187004} (\bibinfo {year} {2021})}\BibitemShut {NoStop}%
\bibitem [{\citenamefont {Denner}\ \emph {et~al.}(2021)\citenamefont {Denner},
  \citenamefont {Thomale},\ and\ \citenamefont {Neupert}}]{Denner2021}%
  \BibitemOpen
  \bibfield  {author} {\bibinfo {author} {\bibfnamefont {M.~M.}\ \bibnamefont
  {Denner}}, \bibinfo {author} {\bibfnamefont {R.}~\bibnamefont {Thomale}},\
  and\ \bibinfo {author} {\bibfnamefont {T.}~\bibnamefont {Neupert}},\ }\href
  {https://doi.org/10.1103/PhysRevLett.127.217601} {\bibfield  {journal}
  {\bibinfo  {journal} {Phys. Rev. Lett.}\ }\textbf {\bibinfo {volume} {127}},\
  \bibinfo {pages} {217601} (\bibinfo {year} {2021})}\BibitemShut {NoStop}%
\bibitem [{\citenamefont {Miao}\ \emph {et~al.}(2021)\citenamefont {Miao},
  \citenamefont {Li}, \citenamefont {Meier}, \citenamefont {Huon},
  \citenamefont {Lee}, \citenamefont {Said}, \citenamefont {Lei}, \citenamefont
  {Ortiz}, \citenamefont {Wilson}, \citenamefont {Yin}, \citenamefont {Hasan},
  \citenamefont {Wang}, \citenamefont {Tan},\ and\ \citenamefont
  {Yan}}]{Miao2021}%
  \BibitemOpen
  \bibfield  {author} {\bibinfo {author} {\bibfnamefont {H.}~\bibnamefont
  {Miao}}, \bibinfo {author} {\bibfnamefont {H.~X.}\ \bibnamefont {Li}},
  \bibinfo {author} {\bibfnamefont {W.~R.}\ \bibnamefont {Meier}}, \bibinfo
  {author} {\bibfnamefont {A.}~\bibnamefont {Huon}}, \bibinfo {author}
  {\bibfnamefont {H.~N.}\ \bibnamefont {Lee}}, \bibinfo {author} {\bibfnamefont
  {A.}~\bibnamefont {Said}}, \bibinfo {author} {\bibfnamefont {H.~C.}\
  \bibnamefont {Lei}}, \bibinfo {author} {\bibfnamefont {B.~R.}\ \bibnamefont
  {Ortiz}}, \bibinfo {author} {\bibfnamefont {S.~D.}\ \bibnamefont {Wilson}},
  \bibinfo {author} {\bibfnamefont {J.~X.}\ \bibnamefont {Yin}}, \bibinfo
  {author} {\bibfnamefont {M.~Z.}\ \bibnamefont {Hasan}}, \bibinfo {author}
  {\bibfnamefont {Z.}~\bibnamefont {Wang}}, \bibinfo {author} {\bibfnamefont
  {H.}~\bibnamefont {Tan}},\ and\ \bibinfo {author} {\bibfnamefont
  {B.}~\bibnamefont {Yan}},\ }\href
  {https://doi.org/10.1103/PhysRevB.104.195132} {\bibfield  {journal} {\bibinfo
   {journal} {Phys. Rev. B}\ }\textbf {\bibinfo {volume} {104}},\ \bibinfo
  {pages} {195132} (\bibinfo {year} {2021})}\BibitemShut {NoStop}%
\bibitem [{\citenamefont {Feng}\ \emph {et~al.}(2021)\citenamefont {Feng},
  \citenamefont {Jiang}, \citenamefont {Wang},\ and\ \citenamefont
  {Hu}}]{Feng2021}%
  \BibitemOpen
  \bibfield  {author} {\bibinfo {author} {\bibfnamefont {X.}~\bibnamefont
  {Feng}}, \bibinfo {author} {\bibfnamefont {K.}~\bibnamefont {Jiang}},
  \bibinfo {author} {\bibfnamefont {Z.}~\bibnamefont {Wang}},\ and\ \bibinfo
  {author} {\bibfnamefont {J.}~\bibnamefont {Hu}},\ }\href
  {https://doi.org/10.1016/j.scib.2021.04.043} {\bibfield  {journal} {\bibinfo
  {journal} {Sci Bull. Fac. Agric. Kyushu Univ.}\ }\textbf {\bibinfo {volume}
  {66}},\ \bibinfo {pages} {1384} (\bibinfo {year} {2021})}\BibitemShut
  {NoStop}%
\bibitem [{\citenamefont {Li}\ \emph {et~al.}(2022)\citenamefont {Li},
  \citenamefont {Wan}, \citenamefont {Li}, \citenamefont {Li}, \citenamefont
  {Gu}, \citenamefont {Yang}, \citenamefont {Li}, \citenamefont {Wang},
  \citenamefont {Yao},\ and\ \citenamefont {Wen}}]{Li2022}%
  \BibitemOpen
  \bibfield  {author} {\bibinfo {author} {\bibfnamefont {H.}~\bibnamefont
  {Li}}, \bibinfo {author} {\bibfnamefont {S.}~\bibnamefont {Wan}}, \bibinfo
  {author} {\bibfnamefont {H.}~\bibnamefont {Li}}, \bibinfo {author}
  {\bibfnamefont {Q.}~\bibnamefont {Li}}, \bibinfo {author} {\bibfnamefont
  {Q.}~\bibnamefont {Gu}}, \bibinfo {author} {\bibfnamefont {H.}~\bibnamefont
  {Yang}}, \bibinfo {author} {\bibfnamefont {Y.}~\bibnamefont {Li}}, \bibinfo
  {author} {\bibfnamefont {Z.}~\bibnamefont {Wang}}, \bibinfo {author}
  {\bibfnamefont {Y.}~\bibnamefont {Yao}},\ and\ \bibinfo {author}
  {\bibfnamefont {H.-H.}\ \bibnamefont {Wen}},\ }\href
  {https://doi.org/10.1103/PhysRevB.105.045102} {\bibfield  {journal} {\bibinfo
   {journal} {Phys. Rev. B}\ }\textbf {\bibinfo {volume} {105}},\ \bibinfo
  {pages} {045102} (\bibinfo {year} {2022})}\BibitemShut {NoStop}%
\end{thebibliography}%

\end{document}